%% file: phd.tex
\documentclass[11pt]{report}
\usepackage{psfig}
\usepackage{picinpar}
\usepackage{amssymb}

\begin{document}
\newcommand{\real}{\mbox{{\bf I}\kern-.15em{\sf R}}}
\newcommand{\forget}[1]{}
\newcommand{\eq}[1]{\begin{equation} #1 \end{equation}}
\newcommand{\ma}[1]{$ #1 $}

\newcommand{\beq}{\begin{displaymath}}
\newcommand{\eeq}{\end{displaymath}}
\newcommand{\beqn}{\begin{equation}}
\newcommand{\eeqn}{\end{equation}}
\newcommand{\beqa}{\begin{eqnarray}}
\newcommand{\eeqa}{\end{eqnarray}}
\renewcommand{\k}{\kappa}
\newcommand{\pd}{\partial}

\newcommand{\diagram}[2]{
  \begin{minipage}{3cm}
    \vspace{3mm}
    \centerline{\psfig{file=pictures/#1.ps,angle=270,height=#2cm}}
     \vspace{3mm}
  \end{minipage}
}
\newcommand{\ch}[1]{
  \begin{minipage}{2cm}
    \vspace{2mm}
   #1
    \vspace{2mm}
  \end{minipage}
}

\include{title}

\tableofcontents

\include{intro}

\include{hyper}

\include{sqg}

\include{summary}

\include{biblio}
\end{document}

%% file: title.tex
\begin{titlepage}

\vspace*{2cm}

\begin{center}
{\Large \bf Studies in Random Geometries:\\
 Hyper-Cubic Random Surfaces\\ and \\
 Simplicial Quantum Gravity}          

\vspace{2.cm}

{\bf Dissertation}\\
zur Erlangung des Doktorgrades\\
der Fakult\"at f\"ur Physik\\
der Universit\"at Bielefeld

\vspace{2.5cm}
vorgelegt von\\
Sven Bilke\\

\vspace{1.0cm}

November 1997\\

\vspace{1.8cm}
{\bf Abstract}
\end{center}
We analyze two models of random geometries~: planar hyper-cubic random surfaces
and four dimensional simplicial quantum gravity.

We show for the hyper-cubic random
surface model that a geometrical constraint does not change the critical
properties of the model compared to the model without this constraint. We
analyze the phase diagram for the model with extrinsic curvature. 

For four dimensional simplicial quantum gravity we find that in the large
volume limit the leading contribution to the entropy does not depend on the
underlying topology. We find for the first time a strong back-reaction of the 
geometric sector for matter coupled to gravity. 

\end{titlepage}

%% file: intro.tex
\chapter{Introduction}
At the beginning of this century two theories forced a new
understanding of physics. Quantum theory on the one hand describes the physics
on small scales, where "small" scales do not only appear on the atomic
level but also for macroscopic objects, such as white dwarfs or 
super-fluid Helium. On the other hand we have general relativity describing 
large distance physics which is dominated by the 
gravitational field.  

As of today all attempts to unify these formalisms failed. This is very 
unfortunate because with such a theory at hand  it may be possible to 
overcome a weak point  of general relativity: it predicts space-time 
singularities and at the same time breaks down at these singularities. 
In particular,
the Big Bang could have produced anything, as far as classical general
relativity is concerned. A theory of quantum gravity on the other hand 
might be able to produce some boundary conditions that can tell us why the 
universe looks like it does.

It is however very difficult to construct such a theory. The scale in
quantum gravity, where quantum effects become important, is set by
the Planck mass \mbox{$M_p \approx 10 ^{19}$ GeV}. Typical energies in
particle physics experiments are, however, several orders of magnitude smaller,
which is why no experimental results for quantum effects of gravity are 
available. Therefore one has no no guidance from experiments, which, for 
example, in the case of the standard model provided very important insight
necessary to construct the theory. 

String theories are one attempt to unify the fundamental interactions in nature
(see for example \cite{polch94}). 
These theories avoid  the singularities 
which usually plague quantum field theories by considering the propagation of 
strings rather than  point like particles. Differently from point particles 
which  propagate along paths strings propagate along the world-sheet, a two 
dimensional surface. Is is very interesting that string theories contain 
spin $2$ particles, i.e.  gravitons, which makes them a promising candidate 
for quantum gravity. 

General Relativity and string theory have in common that they both are in a 
certain sense geometric theories. General relativity describes gravity as a 
curved  four dimensional space time metric, strings propagate along two 
dimensional surfaces. As we will describe in more detail  below quantum 
gravity and  quantum string theory both can be looked at as  a statistical 
system of geometries, random geometries.  Such systems are, unfortunately,
very difficult to deal with analytically, although some success has been made 
in two dimensions for random surfaces. 
However, when coupled to matter with central 
charge $c \geq 1$ these methods break down. Some observables like the intrinsic
Hausdorff dimension are  not accessible at all by analytic methods.
In higher dimensions  almost nothing is known analytically. 
The motivation of many of the conjectures and assumptions used for higher 
dimensional models is  therefore derived from the experience obtained from the
two-dimensional models.

This work addresses these issues.  We consider two models with different
internal local dimension, two and four.  These models 
are namely  hyper-cubic random surfaces, a discretization of bosonic strings,
and four dimensional simplicial quantum gravity. To strengthen the common
bases for the two models we discuss the subject in a somewhat broader context
in the remaining part of this chapter.  The motivation and 
definition of the models, considered in the numerical part of this work, 
is given in chapters \ref{hyper} and  \ref{sqg_chap}.

\section{Discrete random geometries}
Random geometries have attracted a lot of interest since it was realized that
surfaces and higher dimensional  manifolds are central to the understanding 
of the fundamental interactions in nature. In this thesis we are interested 
mainly in quantum gravity and string theory, but they are also of interest
in other fields like biology \cite{peliti} or condensed matter
physics \cite{nelson94}. All of these theories have in common that they 
describe statistical systems where we have to integrate over geometries.

The thermodynamic properties of these models can be analyzed using standard
methods of the theory of critical phenomena. It is generally believed that 
critical phenomena may play a similarly important role for random geometries 
as they do in statistical field theory. There it is known that close
to a phase transition a certain degree of universality emerges  in the
sense that models with  very different microscopic interaction exhibit the
same critical behavior, described by a set of critical exponents. This is
usually related to some common symmetry. Something similar happens for two 
dimensional quantum gravity coupled to matter with central charge $c < 1$. 
It was shown \cite{KPZ} for this type of models that the entropy exponent 
$\gamma $ depends only on the central charge but not on the type of the 
matter coupled to gravity\footnote{The  definition of the exponent $\gamma $ 
is given in section  \ref{rs}.}. In higher dimensions  
one numerically observes  universal behavior  in the sense that some phases 
of very different models have similar geometrical properties. For example 
both types of models considered in this thesis have a phase where the 
geometry is essentially described by a branched polymer. 

Discretization is a very useful tool in analyzing statistical field theories. 
It reduces the number of degrees of freedom from non-countable  to countable. 
A discrete lattice can serve as a regulator for UV-divergences and also opens 
the possibility for non perturbative approaches, such as numerical 
simulations. Typically, models of random geometries are invariant under 
coordinate transformations. Any regulation of the theory has to obey this 
invariance. Obviously the regulator introduced by discretizing on the
geometric level is reparametrization invariant as this procedure does not
refer to any coordinate system. However, one has to ask the question what
happens to diffeomorphism invariance, the basic symmetry of general relativity,
in a discrete space. When taking the continuum limit of the discretized theory
one has to check that this symmetry is restored, at least on large distances.  

Discretization can also help to define a
measure. When performing the integration over distinct physical
states for the partition function the measure has to be chosen so as not to 
over-count surfaces by counting  different parameterizations of the same 
geometry. For two dimensional quantum gravity it was shown that the critical
behavior of the discrete dynamical triangulation model (a comprehensive
introduction can be found in \cite{gp91}), where the integral over geometries 
involving the measure is replaced by a sum over triangulations, and the 
critical behavior of the continuum theory agree. 

The key ingredient that allows to go back from the discrete model  to the 
continuum is universality. If the discretized model  is in the same 
universality class with the continuous version its  critical properties 
describe continuum physics. The physical reason is that if the correlation 
length in units of lattice spacings diverges, for example at a second order 
phase transition, details of the discretization become unimportant and the 
system exhibits continuum behavior.

\section{Random walk \label{rw}}

A random walk describes the propagation of a point like particle in
$d$ - dimensional space. Despite its relative simplicity it contains many
of the aspects common to other random geometries. As this model  is 
well known also to readers from outside this field of research, it is 
perfect to demonstrate these aspects. For the details of the calculations we
refer to \cite{am_hou}.

In a quantum theory the amplitude $G(x,x')$ for a 
particle to go from $x$ to $x'$ is given by the sum over all possible 
trajectories $P_{xx'}$ connecting these points.  In Euclidean space 
we have
\eq{G(x,x') = \int_{P_{xx'}} {\cal D }P \exp(-S[P_{xx'}]), 
\label{rw_corr}
}
where $S[P_{xx'}]$ is the classical action for the given path. As such we
choose the simplest characterization of the path, its length: 
\eq{ S = m_o \int dl = 
       m_0 \int _0 ^1 d\xi \sqrt{\partial x_{\mu} \partial x_{\mu} }. 
\label{rw_act}
   }
In the right-hand expression we introduced a parametrization  
\eq{x(\xi): [0,1] \rightarrow R^D, x(0) = x, x(1) = x'}
of the path. Expression (\ref{rw_act}) is invariant under  
reparametrizations given by
\eq{ x_{\mu }(\xi) \rightarrow x_{\mu}(\phi(\xi)) \label{rw_reparm}}
with the function $\phi(\xi)$ satisfying the conditions
\eq{\phi(0) = 0, \quad \phi(1) = 1, \quad 
   \frac{\mbox{d}\phi(\xi)}{\mbox{d}\xi} > 0. }
We will show that with the action (\ref{rw_act}) the quantum amplitude 
(\ref{rw_corr}) describes the propagation of a free relativistic particle. 
Other terms in the  action, for example higher derivative terms, can be 
relevant in the continuum as well. A non-trivial term is for example a 
coupling to the external curvature $K$. 
\eq{\epsilon \int dl\; |K|}
For finite coupling $\epsilon $ one
finds the critical behavior of the ordinary random walk discussed below. 
In the limit $\epsilon \rightarrow \infty$ one finds a different class,  
smooth random walks \cite{adj86, adj87}.
 
The expression (\ref{rw_corr}) is rather formal. The measure 
${\cal D} P$ has to be defined in such a way  that the integral counts only the
number of distinct paths, i.e. paths which differ only by 
reparametrization (\ref{rw_reparm}) have to be counted only once. 
In continuum formulations one formally defines this as the number of 
all paths modulo reparametrizations,
\eq
{ {\cal D }P \rightarrow \frac{ {\cal D} P}{{\cal D} f}.}

A way to define the measure and at the same time regularize 
the model is to discretize it. This has to be done carefully, though: 
it is surely wrong to cut the interval $0 \leq \xi \leq 1$ into equally small 
pieces, as this  would spoil reparametrization invariance. Instead of
discretizing a certain parameterization one has to discretize the internal
structure of the path. This amounts to considering only piecewise 
linear paths where each step on the path is of length $a$, i.e.  we 
refrain from discussing structures smaller than $a$. This cut-off is 
invariant by definition.  The possible length of the path is now $l = n a$ 
and the action is $S = m_0 l$. With this discretization the expression 
for the amplitude 
(\ref{rw_corr}) is 
\eq{
G_a(x,x'; m_0) = \sum _{n=1}^{\infty} \exp(-m_0 a n) 
                 \int \prod _{i=0}^n \mbox{d} \vec{e}_i  \delta \left (
                 a \sum \vec{e}_i -(x-x') \right ),
}
where the integral is over unit-vectors in $R^d$. After  Fourier transforming
this and  some algebra one finds
\eq{ G_a(p; m_0) = \frac{1}{1 - \exp(-m_0 a) f(pa) } \label{disc_amp} }
with
\eq {
f(pa) =    \int \mbox{d}\vec{e} \exp(- i p \vec{e}) 
   \approx f(0) (1 - c^2(p a)^2  + \cdots ).
}

To recover the continuum from the discrete description the lattice spacing
has to go to zero: $a \rightarrow 0$. One gets the propagator for the free 
relativistic particle if at the same time the bare mass $m_0$ is renormalized 
such that 
\eq{ \exp(-m_0 a) f(pa) \rightarrow 1 - c^2 m_{ph}^2 a^2, \quad  
\mbox{i.e.} \quad m_0 = \frac{\log(f(0))}{a} + c^2 m_{ph}^2 a,}
where $m_{ph}$ is the physical mass.  
Putting the last three formulas together one gets 
\eq{
G_a(p; m_0) \rightarrow \frac{1}{c^2a^2}\frac{1}{p^2 + m_{ph}^2} 
                =       \frac{1}{c^2a^2} G_c(p; m_{ph})
}
when one collects the leading order $a^2$ terms.
The discrete amplitude $G_a$  in this limit is related to the continuum 
propagator $G_c$ of the free relativistic particle by the divergent
factor $\frac{1}{c^2 a^2}$, which can be interpreted  as a 
wave function renormalization. 

It is interesting to note that these results are quite  universal.
They do not depend on the details of the discretization. For example one finds
the same continuum limit for the random walk on a discrete lattice $Z^D$, 
where one has to sum over unit-vectors of $Z^D$ in (\ref{rw_corr}) rather than
integrating over the unit-sphere. One also finds universality concerning the 
action. Consider for example the covariant action of one dimensional gravity~:
\eq{
S[x,g] = \frac{1}{\alpha} \int _0^1 \mbox{d}\xi \sqrt{g(\xi)} 
         \left [ g^{ab}(\xi) 
             \frac{\partial x^{\mu}}{\partial \xi^a}
             \frac{\partial x^{\mu}}{\partial \xi^b}
             + \mu \right ],
\label{rw_qg_act}
} 
where an independent internal metric "tensor" $g_{ab}, a=b=1$ has been 
introduced. The partition function 
\eq{
\label{rw_qg_path}
Z = \int  \frac{{\cal D}[g]}{{\cal D} f} {\cal D} x e^{-S}[x,g].
}
contains in this case an integration over equivalence classes of metrics. 
The classical equations of motion obtained  from (\ref{rw_qg_act}) by variation
of $g_{ab}$ and $x$ independently agree with those one gets from the action
(\ref{rw_act}). But it is a non-trivial observation that they agree on the
quantum level because in principle the metric field could induce different 
quantum anomalies.

\section{Random surface \label{rs}}
While point particles propagate along path a string sweeps out a surface when
propagating. It is natural to expect that the generalization of the quantum
amplitude (\ref{rw_corr}) to surfaces describes the propagation of a free 
relativistic string if one chooses similar to (\ref{rw_act}) an action
proportional to the area of the surface. Higher derivative terms, like
curvature-squared terms or extrinsic curvature terms may also be relevant in 
a possible continuum limit. It is, however, best to start with a  
simple action and check the relevance of other terms only afterwards.

One such action is the  Nambu-Goto action 
\eq {
S_{NG} [S(l_i)] = \mu \int _{S(l_i)} dA
          = \mu \int _{M(l_i)}\mbox{d}^2\xi \sqrt{
            \left ( \frac{\partial x^{\mu}}{\partial \xi^1} \right )^2 +
            \left ( \frac{\partial x^{\mu}}{\partial \xi^2} \right )^2
            - \left ( \frac{\partial x^{\mu}}{\partial \xi^2}
                      \frac{\partial x^{\mu}}{\partial \xi^1} \right )^2
            }
.\label{nambu_goto_string}
}
The  metric of the surface is completely determined by the 
embedding of the surface $x^{\mu}$ into a $D$-dimensional space. 
The $2$-loop quantum amplitude is
\eq{G(l_1, l_2) = 
\int_{s_{l_1, l_2}} {\cal D S} \exp(-S[S_{l_1, l_2}]),\label{rs_ng_ps}}
where the integration extends over all surfaces $S_{l_1, l_2}$ which have 
$l_1$ and $l_2$ as a boundary. 

The non-polynomial action (\ref{nambu_goto_string}) makes the analytic
treatment of the Nambu-Goto string very difficult. 
Therefore Polyakov suggested  \cite{polyakov_book}  a 
different string model.  He introduced an internal 
metric  $g_{ab},a,b=1,2$ which is independent of the embedding. 
The action for this model is:
\eq{
S[g,x] = \alpha \int _{M(l_i)} \mbox{d}^2\xi \sqrt{g} \left [
          g^{ab}  \frac{\partial x^{\mu}}{\partial \xi^a} 
                  \frac{\partial x^{\mu}}{\partial \xi^b} + \mu \right ]
.\label{polyakov_action}
}
The $2$-loop Greens function now also integrates over the $g_{ab}$ modulo 
reparametrizations~:
\eq{G(l_1,l_2) = \int_{s_ {l_1, l_2}} \frac{{\cal D}g_{ab}}{{\cal D} f}
                         \int {\cal D }_g x \exp(-S[g,x])
.\label{rs_ps_part}
}
The partition function for fixed area $A$ is 
\eq{Z(A) = \int_{s} \frac{{\cal D}g_{ab}}{{\cal D} f}
                         \int {\cal D }_g x e^{-S[g,x]} 
                    \delta \left ( \int\mbox{d}^2\xi\sqrt{g} - A \right ).
\label{canonic_Z}}
We define the entropy exponent $\gamma $ by the asymptotic behavior 
of the partition function for large $A$,
\eq{Z(A) \approx e^{\mu _c A} A^{\gamma - 3}, }
where $\mu _c$ is the critical coupling of the model. 

The classical equations of motion derived from  the actions 
(\ref{nambu_goto_string}) and (\ref{polyakov_action}) agree because in the
saddle point solution the metric $g_{ab}$ takes the form induced by the
embedding. But it is not at all obvious that the quantum theories are 
identical. It has been demonstrated \cite{M90} that the two quantizations are
equivalent in the critical  dimension $D=26$. In lower dimensions it is not
clear that the quantization 'based' on the two actions are indeed the same. 
It is, for example,  not excluded that the $c=1$ barrier, which we discuss 
below, is absent for the model defined by (\ref{rs_ng_ps}).

In defining a string theory using the the Polyakov action  one has to deal 
with the problem of avoiding over-counting of surfaces. The measure 
\eq{\frac{ {\cal D }g_{ab}(\xi)}{{\cal D}f(\xi)}}
is a functional integral over the space of metrics divided by the 
volume of the diffeomorphism group. In the continuum one can deal with this
problem by gauge fixing.
\forget{
To  make use of this formal expression we change to integration over new 
variables $\sigma$ and $f$ such that
\eq{
  \frac{ {\cal D} g_{ab}}{{\cal D} f} \rightarrow 
  \frac{{\cal D}\sigma (\xi ) {\cal D} f }{{\cal D}f} \times (\mbox{Jacobian}) 
  = {\cal D}\sigma (\xi )  \times (\mbox{Jacobian}) 
\label{strat} 
}
is a well-defined expression. The Jacobian will later reappear as
a "ghost" in the model. 
}
In the conformal gauge the metric $g_{ab}$ can be written as
\eq{ g_{ab} = e^{\sigma } \delta _{ab} \label{c_xform} }
with only one parameter $\sigma $. A naive argument which explains this result 
is that the four components of the metric tensor are not independent; a metric
tensor is symmetric, i.e. the off-diagonal elements are identical. 
Reparametrization invariance kills another two degrees of freedom, so one ends
up with only one free parameter. This, however, is a bit too naive because the
conformal gauge is accessible only if the required coordinate transformation 
is non singular. It can be shown \cite{polyakov_book} that on a sphere this 
is always the case. 

The action in (\ref{canonic_Z}) is invariant under the conformal deformation 
$g_{ab}(\xi) \rightarrow e^{\sigma(\xi) } g_{ab}(\xi)$ of the
metric. Therefore the classical action does not depend on $\sigma $ in the
conformal gauge which reflects the triviality of the classical model.
In the quantum theory, however, the measure ${\cal D} _{g} x $
transforms in a non-trivial way. This so-called conformal anomaly brings back 
the $\sigma $ dependence into the the effective action 
\eq{S_l[\phi] = - \frac{D}{48 \pi } 
   \int \mbox{d}^2 z (2 \partial _z \sigma \partial _z \sigma + 
                      \mu ^2 e^{\sigma})
\label{liouville_action},}
which is known as  Liouville action. The gauge-fixing of the measure
${\cal D} g$ introduces a ghost $\frac{26}{D} S_l$. The 
partition function for the Polyakov string in the conformal gauge is~: 
\eq{Z = \int {\cal D} \sigma \exp \left \{ - \frac{26 - D}{48 \pi} 
        \int \mbox{d}^2 z (2 \partial _z \sigma \partial _z \sigma + 
                      \mu ^2 e^{\sigma}) \right \} .}

The $x^{\mu}$ in the Polyakov action (\ref{polyakov_action}) can  be 
interpreted as $D$ scalar fields coupled to the  surface described by the 
metric $g_{ab}$. Note that with this interpretation the case $D=0$, a 
surface embedded into a zero dimensional space, does make sense. In  section 
\ref{intro_quantum} we will show that this case describes pure gravity 
without matter. In \cite{KPZ} it was shown that the entropy 
exponent $\gamma $ for two dimensional quantum gravity coupled to conformal
matter with central charge $c$ is given by
\eq{\gamma = \frac{c - 1 - \sqrt{(1 - c) (25-c)}} {12}. \label{kpz_form}}
The central charge $c$ in a sense counts the number of degrees of freedom of
the conformal field theory. For a free bosonic strings it can be identified
with the dimension of the embedding space.
If one considers only unitary models  with 
$c < 1$ the exponent $\gamma $ takes only a discrete set of  negative values
\eq{\gamma = -\frac{1}{m}, m=2,3,\ldots \label{kpz_series}.} 
Note that the formula (\ref{kpz_form}) is not valid for $c > 1$ because it 
gives  unphysical complex results; in the literature this is often referred to
as the "$c=1$ barrier". A more detailed discussion can be found in section
\ref{hyper_introduction}.

\section{Quantum gravity \label{intro_quantum}}
The classical description of gravity is given by Einsteins theory of general
relativity (for an introduction see for example \cite{L39,we71}). The
space-time metric describes the gravitational field. The field equations 
for the metric field are given by
\eq{
\label{eingl}
R_{ik} - \frac{1}{2}g_{ik} R = \frac{8 \pi G_N}{c^4} T_{ik}.
}
where $G_N = 6.67 10^{-11}\mbox{m}^3 kg^{-1} s^{-2}$ is Newton's constant,
$g_{ik}$ the metric tensor and $T_{ik}$ the energy-momentum tensor of a 
matter field coupled to gravity.  The Ricci tensor $R_{ik}$ and the 
scalar curvature  $R = g_{ik}R^{ik}$ are functions of the metric tensor and
contain first and second derivatives. The dynamical variable of the theory 
is the metric of space time. 

The classical equations of motion for this metric  can be derived 
using the principle of least action, from the classical action. 
It is generally believed that it is possible to define a quantum theory of
gravity using the path integral formalism.  One assumes that not only the
classical trajectory  but all trajectories weighted by
$e^{-S[g]}$ contribute to quantum mechanical transition probabilities,
giving
\eq{
<g^{(f)} | g^{(i)}> \; \propto \int_{g^{(i)}\rightarrow g^{(f)}}
                             \frac{{\cal D}[g]}{{\cal D} f} e^{-S[g]}
.\label{qg_corr}}
The integral is over all $D$-dimensional Euclidean space time metrics 
which satisfy the boundary conditions $g^{(f)}, g^{(i)}$. As usual in quantum
field theory one has to switch to the Euclidean formulation. Unlike for
conventional field theories it is still an open issue whether the Minkowski
formulation can be recovered by a Wick rotation \cite{hh83}.
\forget{The reason why 
Euclidean rather than a Minkowski metrics are used is mainly the lack of
understanding of how to define the theory with the Minkowski metric.
One hopes \cite{hh83} that gravity in  Minkowski space can be obtained
from the Euclidean formulation by a Wick rotation.}
In $D \le 4$ dimensions the action in Euclidean space has the form
\eq{
S=\frac{1}{G_N}\int\mbox{d}^D \xi \sqrt{|g|} (2 \Lambda - R)  + S_{Matter},
\label{qg_cact}
} where we have introduced the cosmological constant $\Lambda $.

The quantum amplitude (\ref{qg_corr}) describes a system of random geometries.
We define  the partition function of quantum gravity by
\eq{
\label{qg_path}
Z = \int  \frac{{\cal D}[g]}{{\cal D} f} e^{-S[g]}.
}
The case $D=1$ has already been discussed in the context of random walks
(\ref{rw_qg_path}).The distance $l = \int \sqrt{|g|} d\xi$ is  
an invariant internal characteristic of the metric, but  we cannot define an 
internal curvature. Therefore the scalar curvature $R$ term in 
(\ref{qg_cact}) drops out. Comparing  with the action (\ref{rw_qg_act})
 we see that the two expressions differ by the term
\eq{
S[x,g] = \frac{1}{\alpha} \int _0^1 \mbox{d}\xi \sqrt{g(\xi)} 
         \left [ g^{ab}(\xi) 
             \frac{\partial x^{\mu}}{\partial \xi^a}
             \frac{\partial x^{\mu}}{\partial \xi^b}
             \right ],
\label{qg_cont_act}
}
which describes the embedding of the random walk in a $\Delta $-dimensional 
space. For gravity is does not make sense to interpret the 
$x^{\mu}, \mu = 1,\ldots, \Delta $ as an embedding, but one can interpret this
term as the contribution of $\Delta$ scalar fields. The case $\Delta = 0$,
i.e. the random walk embedded in a zero dimensional space, describes 
one dimensional quantum gravity without matter fields.

For $D=2$ quantum gravity is described by the Polyakov string
(\ref{rs_ps_part}) embedded in zero dimensions. Note that for surfaces
the curvature $R$ is a local invariant, so we have a new degree of freedom
compared to the one dimensional model. On the other hand it is known from the 
Gauss-Bonnet theorem, 
\eq
{
\oint\mbox{d}^2 \xi \sqrt{|g|} R = 4 \pi \chi_E,
\label{gauss_bonnet}
}
that the integrated curvature is a topological invariant with
$\chi _E=2-2h$, where $h$ is the number of handles in the surface.
The Euler character is $\chi _E = 2$ for the sphere, $\chi _E = 0$ 
for a torus and so on. For fixed topology the curvature term 
in the action (\ref{qg_cact})  gives therefore no dynamical contribution,
which makes the classical model trivial. One might suspect that
this is also the case for the quantum model. In fact the two dimensional
model does not contain gravitons,   however the conformal quantum-anomaly  
generates the non-trivial behavior discussed in the previous section. 

Quantum gravity in physical four dimensions is one of the models considered in
the numerical part of this work. Therefore we discuss this case in detail in
chapter \ref{sqg_chap}. 
Here we only want to mention that for the time being no 
continuum solution for the model is known. The discrete approach, which 
is very successful in two dimensions can be generalized to four 
dimensions. With this method at hand numerical simulations allow us to survey
properties of quantum gravity in this interesting case. 

\section{Matrix models}\label{matrix}

\begin{figure}[htbp]
\centerline{{\psfig{file=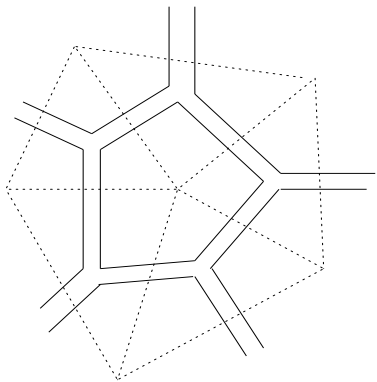,angle=270,height=5cm,width=5cm}}}
\caption{A part of a $M^3$-Graph. The dotted line depicts the triangulation 
         dual to the graph. }
\label{dualt}
\end{figure}
In dynamical triangulations \cite{adf85, d85, k85}, the path integral 
(\ref{qg_path}) over two
dimensional equivalence classes of metrics is replaced by a sum over 
triangulations $T$~:
\eq{
\int \frac{{\cal} D g_{ab}}{{\cal D} f} \rightarrow \sum _T \frac{1}{\sigma(T)}
\label{dt_meas}
}
The symmetry factor $\sigma (T)$ is a discrete
remnant of the reparametrization group. It is mainly important for small 
triangulations because typically larger triangulations are not
symmetric. A triangulation is a surface obtained by gluing together 
equilateral triangles along their edges so that exactly two triangles 
meet on each edge with possible exceptions on the boundary. As in the 
continuum we choose the action to be proportional to the area. In the 
discrete model the area is  given by the number $n$ of triangles in the 
triangulation. 

In the partition function
\eq{Z(\mu,G) = \frac{1}{\sigma} \sum _{\cal T} \exp (-\mu n)  =  
        \sum _\chi \exp \frac{\chi}{G} \sum _n\exp (-\mu n) {\cal N}_\chi^{(c)}(n).
\label{dyty}
}
the topology is not a priori fixed. However the number of graphs with free
topology grows factorially with $n$ and therefore the naive topological 
expansion in the number $h$ of holes in the surface is as stands ill defined.
Therefore calculations and numerical simulations are in most cases done for 
ensembles with fixed topology. At the end of this section we discuss the 
double scaling limit, which can be used to define the model with coherent 
contributions from all topologies.

The dynamical triangulations model can be formulated as a matrix model, 
defined by the generating 
function
\eq{
\label{zmatrix}
\overline{Z} = \log \int dM \exp(-Tr M^2 + \frac{g}{\sqrt{N}} Tr M^3),
} where $M$ is a  $N\times N$ hermitian matrix and $dM$ the flat
measure
\eq{
dM = \prod _{i=1}^{N}dM_{ii} \prod _{1\leq i<j\leq N}
  d(Re M_{ij}) d(Im M_{ij}).
}
Doing perturbative expansions in $g$ one gets Feynman diagrams similar to
those depicted in figure \ref{dualt}. Note that the dual graph, drawn in
dotted lines, is a triangulation. Because (\ref{zmatrix}) generates all
possible graphs it  sums over all triangulations. It should be
emphasized that the results presented below do not rely on using 
triangulations. In \cite{k89} it was shown that the continuum limit 
for the model with a lattice built from squares, pentagons, etc. agrees 
with the result obtained for triangulated surfaces.

Differently  from a standard Feynman graph, the propagator 
\beq
<M_{ij} M^*_{kl}> = \delta _{il}\delta _{jk} = \ 
\begin{minipage}{4cm}
\centerline{{\psfig{file=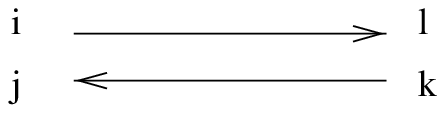,angle=270,height=1cm,width=4cm}}}
\end{minipage}
\eeq
carries two indices. This is necessary to interpret the graph as a two
dimensional surface; a priory Feynman graphs do not have any "dimension". 
Moreover, (\ref{zmatrix}) generates orientable surfaces 
because  $M$ is hermitian.
Due to the Wick theorem one gets the $n$th term of the perturbative expansion
by summing over all diagrams with $n$ vertices, where each vertex has a weight
\eq{
\frac{g}{\sqrt{N}} \Phi _{ij}\Phi _{jk} \Phi _{ki} = \  
\begin{minipage}{4cm}
\centerline{{\psfig{file=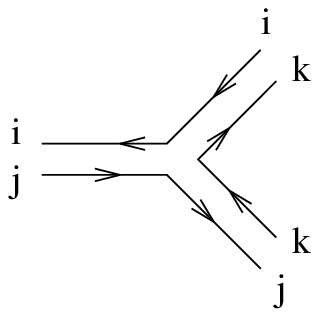,angle=180,height=3cm,width=3cm}}}
\end{minipage}
.}
A closed index loop contributes a factor $N$ because the trace sums over the 
indices of the matrix. A diagram  $D$ with $V=n$ vertices, $L$ closed  
index-loops and $P$ propagators therefore contributes
\eq{
\label{beitrag}
\frac{1} {\sigma (D)}(g N^{-1/2})^V N^L = g^V \frac{1}{\sigma (D)} N^{\chi}, 
} where $\sigma$ is the symmetry factor of the diagram. 
The expression on the right hand side is obtained by noting  
$3V=2P$ and using the  Euler theorem $V+P-L = 2-2h = \chi$, where $\chi $ is
the Euler character of the surface, which uniquely identifies the topology of
the graph. 

In the diagrammatic expansion the partition function of the one matrix model 
reads
\eq{
\overline{Z}(g,N) = \sum _hN^{\chi}\sum_ng^n\sum _{D\in{\cal D}_{h,n}} 
                         \frac{1}{\sigma(D)} = 
                    \sum _h N^{\chi }\sum_n g^n {\cal N}_\chi^{(c)}(n).
\label{matrix_topol_exp}
}
The number ${\cal N}_\chi^{(c)}(n)$ of graphs  can be calculated 
with the method of orthonormal  polynomials \cite{m81}, one
finds 
\eq{ {\cal  N}_\chi^{(c)}(n) \propto n^{\gamma _\chi -3} g_c^{-n}. 
\label{n_conf}}
The entropy exponent  
\eq{\gamma _\chi - 2=   \bar{c} \chi, \quad \bar{c} = -\frac{5}{4}
\label{gamma_matrix} } 
is linear in $\chi$.  The constant  $g_c$, the critical coupling, does not 
depend on the topology. 

In the limit $N \rightarrow \infty$, the so-called planar limit, only 
diagrams with spherical topology $\chi=2$ contribute to the sum over 
topologies in the partition function ({\ref{matrix_topol_exp}}). 
Contributions from higher genus diagrams are suppressed by $N^{-2h}$. 
On the other hand these contributions  are enhanced 
as $g \rightarrow g_c$, which can be seen from the asymptotic behavior of 
the partition function with fixed topology~:
\eq{
Z_{\chi}(g) = \sum _n g^n {\cal N}_{\chi}^{(c)}(n) 
            \approx (g - g_c) ^{2 - \bar{c} \chi}
}
This suggests that if the limits $N \rightarrow \infty$ and
$g \rightarrow g_c$ are taken together in a correlated way the large $N$ genus
suppression can be compensated by the $g \rightarrow g_c$ enhancement.
This would result in a coherent contribution from all genus surfaces
\cite{do1, do2, do3}. This so-called double scaling limit is defined by
taking the limits $N \rightarrow \infty, g \rightarrow g_c$ while holding 
fixed the the 'renormalized' string coupling 
\eq{\kappa ^{-1} = N (g - g_c)^{-\bar{c}}.}

By comparison of (\ref{matrix_topol_exp}) with the partition function 
of dynamical triangulations (\ref{dyty}) we find the relations
\eq{
N = \exp \frac{1}{G}, \quad g = \exp (-\mu )
}
for the coupling constants.
For spherical topologies we can compare the exponent $\gamma _2$ with the the 
continuum results (\ref{kpz_form}). The value $\gamma_2 = -\frac{1}{2}$ 
obtained from (\ref{gamma_matrix}) agrees with the value 
$\gamma = -\frac{1}{2}$ obtained in the continuum for pure gravity, 
i.e. $c=0$. Some variants of the matrix model
discussed here can be used to describe discrete gravity coupled to matter. 
For example a particular two  matrix model describes the Ising model 
($c = \frac{1}{2}$) coupled to gravity \cite{kk86}. One finds 
$\gamma = - \frac{1}{3}$  in this case which is in agreement with the 
results obtained in the continuum.

\section{Branched Polymers}
\begin{figure}[t]
\hspace{1.5cm}
\psfig{file=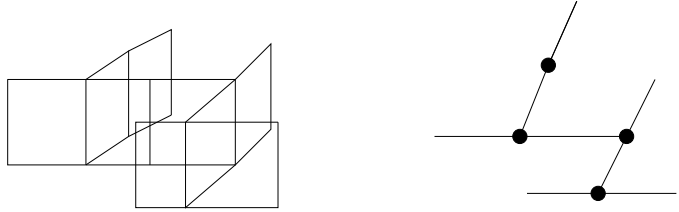,angle=270,height=2.8cm}
\caption{A part of a hyper-cubic surface and the representation of the surface
as a branched polymer}
\label{bp1}
\end{figure}
\begin{figure}[h]
\hspace{0.3cm}
\psfig{file=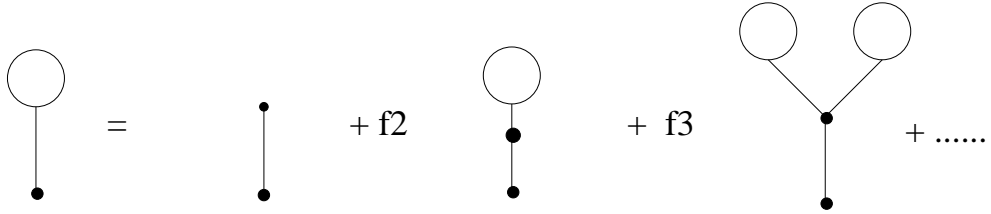,angle=270,height=2.8cm}
\caption{The self-consistent equation for rooted trees.}
\label{bp2}
\end{figure}

Many models of random geometries, for example the two types of models 
considered in this thesis,  have a phase which can essentially be
described by branched polymers. Following \cite{am_hou} we want to derive the 
result $\gamma = 1/2$ for the entropy exponent because we frequently refer 
to this result. As an example for a branched polymer structure we show 
in figure \ref{bp1} a small hyper-cubic surface, which can be decomposed 
into small components by cutting open loops of length two. The individual 
components are not critical, their size is of the order of the lattice 
spacing.  The only way the number of plaquettes can grow to infinity is by 
successive gluing of branches, all the dynamics lies in this gluing. This 
dynamics is described by the branched polymer model~: in the right hand side 
of figure \ref{bp1} the individual components are represented as links $l_i$ 
with an associated  chemical potential $\mu $. We also introduce a weight 
$f_n$ for the joining of $n$ links $l_i$ at a vertex $v_n$. The partition 
function for the branched polymer model is
\eq{ Z(\mu) = \sum _{\mbox{BP}} e ^{- \mu N } \prod _{v} f_n(v), } 
where the sum runs over all branched polymers, i.e. tree graphs if we 
restrict ourselves to spherical topology. To extract $\gamma $ we consider the 
one-point function
\eq{G(\mu ) \approx c (\mu - \mu _c) ^{1 - \gamma}, \label{bp_one}}
the amplitude of rooted trees with one marked vertex. Such a tree can be 
built recursively as seen in figure \ref{bp2}, where the
blobs represent a rooted tree. This can be used to write down 
the self-consistent equation \cite{adf86} 
\eq{
G_{\mu} = e^{-\mu} \left ( 1 + f_2 G _{\mu} + f_3 G_{\mu}^2 + \ldots \right )
.}
Solving for  $e^{\mu}$ gives~:
\eq{
e^{\mu} = \frac   {1 + f_2 G_{\mu} + f_3 G_{\mu}^2 + \ldots}
                  {G_{\mu}}
        = F(G_{\mu})
\label{konse}
}
We conclude that the lowest value  $\mu _c$ of $\mu $ for which eq.
(\ref{konse}) has a solution is the minimum of $F(G_{\mu})$ if the weights
$f_n$ are positive~: 
\eq{ \frac{d}{dG_{\mu}}  F(G_{\mu})| _{\mu = \mu _c} = 0}
Therefore the linear term drops out if one expands $(\mu - \mu _c)$ in 
$G_{\mu}$~:
\eq{\mu - \mu _c \approx (G_{\mu _c} - G_{\mu})^2 \quad \mbox{i.e. }\quad
    G_{\mu} \approx G_{\mu _c} - c \sqrt{\mu - \mu _c}.
}  
Comparing with eq. (\ref{bp_one}) one finds $\gamma = \frac{1}{2}$.

\section{Monte Carlo methods\label{intro_algo}}
The algorithms used to simulate the two different models, hyper-cubic random
surfaces and four dimensional simplicial quantum gravity, are both based on
the same numerical technique, namely Monte Carlo methods.  We give a
detailed description of the algorithms separately for each model 
in section \ref{hyper_algorithm} and  \ref{sqg_algorithm}. Here we want to
summarize briefly some well-known facts about Monte Carlo algorithms which we
use as a foundation for the more detailed discussion. 

Despite the apparent simplicity of a discrete partition function in the 
form
\eq{ Z = \sum _{\Gamma} \exp ( -S[\sigma] ) }
it is only for a few simple systems possible to actually perform the 
summation. Even with a computer it is only for very small systems possible
to sum over {\em all} states $\sigma \in \Gamma$ because the number of states 
${\cal N}_\sigma (V)$ in general grows exponentially with the volume. However,
only comparatively few states give a significant contribution to the partition
function. The overwhelming majority of the $\sigma $ are either entropically 
suppressed or damped by the Boltzmann factor $\exp( - S[\sigma])$. Monte Carlo
algorithms are a  method to sample the important contributions
by choosing at random $n$ samples $\sigma _i \in \Gamma$ with the
probability distribution
\eq{\pi (\sigma _i) \propto \exp -S[\sigma _i].}
This is the thermodynamical equilibrium distribution. One can therefore 
approximate observables by their expectation value
\eq{<{\cal O}> = \frac{\sum _{\{\sigma _i\}} {\cal O}(\sigma _i)}{n}.}
It follows from the central limit theorem that  the error depends on the number
$n$ of measurements as $\sigma({\cal O}) \propto 1 / \sqrt{n}$ if the 
$\sigma _i$ are independent.

A problem is of course how to choose the $\sigma _i$ at random from the state
space $\Gamma $. Dynamical Monte Carlo methods simulate a Markov process which
generates a sequence of $\sigma _i$. The main ingredient of the Markov chain is
a set of  transformation
\eq{\sigma _i \rightarrow \sigma _{i+1} \label{intro_mc_transform}}
and a probability matrix $({\cal P}(\sigma _a, \sigma _b))_{ab}$.
One can show that the 
$n$-step transition probability ${\cal P}^{(n)}(\sigma _a, \sigma _b)$
converges to a equilibrium distribution 
\eq{\lim _{t \rightarrow\infty} {\cal P}^{(t)} \rightarrow \pi (\sigma _b)}
if the following conditions are satisfied: 
\begin{itemize}
\item {\bf Ergodicity} For each pair $\sigma _1, \sigma _2 \in \Gamma$ 
exists a sequence of transformations which as a whole transform 
$\sigma _1$ into $\sigma _2$. 
\item {\bf Stationarity } The following equation holds for the components
of the transition matrix and for all $\sigma _b \in \Gamma$:
\eq{\sum _{\sigma _a} \pi(\sigma _a) {\cal P}(\sigma _a, \sigma _b) = \pi(\sigma _b) 
\label{stat}} 
\end{itemize}

The solutions of the detailed balance equation
\eq{\pi(\sigma _a) {\cal P}(\sigma _a, \sigma _b) = 
     \pi(\sigma _b) {\cal P}(\sigma _b, \sigma _a) \label{detailed_balance}}
satisfy equation (\ref{stat}). This condition is stronger than (\ref{stat})
but  practically more convenient.  In praxis the
following solutions of this equation are important. With the heat bath weight
\eq{ {\cal P}(\sigma _a, \sigma _b) = \pi(\sigma _b)}
the transition probability to the new state $\sigma _b$ is independent from
the old state $\sigma _a$. The Metropolis weight accepts all transformation for
which $\pi(b) > \pi(a)$. The weight for the inverse transformation can be
deduced from (\ref{detailed_balance}). The shorthand notation for both
directions 
\eq{{\cal P}(\sigma _a, \sigma _b) = \mbox{min} 
  \left \{ 1, \frac{\pi(\sigma _b)}{\pi (\sigma _a)} \right \}
\label{metropolis}. }
is especially useful for a computer implementation.

%% file: hyper.tex
\chapter{Hyper-cubic random surfaces\label{hyper}}

\section{Introduction \label{hyper_introduction}}
The Nambu-Goto action (\ref{nambu_goto_string})  describes a bosonic string 
embedded in a $D$ dimensional space $\cal R$. The world-sheet metric is, 
different from a theory of quantum gravity using the Polyakov string,
completely determined by the embedding 
of the string in $\cal R$. One can therefore attempt
to discretize the model by discretizing the embedding space. 
This idea was used long ago in \cite{dw80} to define a model of 
hyper-cubic random surfaces.  However, in \cite{dfj84} it was 
shown that this discretized model is trivial because the surfaces which 
dominate the partition function have the structure of branched polymers, 
basically one dimensional highly branched thin tubes. It was generally 
believed on grounds of renormalization group arguments that one cannot cure 
this behavior by adding local terms to the action. It therefore came as a 
surprise when some years later  numerical results for a slightly modified 
model with a local 
constraint indicated non-trivial behavior \cite{bb86}. In $D=4$ the 
anomalous scaling dimension $\eta \approx 1$ was measured to be different 
from zero and for the  entropy exponent $\gamma \approx 0.25$ was obtained. 

In today's language of random surface theory one can interpret this model as a
random surface model coupled to matter with central charge $c = D > 1$. 
The KPZ-formula (\ref{kpz_form}) classifies $2d$-gravity coupled to matter 
with $c \leq 1$. The $c > 1$ regime however, where (\ref{kpz_form}) gives an
unphysical imaginary exponent $\gamma $, is still not completely understood.
Discretized models of $2d$-gravity are, on the other hand, well defined for
$c>1$. Numerical simulations clearly indicate  branched polymer 
behavior  for $c \lessapprox 5$, although in the regime 
$1 < c \gtrapprox 5$ the numerical evidence is not so obvious.  
They seem to indicate a smooth cross-over to the branched polymer 
phase as $c$ increases.  Based on the numerical evidence one might
 conclude that there is a different critical behavior for 
$1 < c \lessapprox 5$, although it has been argued \cite{david}, 
based on a renormalization group analysis, that the universal behavior for 
$c > 1$ models is always a branched polymer structure.
This behavior, however,  is masked by the presence of the 
$c=1$ fixed point, which becomes complex for $c>1$. This would lead to 
exponentially enhanced finite size effects.  A recent numerical work 
\cite{bg97} supports this conjecture. 

In view of this  result $\gamma \approx 0.25$ for the modified hyper-cubic 
random surface model, considered in \cite{bb86} is, if true,  very
interesting \cite{a94}. It would provide a very simple model with  a 
non-trivial value, besides the mean field result $\gamma = \frac{1}{3}$,
of the series 
    \eq{\gamma = \frac{1}{n}, n = 2,\cdots \label{hyper_durhuus}} 
of positive $\gamma$ values discussed by Durhuus \cite{du94}\footnote{see
also section \ref{ext_intro}}. 
However one may ask how a local constraint, which disappears after one 
coarse graining step in a renormalization group analysis, could influence 
the large scale properties of the model. On the other hand, it is just the
fact that the coarse grained surfaces in general does not belong to the
constrained class of surfaces which invalidates the triviality proof 
\cite{dfj84} for the unrestricted model. In addition a different universal 
behavior would indicate the breakdown of universality because in a sense the
constraint is nothing but a detail of the discretization.

A mechanism which might induce non-trivial behavior into the model  is
suggested by the renormalization  argument used to derive the series 
(\ref{hyper_durhuus}). In this argument the  surface is decomposed into 
shortest loop irreducible components, {\em i.e.} components which cannot be
cut into two parts along a (model dependent) shortest possible loop.
For systems with $ n > 2$ the entropy of the surface is still dominated by 
branching. But differently from branched polymers the components of the 
surface which form the branches are themselves 
critical with $\bar{\gamma} < 0$.  Such a behavior is induced by 
matter field(s) with the couplings tuned to their critical values. 
The local constraint introduced into the model (\ref{restrict}) can be 
written as an additional term in the action (\ref{soft_restrict}) as a
coupling to a external curvature. For the (actually infinite) coupling, 
introduced by the constraint, the model could be critical and above scenario 
could apply.

We will address the question if a local constraint induces really a 
different critical behavior. We furthermore generalize the unrestricted 
hyper-cubic random surface model by adding to the action a term coupled 
to the external curvature of the  surface. This model allows an interpolation 
between the  \cite{dfj84} unconstrained  and the 
potentially non-trivial \cite{bb86} constrained version of the model. 
In a  optimistic scenario one may hope that by tuning the
couplings one may induce a non-trivial phase structure with a weak coupling
branched polymer phase and a strong coupling flat phase. At the phase 
transition the model might have a another value of $\gamma $, just as in the 
scenario described by Durhuus. 

\pagebreak

\section{The model}
\begin{figwindow}[2,r,{\hspace{5mm}\psfig{file=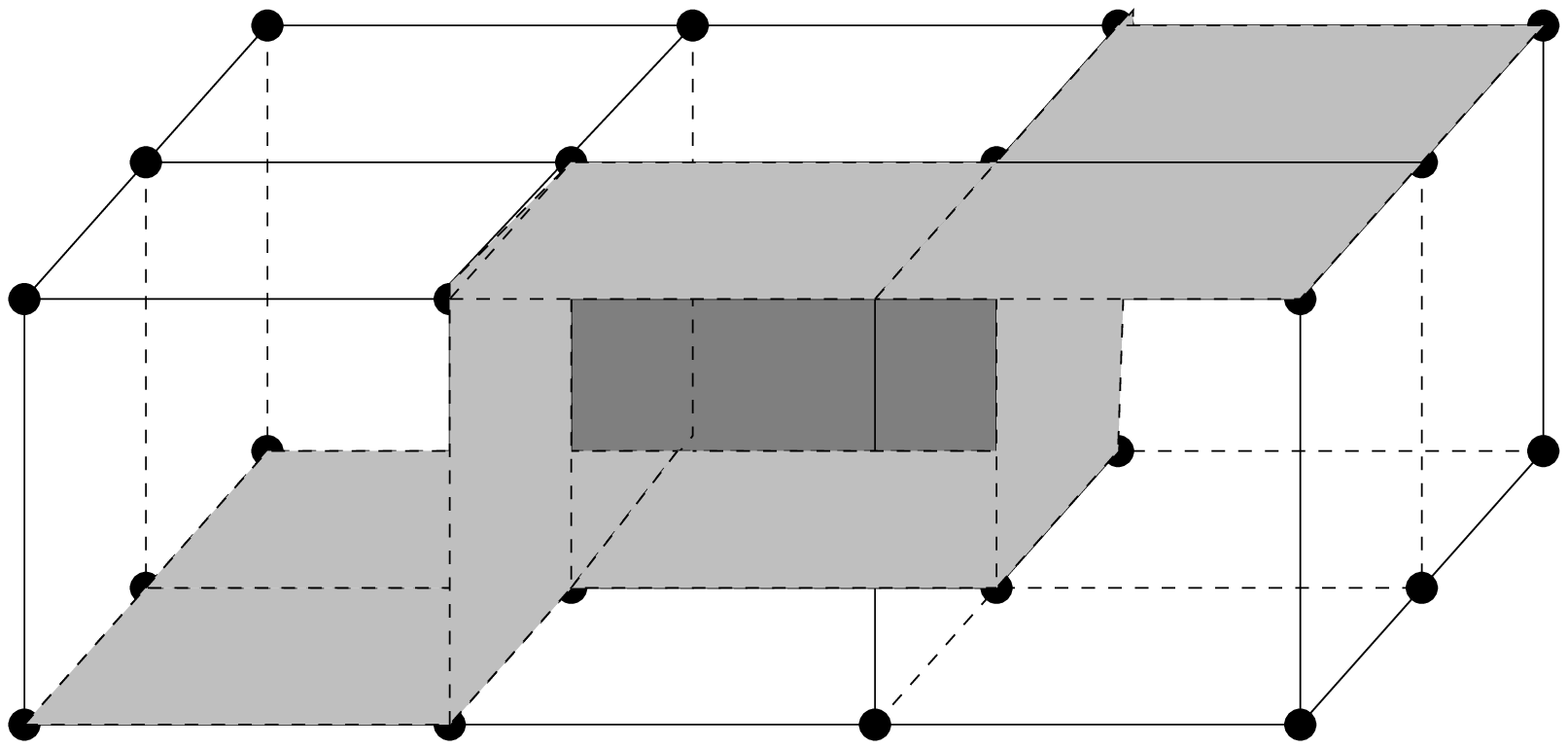,angle=270,height=3.5cm,width=4.5cm,rwidth=5.5cm}},{A part of a hyper-cubic surface in D=3 dimensions}]
A hyper-cubic surface is an orientable surface 
embedded in $Z^D$, obtained by gluing pairwise together plaquettes $P_{x,y}$ 
along their links until no free link is left. A plaquette $P_{x,y}$ is a 
unit-square which occupies one of the unit-squares 
$S_{z^{\mu}}, \mu = 1 \cdots D$  in the embedding lattice. 
In the following we  use the word square to refer to the embedding lattice 
and the words link or plaquette to describe the internal
connectivity of the surface. A link is shared by exactly two plaquettes, which
are glued together along each one of their edges. Note that a link or a 
square in the embedding lattice can be occupied  more than once. In other 
words, the surface is not self avoiding; touching and self intersection are
allowed.  
\end{figwindow}

The  hyper-cubic random surface model is defined by the 
partition function:
\eq{
{\cal Z}(\beta)= \sum _{E \in \cal S} e^{- S} 
        = \sum _A e^{-S} {\cal N}(A), 
\label{hyper_Z}
}
where the sum runs over an ensemble $\cal S$ of hyper-cubic surfaces 
with spherical topology. In the large volume limit, the number ${\cal N}(A)$
of surfaces  with a given area $A$ is expected to grow like 
\eq{ 
{\cal N}(A) \approx e^{\mu _{c} A} A ^{\gamma - 3}, 
\label{hyper_largeN}
}
where $\gamma $ is the entropy exponent. Let us for the moment consider 
the action with only one term:
\eq{S(\mu ) = \mu A \label{hyper_area_action}.}  
The partition function (\ref{hyper_Z}) is well defined 
only if the coupling $\mu$ is larger than the critical value $\mu _c$. 
The continuum limit is obtained by taking the lattice spacing $a$ to zero 
and at the same time $\mu \rightarrow \mu_c^+$. 
\forget{At this point the average 
number $<A>$ of plaquettes diverges and with  $a\rightarrow 0$ the 
discrete surface can smoothly approximate a continuous surface. }

There are three possible ways to embed two neighbor plaquettes $p_1, p_2$
in a hyper-cubic lattice $Z^D$. The external angle $\Theta _l$, which we
assign to link $l$, can therefore take three possible values:
\begin{itemize}
\item[0: \hspace{3mm}] 
             $p_1$ and $p_2$ occupy two neighboring squares in the same 
             coordinate plane,
\item[$\frac{\pi}{2}$: \hspace{3mm}]  
             $p_1$ and $p_2$ occupy two neighbor squares in 
             different coordinate planes,
\item[$\pi$: \hspace{3mm}] 
             $p_1$ and $p_2$ occupy the same square. We call such a
             configuration self-bending. 
\end{itemize}

The model defined so far  describes what we
called the unconstrained model. The constrained model considers the 
ensemble of hyper-cubic surfaces 
\eq {
{\cal S}' = \{ S \in {\cal S} | \mbox{$S$ does not contain self bendings} \}
\label{restrict}.
}
This restriction can be formulated by introducing a potential 
\eq{
S^{(1)}_{ext}[E](\epsilon) = \epsilon 
        \sum _{l \in E} \delta_{\pi, \Theta _l} 
\label{soft_restrict}
}
to the action with an infinite coupling $\epsilon $. The angle $\Theta _l$ is
the angle of two plaquettes, which have the edge $l$ in common, measured in the
embedding space. Therefore this term couples to the external curvature of the
surface.

It seems a bit artificial that this potential
only contains  contributions from links $l$ with $\Theta _l = \pi$. 
We therefore also use the potential
\eq{
S^{(2)}_{ext}[E](\epsilon)  
      =  \frac{\epsilon}{2} \sum _{l \in E} \left ( 1 - \cos \Theta _l \right ) 
      =  \epsilon \sum _{l \in E} \left ( \delta _{\pi, \Theta _l} + 
                      \frac{1}{2} \delta _{\pi / 2 , \Theta _l} \right )
\label{ext_curv}.}
In summary: The action of our model is
\eq{S[E](\mu, \epsilon) = \mu A + S^{(i)}_{ext}[E], \quad i = 0 \ldots 2
\label{hyper_action}}
where 
\eq{S^{(0)}_{ext}(\epsilon) = 0} formally denotes the action without a 
curvature term. 
The canonical partition function is
\eq{
Z(A, \epsilon) = \sum _{E \in {\cal S}_A} e^{-S[E](A,\epsilon)} 
\label{hyper_z},
} where the action $S$ contains one of the external curvature terms discussed
above. The sum runs over hyper-cubic- surfaces with fixed area $A$.  
\forget{
Observables are  defined by
\eq{<{\cal O}> = Z^{-1}  \sum _{E \in \cal S} e^{- S_A} 
        = \sum _A e^{-S_A} {\cal N}(A).}
}

\clearpage

\section{Universality}
In this section answer the question of whether  the constrained 
hyper-cubic random surface model has a critical behavior different from the
unrestricted model  as indicated by earlier numerical results \cite{bb86} 
in four embedding dimensions. 
With the algorithms and the improved computer hardware available today we 
can simulate systems up to two orders of magnitude larger then those used in 
earlier works.

The model is defined by the partition function (\ref{hyper_Z}) where the
summation runs over the constrained ensemble (\ref{restrict}). The action
(\ref{hyper_action}, $i=0$) contains no external curvature term. We measure 
the universal entropy exponent $\gamma $ defined in (\ref{hyper_largeN}),
which  is a good indicator for different phases of the model. For branched polymers 
it takes the value $\gamma = \frac{1}{2}$. A different value indicates a 
different geometry. This will become clearer in section \ref{shake_it}
where we use a geometrical decomposition to measure the exponent $\gamma $.

\subsection{Old method}
As the  first step we want to make sure that our program, which uses an
algorithm different from the one used in the earlier works, gives the same
answers if one uses a similar numerical setup. We  simulate the
grand canonical ensemble to extract the distribution $N_{\mu}(A)$ of 
surface areas measured with coupling $\mu$. Taking  into
account the Boltzmann factor one can fit $N_{\mu}(A)$  to the theoretical
distribution ${\cal N}(A)$ eq. (\ref{hyper_largeN}). Besides an uninteresting
normalization constant one gets an estimate for $\gamma$ and  
$t = \mu_c - \mu$. However, one has to fine-tune $\mu $ very precisely so that
$t = 0$, because for large $(t A)$ the distribution is dominated by the leading
exponential factor and therefore the sub-leading behavior is difficult to observe in
this case. We have  used this method anyhow for small surfaces and got 
a value for $\gamma $ which is compatible with the results presented below. 

In \cite{bb86} a clever way to avoid this problem is used. The unknown 
parameter $\mu _c$ can be absorbed into a constant if one divides  two
histograms $N(A_1), N(A_2)$ with $\delta = A_1 - A_2$ kept constant~:  
\eq{
\frac{N^{\mu _{1}}(A_{1})}{N^{\mu _{2} }(A_{2})}
= \frac{e^{(\mu _{c} - \mu_1) A_1}}{e^{(\mu _{c} - \mu_2) A_2}}
  \frac{ A_1 ^{\gamma - 3}}{ A_2 ^{\gamma - 3}}
= c_{\mu _{1}}^{\mu _{2}}  e^{(\mu _{2} - \mu _{1}) A _{1}}
\left ( \frac{A_{1}}{A_{2}} \right ) ^{\gamma - 3}.
\label{bb_gamma_mamma}
}
In this case the  factor
\eq{
c_{\mu _{1}}^{\mu _{2}} = e^{(\mu _c - \mu _2) \delta}
} is just a constant. 

\begin{table}
\begin{displaymath}
\begin{array}{c|c|c} \hline
               &  A_{max} = 300        &   A_{max} = 5000\\ \hline
\mu   = 1.60   &                       &   3 \times 10^{10}\\
\mu   = 1.65   &  1 \times 10^{10}     &   3 \times 10^{10}\\
\mu   = 1.70   &  2 \times 10^{10}     &   3 \times 10^{10}\\
\mu   = 1.75   &  1 \times 10^{10}     &   \\ \hline
\end{array}
\end{displaymath}
\caption{Monte Carlo statistics in  iterations collected for parameters 
$A_{\mbox{max}}$ and $\mu$}
\label{hyper_bb_statistic}
\end{table}

\begin{figure}[t]
\hspace{1.5cm}
\psfig{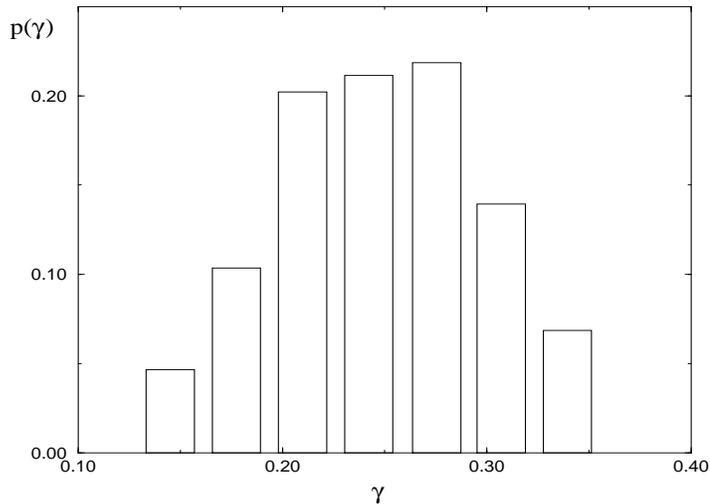}
\caption{The distribution of $\gamma $ values}
\label{bbhist}
\end{figure}

In the numerical simulations we use our  Monte Carlo program to generate the 
distributions $N_\mu(A)$ for 
$\mu < \mu_c$. We prevent the surface from growing to infinity by introducing
an upper bound $A_{max}$ for the area. 
This does not spoil the detailed balance condition but ergodicity may in 
principle be affected for surfaces not much smaller then $A_{max}$. In the
verification of the algorithm (section \ref{hyper_verify}) we, however, did 
{\em not} observe any effect on the distribution of surfaces induced by the
upper bound. Following \cite{bb86} we have anyhow discarded the data for 
surfaces of size $[A_{max} - 20, A_{max}]$ when analyzing  the data. 
We have measured 
$N_{\mu}(A)$ with two different $A_{max}$ and for different couplings. 
In table 
\ref{hyper_bb_statistic} we summarize the number of iterations collected for 
six different parameters $A_{max}$ and $\mu$. After dividing each dataset into 
three parts there are $9 * 9 = 81$ ways to combine data for $A_{max} = 5000$ 
in formula (\ref{bb_gamma_mamma}) with data for $A_{max} = 300$. We have
extracted $\gamma $ for the $81$ combinations. In figure {\ref{bbhist}}
we show the distribution of $\gamma $ values obtained with this procedure.
Taking the average we get $\gamma =  0.26(2)$. This is in agreement with the
result $\gamma = 0.24(3)$ in \cite{bb86}. We have repeated the procedure 
for the model which allows for self-bending surfaces and found 
$\gamma = 0.44(3)$ which is in reasonable agreement with  the expected result 
$\gamma = \frac{1}{2}$ \cite{dfj84}. 

\subsection{The Baby universe method \label{shake_it}}

\begin{figwindow}[4,l,{\hspace{5mm}\psfig{file=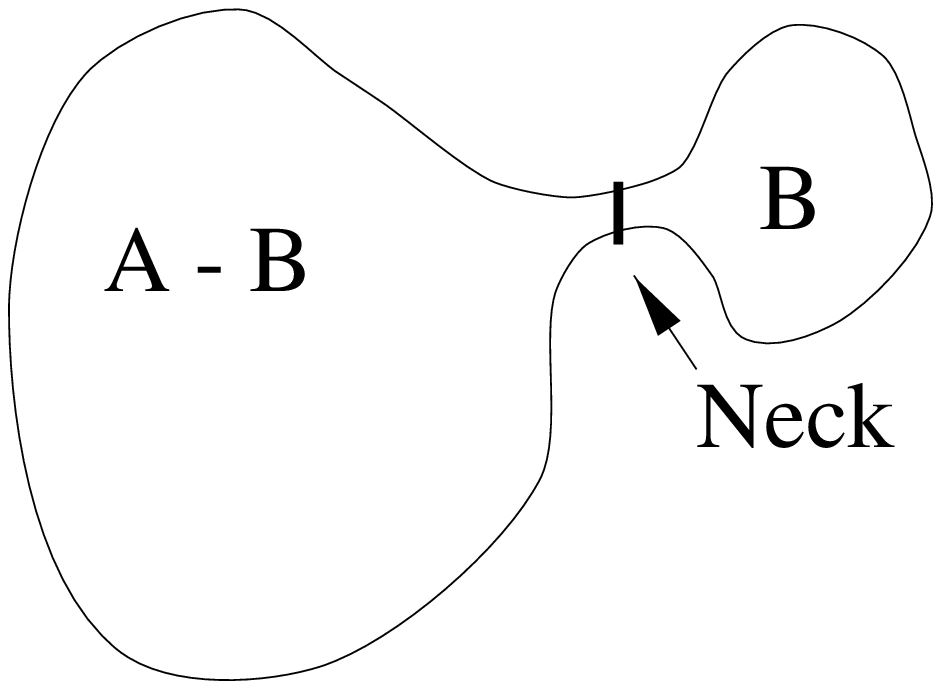,angle=270,height=2.8cm}},{A
baby surface}]
\label{comon_baby}
The baby universe method  \cite{ajt93} has become a  standard tool for 
measuring the entropy exponent $\gamma $ in numerical simulations of random 
surfaces. 
It is very convenient because it does not require fine tuning of parameters 
and allows the use of large systems in the simulation. In this section we 
present the numerical results obtained with this method for hyper-cubic 
random surfaces with up to $80000$ plaquettes.

A baby-universe is a part of a surface $S$ with the  topology of a disc. 
It is connected with the rest of $S$  in such a way that the 
disc-perimeter $l$ is much smaller than the disc area $B$: $l^2 \ll B$. 
In  figure \ref{comon_baby} this is depicted graphically: a baby universe
is a part of $S$ connected to the rest of $S$ by a bottle-neck.
If one looks at this neck from  far enough away one sees it as a point at 
which two surfaces are joined, one with area $B$, the other with area 
$A-B$, where $A$ is the area of the whole surface $S$.
\end{figwindow}

One can approximate the baby universe distribution \cite{jm92} by ~:
\eq{
N_A(B) \sim {\cal N}_l(B) {\cal N}_l(A-B)
\label{minbu_jain}
}
where ${\cal N}_l(A)$ is the number of discs with loop $l$ and area $A$. 
This number  can be expressed by the number of spherical
surfaces ${\cal N}(A)$ eq. (\ref{hyper_largeN}). Consider for example the disc
with area $A$ and a unit-square as the boundary. Such a disc can be created
by removing one plaquette from a spherical surface with area $A + 1$. Taking
into account the factor $(A+1)$ which counts the number of plaquettes that can
be removed from the sphere one gets
\eq{ {\cal N}_l(A) = (A+1){\cal N}(A+1).}
This formula is approximately true also for different boundary shapes.  
Substituting this expression and (\ref{hyper_largeN}) 
in (\ref{minbu_jain}) gives 
\eq{N_A(B) \sim c(A) (B(A-B))^{\gamma-2} (1 + \mbox{corr.})
\label{hyper_mindis}}
for the distribution of baby surfaces with area $B$ on a surface with total
area $A$.  The leading exponential contribution 
$c(A) = c \exp \left (\mu _c (B + \{A-B\}) \right )$ depends only on the area
$A$ of the whole surface and not on $B$. 

\begin{figure}
\hspace{1.0cm}
\psfig{file=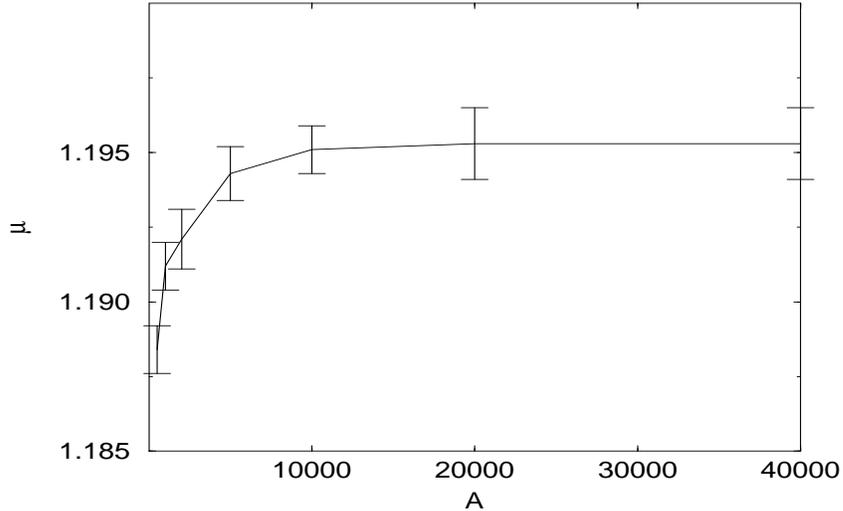,angle=270,height=8cm,width=12cm}
\caption{Pseudo critical coupling $\mu_{pc}(A)$}
\label{hyper_pc_mu}
\end{figure}

This is very convenient in numerical simulations where one can 
use (\ref{hyper_mindis}) to measure the exponent $\gamma $,  which is part 
of a volume correction term in the canonical (fixed area) ensemble. This means
that one does not need to fine-tune $\mu$ to  $\mu_c $. However, for technical
reasons, we still have to tune with comparatively low precision the coupling
$\mu $ to its pseudo-critical value $\mu_{pc}$. Our algorithm is inherently
grand canonical since ergodicity requires volume fluctuations. We extract the 
information for the canonical ensemble with $A = A_0$ from the simulated grand
canonical ensemble by measuring  only if $A = A_0$. Inefficient large 
volume fluctuations about $A_0$ are suppressed by a  potential
\eq{U(A) = \frac{\delta}{2} (A - A_0)^2 \label{hyper_multi_c}}
which we add to the action (\ref{hyper_action}) with an appropriate choice of
$\delta $.  We used $\delta = 0.002$ in the simulations. 
For the simulation the coupling 
$\mu $ has to  be tuned so that $<A> \approx A_0$. The Gaussian form of the
grand-canonical ensemble allows us to estimate $\mu - \mu_{pc}$ 
by measuring $<A>$, i.e. 
\eq{\mu - \mu_{pc}(A_0) \approx \delta \left ( <A> - A_0 \right ) 
\label{mu_imp}.}
We use this formula to improve, recursively, the estimate for $\mu $ in the
simulations. In figure \ref{hyper_pc_mu} we show the pseudo-critical coupling
obtained with this procedure.

\begin{figure}
\hspace{1.5cm}
\psfig{file=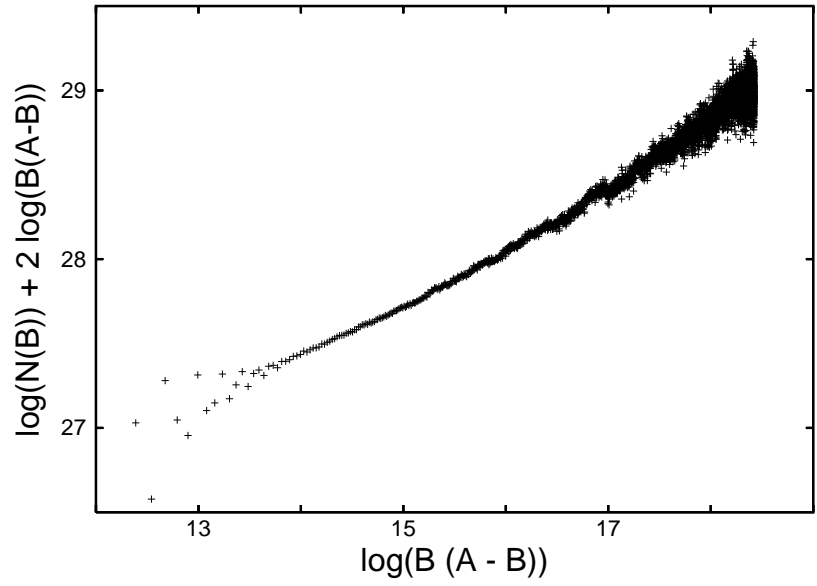,angle=270,height=11cm,rheight=8cm,rwidth=11.2cm}

\caption{The distribution of baby-surface areas for $A=20000$}
\label{mindisf}
\end{figure}

\begin{figure}
\hspace{1.5cm}
\psfig{file=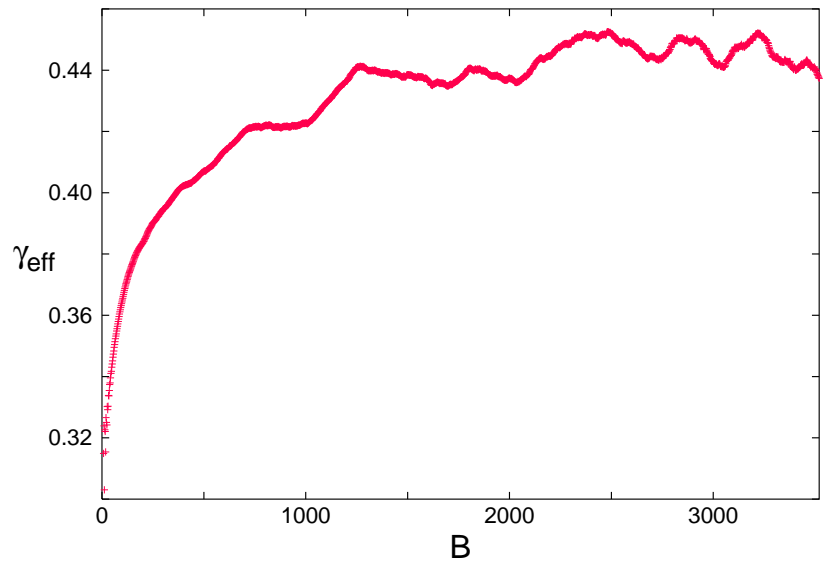,angle=270,height=11cm,rheight=8cm,rwidth=11.2cm}
\caption{The effective  $\gamma _{eff}$ as a function of the lower cut-off $B$
for $A = 20000$.}
\label{geff}
\end{figure}

We choose the coordinates
\eq{
\begin{array}{rcl}
y & = & \ln \left [ ( B (A-B) )^2 N_A(B) \right ] \\
x & = & \ln \left [ B(A - B) \right ]
\end{array}
\label{plot}
}
to analyze the measured baby universe distribution $N_A(B)$. In these
coordinates the exponent $\gamma $ can be extracted from the numerical data
with a linear fit to the equation $y = \gamma x + b$. In figure
\ref{mindisf} we show a typical baby universe  distribution using the 
coordinates (\ref{plot}).
The total  surface area is $A = 20000$. For small values of $B$ 
we see clear evidence of finite size effects; the data-points do not lie on a
straight line. On the other hand for large values of $B$ one can see strong 
statistical noise which comes from the low frequency of 
large baby universes. For example, for $A=10000$ the probability of finding 
a baby universes of area $B=4999$ is $7 \times 10^{-4}$.
When estimating the value of $\gamma $ one has to deal with the finite size
effects. In our analysis we discard data for small baby universes.
To decide, which part should be kept, we look at the behavior 
of $\gamma _{eff}(B)$  obtained from the fit to the data 
for baby surfaces larger than $B$ as a function of $B$.
As one can see in figure (\ref{geff}), where we show as an 
example $\gamma _{eff}(B)$ measured for $A=20000$, that $\gamma _{eff}(B)$ 
grows to a plateau value. We use this value to estimate $\gamma _{eff}$.

\begin{figure}
\hspace{1.5cm}
\psfig{file=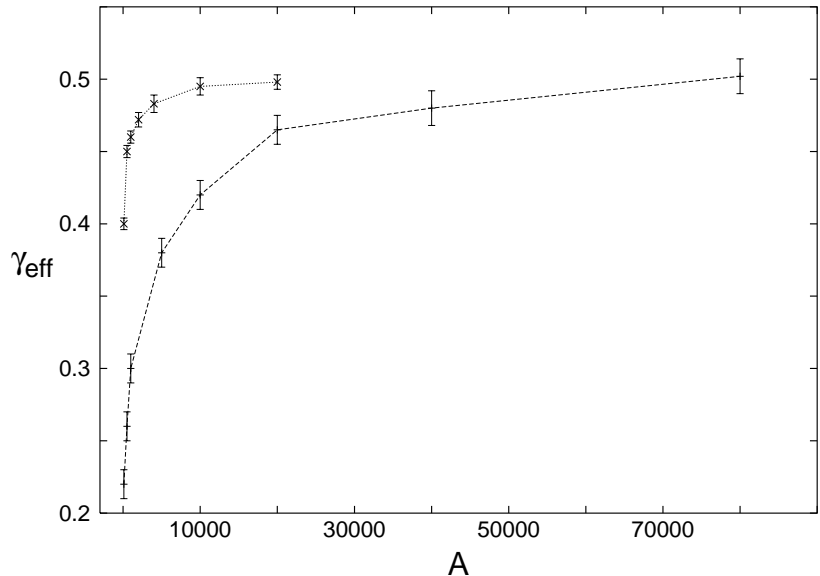,angle=270,height=11cm,rheight=8cm,rwidth=11.2cm}

\caption{The effective entropy exponent $\gamma _{eff}$ as a function of the
canonical volume $A$ for  ensembles  including (upper curve) 
or excluding (lower curve) self--bending, respectively.}
\label{gamma}
\end{figure}

We analyzed the baby universe distribution for different canonical volumes. 
In figure (\ref{gamma}) we show $\gamma _{eff}$ as a function 
of the canonical volume $A$.
For comparison we also  performed simulations for the unrestricted ensemble.  
As discussed above, for this case one can 
show analytically that $\gamma = \frac{1}{2}$.
For the constrained model without self--bendings (lower curve) one 
observes stronger finite size effects. The asymptotic  value  of 
$\gamma _{eff} $, however,  is the same.
We conclude that the constrained hyper-cubic random surface model 
possesses the same  critical behavior as the unconstrained model. This means
that the model is universal in the sense that the details of the 
discretization are unimportant. 
However the result $\gamma = \frac{1}{2}$  contradicts the results obtained 
with the method used in \cite{bb86}, presented in the previous section. 
The reason for this contradiction is easy to understand if one looks at 
figure \ref{gamma}. In the last section surfaces with an area $200 < A < 280$ 
were used to extract an estimate for $\gamma $. From figure \ref{gamma} 
one sees that for surfaces of this size the effective 
$\gamma _{eff} \approx 0.25$;
on such small surfaces the finite size effects are still dominating.

\enlargethispage{3mm}
\subsection{Discussion}
We have measured the value of the string
susceptibility exponent $\gamma$ for hyper-cubic
surfaces embedded in a $4$ dimensional target space for
an ensemble of spherical surfaces with and  without
self--bendings. In both cases we find numerically, 
for large enough surfaces, $\gamma \approx 1/2$, which is compatible with a
branched polymer behavior.  The local constraint first discussed in 
\cite{bb86} is not sufficient to change the critical behavior of the model.
However, in the next section we will show that a generalized 
external-curvature term can change this behavior.

It is intriguing to note that the discretized model of Nambu-Goto strings
suffers from the same pathological behavior as that of Polyakov string theory.
It collapses down to a branched polymer, i.e. essentially a one-dimensional
object. This result suggests that both models have the same universal behavior.
Unfortunately the Nambu-Goto string is not defined in $D<2$, so it is
not possible to compare the discretized string models in $D < 1$ where one
finds non-trivial behavior for the Polyakov string.

\clearpage

\section{Hyper-cubic random surfaces with extrinsic curvature}
\subsection{Introduction \label{ext_intro}}
The branched polymer behavior of  the constrained hyper-cubic random surface 
does not rule out the possibility that one finds non-trivial
behavior for the model with  the generalized extrinsic curvature action 
(\ref{ext_curv}). With infinite coupling $\epsilon $
the generic surface is, up to finite size effects, flat. For finite couplings
one may hope to find a transition from the branched polymer to a flat phase.

However it was shown in \cite{dj86} for the model with extrinsic curvature 
that "the critical exponents take their mean field value if the 
susceptibility of the model and a coarse grained version of it both diverge". 
In other words under the assumptions
\eq{\gamma > 0 \quad \mbox{and }  \bar{\gamma} > 0\label{assumptions}.} 
the model is always in the branched polymer phase.
This result was obtained with a renormalization group argument where the
branched polymer was decomposed into "blobs", components which can not be cut
into two parts along a loop of length two. The exponent $\bar{\gamma }$ is 
the entropy exponent for these blobs. The equivalence of 
(\ref{assumptions}) with the statement given above can be seen from the 
expected behavior of the susceptibility 
\eq{\chi(\mu ) \approx (\mu - \mu _c)^{-\gamma}, }
which diverges for  $\mu \rightarrow \mu _c$ only if $\gamma > 0$.  

A key step in the derivation of this result in \cite{dj86} is the formula
\eq{\chi (\mu ) = \frac{\bar{\chi}(\bar{\mu})}{1-\bar{\chi}(\bar{\mu})}
\label{durhuus_grund}} 
which relates the susceptibility $\chi(\mu)$ of the original model to the 
susceptibility $\bar{\chi}(\bar{\mu})$ of the decomposed blobs with
renormalized coupling $\bar{\mu}$. Obviously $\bar{\chi} = 1$ if the 
susceptibility $\chi $ diverges. Under the assumptions stated above, 
i.e. $\bar{\gamma} > 0$ for the blobs  one concludes that 
they are not critical $\bar{\mu} > \bar{\mu}_c$ and $\bar {\chi}$ is 
analytic at this point. With a Taylor expansion one gets the self consistency 
relation $ \gamma = 1 - \gamma $ which is solved by $\gamma = \frac{1}{2}$,
the generic value for branched polymers.  

However, Durhuus emphasized in \cite{du94} that the condition 
$\bar{\chi} = 1$ alone does not imply that the blobs are  
non-critical. If $\bar{\gamma} < 0$ the susceptibility $\bar{\chi}$
does not diverge and $\bar{\chi} = 1$ can be satisfied {\em at} the critical 
point $\bar{\mu} = \bar{\mu}_c$ of the coarse grained system. At such a point 
the entropy is still dominated by the branching of the surface but the 
branches themselves are critical, which in effect changes the exponent $\gamma
$ for the whole system.  If one assumes that $\bar{\gamma}$ takes the 
KPZ-values (\ref{kpz_form}) one can derive from  (\ref{durhuus_grund}) the 
series (\ref{hyper_durhuus}). 

The derivation of that formula is rather formal. It does not provide a
description how to construct a system which has this property. We want to
check numerically if the hyper-cubic random surface model exhibits non trivial
behavior at an eventual crumpling transition. In other words we check
if the assumptions (\ref{assumptions}) are fulfilled in the whole
range of couplings $\epsilon$. Non-trivial behavior can be expected only if 
these assumptions are not satisfied, which we expect if 
$\epsilon = \epsilon_c$ is at a critical point.

\begin{figure}[p]
\hspace{1.5cm}
\psfig{file=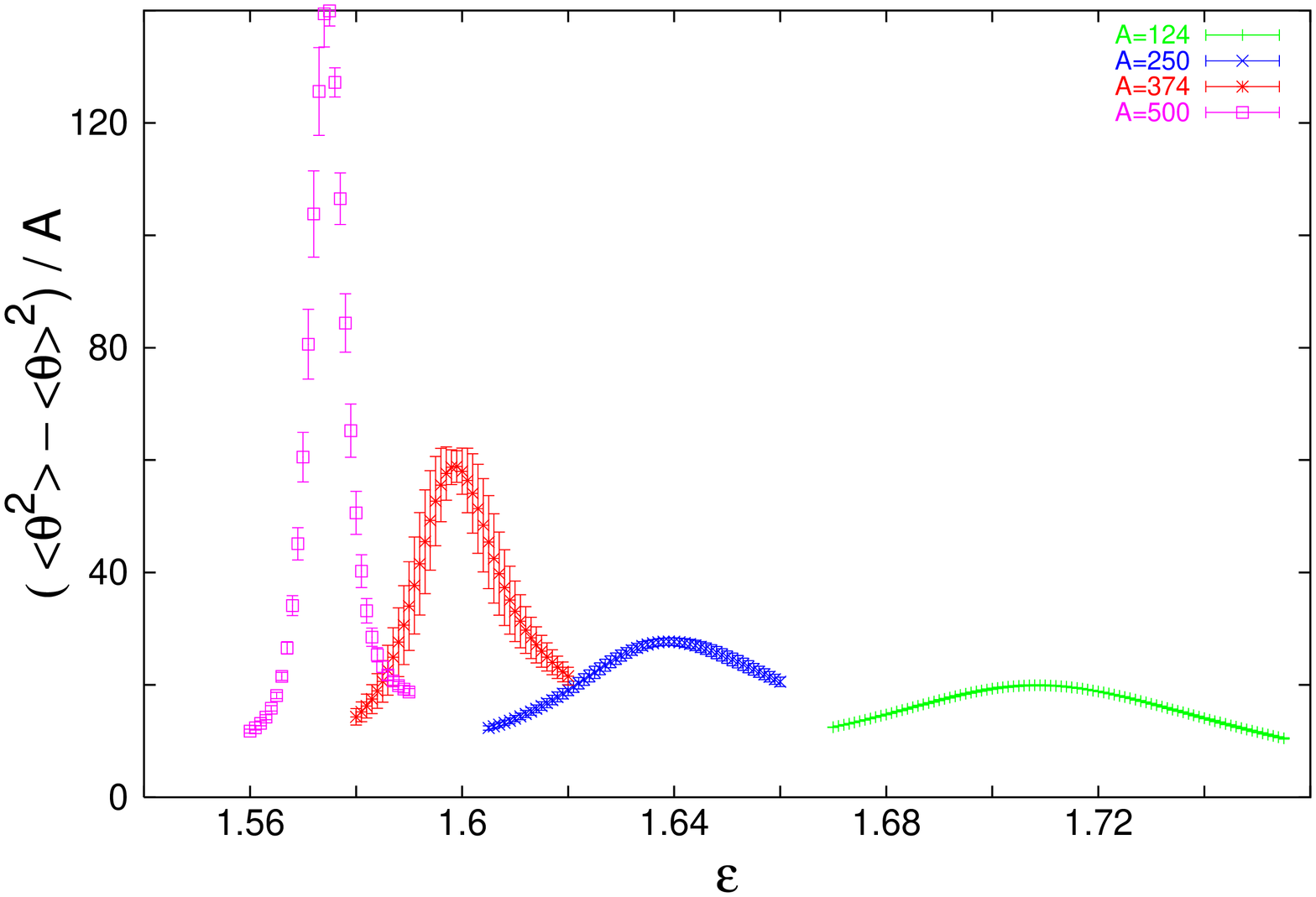,angle=270,height=11cm,rheight=7.5cm}
\caption{The susceptibility $\chi _{ext}$ in $D=3$ dimensions}
\label{chi_ext3}
\end{figure}

\begin{figure}[p]
\hspace{1.5cm}
\psfig{file=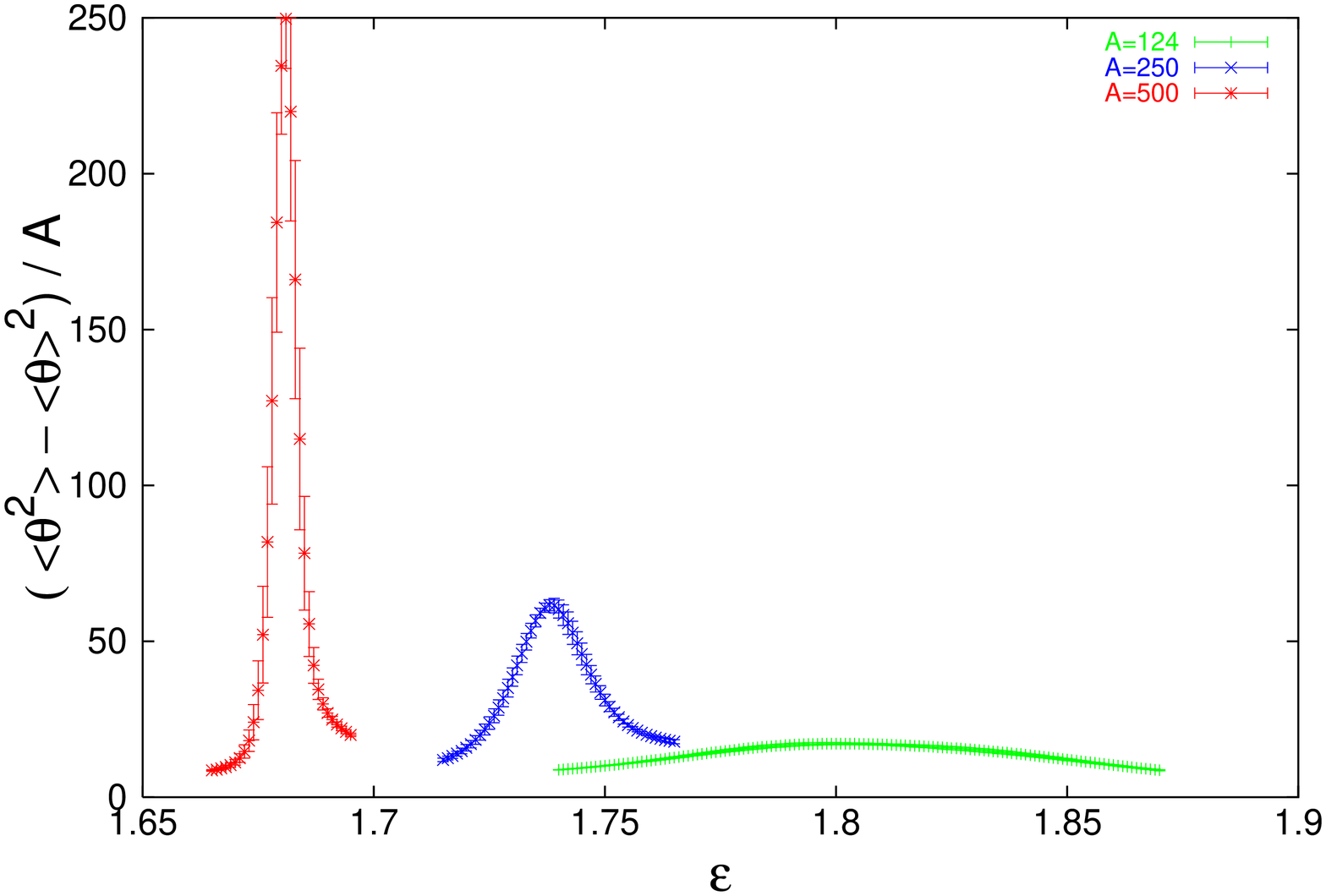,angle=270,height=11cm,rheight=7.5cm}
\caption{The susceptibility $\chi _{ext}$ in $D=4$ dimensions}
\label{chi_ext4}
\end{figure}

\subsection{Numerical results}
\begin{figure}[t]
\hspace{1.5cm}
\psfig{file=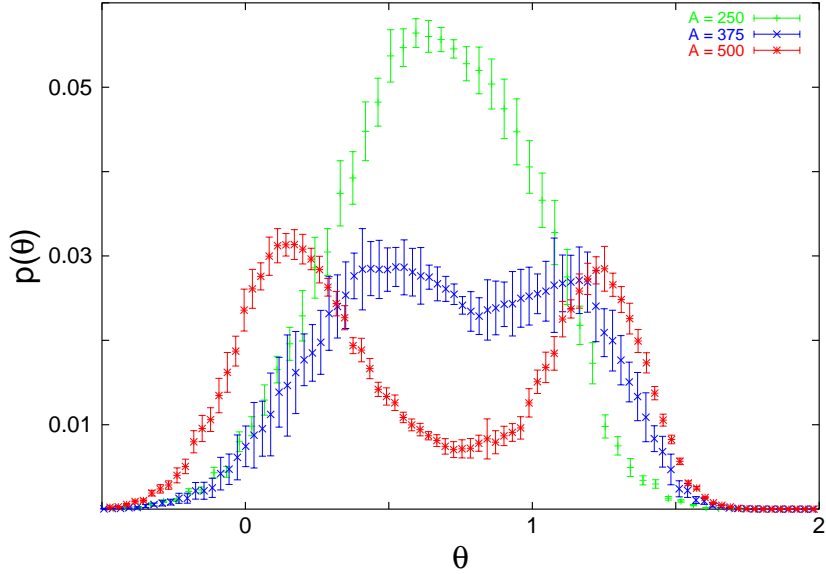,angle=270,height=11cm,rheight=7.5cm}
\caption{The distribution of the average external curvature $\bar{\Theta}$ in
$D=3$ dimensions obtained with re-weighting. The non zero probability for 
unphysical negative values $\bar{\Theta} < 0$ is an artifact of the 
extrapolation to this region in the re-weighting process.
}
\label{eden}
\end{figure}
The simulations are done in the same quasi-canonical way already discussed 
for the constrained model eq. (\ref{hyper_multi_c}).
To localize  possible transitions we measure the energy fluctuation 
of the external curvature field
\eq{\chi_{ext} = \frac{1}{A} \frac{\partial ^2}{\partial ^2 \epsilon} 
           \log Z(A,\epsilon) = \frac{<\Theta^2> - <\Theta>^2}{A}} 
and search for a peak in this observable. The average external curvature is
proportional to the first derivative of the free energy~:
\eq{ <\Theta> =  \frac{\pi}{2 A} \frac{\partial }{\partial \epsilon} 
                     \log Z(A, \epsilon) 
= \frac{1}{2 A} \sum _S \sum _l \Theta _l e^{-S[E]}.
}

We simulated in both three and four  embedding dimensions and scanned the 
coupling range $0 \le \epsilon \le 3$ for peaks of $\chi _{ext}$. 
In both cases we found a single peak. 
We used re-weighting methods \cite{FS88} to extract the shape of the
peaks shown in figures \ref{chi_ext3} and \ref{chi_ext4} from four independent
measurements at $\epsilon \approx \epsilon _{pc}$ per volume.
 
The peaks grow quickly,  in fact a bit faster than
linearly, with the volume. This indicates a first order phase
transition, which is confirmed by a look at the distribution of 
external curvature $\bar{\Theta}$, where the bar indicates the average taken 
over a given lattice. In figure \ref{eden} we show  this distribution
for $D=3$ and the coupling $\epsilon \approx \epsilon_{pc}$ close
to the pseudo-critical coupling. One observes a clear signal of a first order 
phase transition, namely  two maxima separated by a minimum which becomes
deeper as  the size of the surfaces is increased. 
This clearly indicates two separate phases. 
In one phase the average external curvature  is close to zero, the typical
surface is flat. 

\begin{figure}[t]
\hspace{1.3cm}
\psfig{file=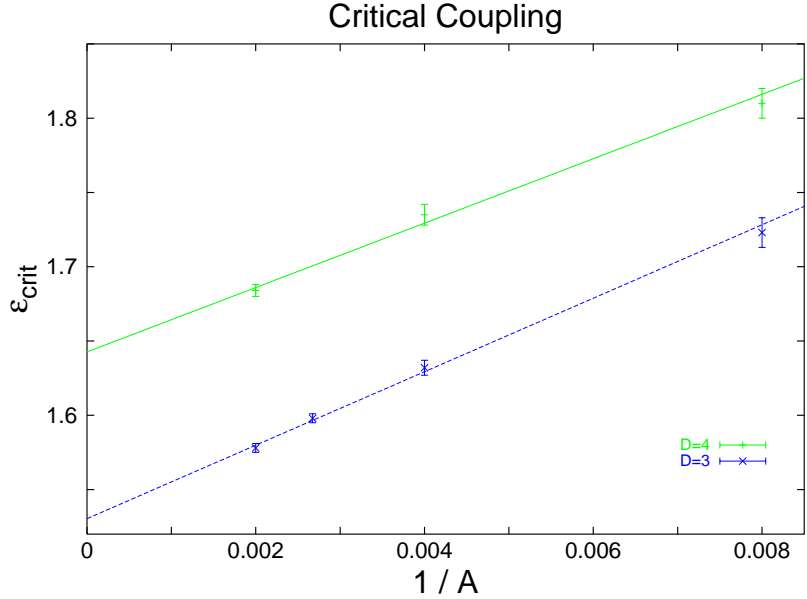,angle=270,height=11cm,rheight=8cm}

\caption{The pseudo-critical coupling $\epsilon _{crit}(1/A)$. }
\label{eta_crit}
\end{figure}
\noindent
To estimate the critical coupling we use that for a first order phase 
transition one can expect 
\eq{\epsilon _{pc}(A) = \epsilon_c + \frac{c}{A} + O(\frac{1}{A^2}) 
\label{SKL}}
for the scaling of the pseudo-critical couplings $\epsilon _{pc}$. 
In figure \ref{eta_crit} we show the numerical estimates  for  
$\epsilon _{pc}$. Note that the pseudo critical coupling decreases with 
the volume which indicates  that $\epsilon _c$ is finite in the 
thermodynamical limit. 
With a linear fit to equation (\ref{SKL}) we find for the critical coupling

\begin{displaymath}
\begin{array}{cc}
  D = 3 & \epsilon _c = 1.530(3) \\
  D = 4 & \epsilon _c = 1.643(4) 
\end{array}
\end{displaymath}

\begin{figure}[p]
\psfig{file=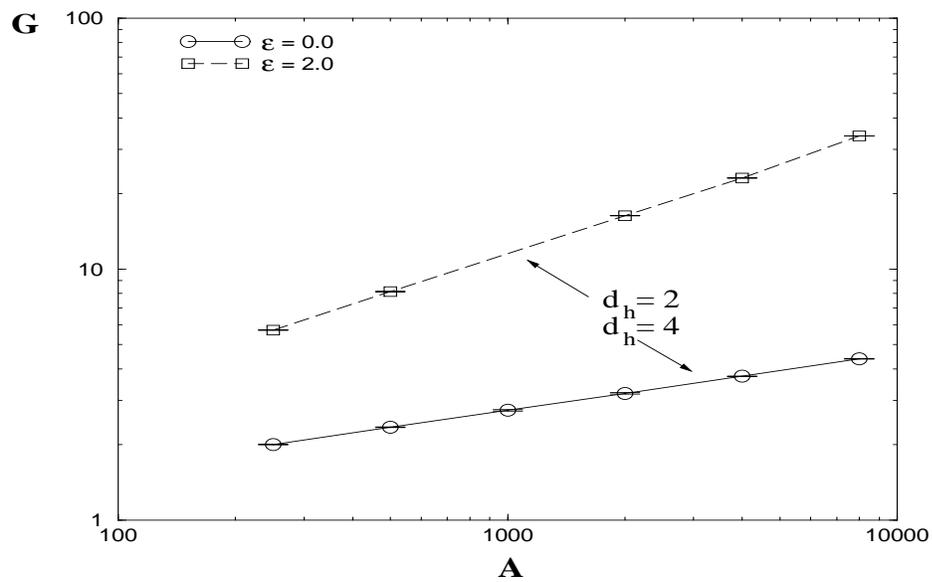,angle=270,height=9cm,width=14cm,rheight=8.7cm}
\caption{The scaling of the radius of gyration. }
\label{hausdorff}
\end{figure}
\begin{figure}[p]
\hspace{1.5cm}
\psfig{file=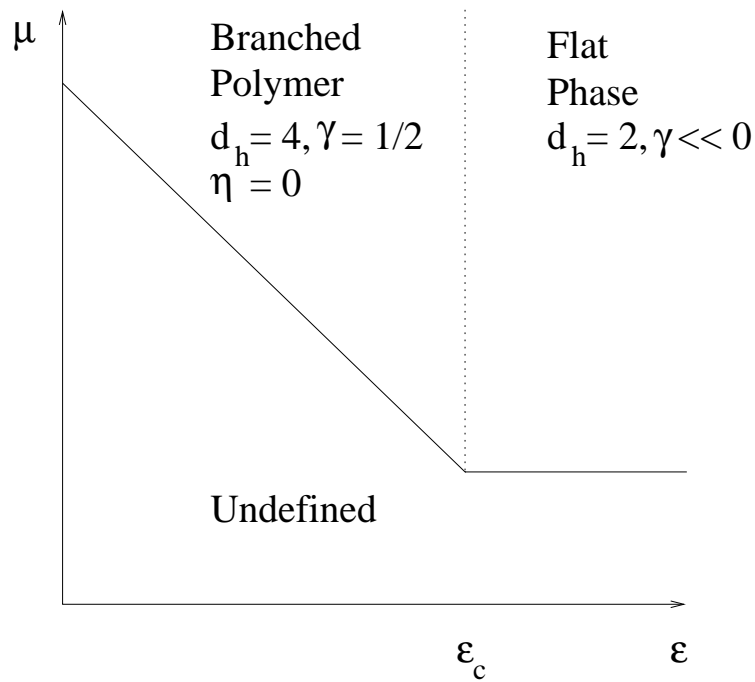,angle=270,height=9cm}
\caption{The phase diagram for the Hyper-cubic random surface model with
extrinsic curvature. One finds qualitatively the same behavior in $D=3$ and 
$D=4$ embedding dimensions.} 
\label{hyper_phase}
\end{figure}

To demonstrate the change in the geometrical behavior we consider  the
radius of gyration
\eq{G^2(A, \epsilon) = < \frac{1}{A} 
          \sum _{p_{x,y}} (S_{\mu}(p_{x,y})  - S_{\mu}(p_0))^2>, }
which is the average squared distance of the plaquettes to some reference
plaquette $p_0$.  The $S_{\mu}$ are the coordinates of the plquette in the
embedding space. In the numerical simulations we do a stochastic average over
the $p_0$ by choosing about 10 \% of the plaquettes at random as $p_0$ 
per measurement. One can use the radius of gyration  to define the 
external Hausdorff dimension $d_h$
\eq
{G^2(A, \epsilon) \approx  A^{2 / d_h}. \label{how_hau} }
In a sense the Hausdorff dimension is the largest possible  dimension for 
the embedding space  which can still be completely filled by the surface. 
One has for  example $G^2 \propto A$ for a flat surface and therefore 
$d_h = 2$. In figure \ref{hausdorff} we show $G(A, \epsilon)$ measured in
the branched polymer phase, at $\epsilon = 0$, and in the strong coupling
phase, at $\epsilon = 2.0$. With a fit to (\ref{how_hau}) we extract the 
Hausdorff dimension in the two phases:
\begin{displaymath}
\begin{array}{cc}
\epsilon = 0.  & d_h = 4.07(3) \\
\epsilon = 2.  & d_h = 2.00(3).   
\end{array}
\end{displaymath}
The value found for $\epsilon = 0$ is two standard deviations away from the
expected result \cite{bb86} but this should presumably be attributed to 
finite size effects.

The results are summarized in the phase diagram figure \ref{hyper_phase}.
The model is defined only for $\mu > \mu _c(\epsilon )$ which defines 
the critical line depicted by the solid line in the phase diagram. 
Above this line we find two phases. 
For $\epsilon < \epsilon _{crit}$ the system is in the branched polymer phase 
with $d_h \approx 4$. If we assume
hyper scaling 
\eq{ \nu = d_H^{-1}, \quad \gamma = \nu (2 - \eta)}
the anomalous scaling dimension is $\eta = 0$ as expected for a branched
polymer. The strong coupling phase is flat with $d_h = 2$. In this phase 
basically no baby universes  are present. 
For example, on a surface of size $A=500$ 
we found at $\epsilon = 2$ on average only $0.06$  baby universes  of 
size $B=5$. The largest one found in this phase is of size 
$33$ and appeared only once in $200 000$ measurements. For comparison in the 
branched polymer phase at $\epsilon = 0$ we found in average $10.6$ baby
universes  of size $5$ on the surface, the probability for the largest 
possible one with $B=249$ is about $0.02$.  Therefore one can not use the 
baby universe distribution to measure $\gamma$ in the flat phase. However the 
fact  that there effectively are no baby surfaces means that $\gamma $ is 
large negative. Therefore the existence of this phase is not 
in contradiction with \cite{dj86} because the assumptions (\ref{assumptions}) 
are not met. 

We hoped to find $\bar {\gamma} < 0$  for the branches 
at the critical point. Instead 
we found $\gamma$ for the whole system large negative in the strong
coupling phase. At the transition point the exponent $\gamma $ is not well
defined because the transition is first order with different values for
$\gamma $ in both phases.  The  close to horizontal critical 
line in the phase diagram fig. \ref{hyper_phase} for the flat phase
reflects  the fact that the curvature field is almost  
frozen at $<\Theta > = 0$. A increase of the coupling $ \epsilon $ 
has no further effect on the geometry and therefore also no effect on
the entropy $\log(\mu _c)$ of the surface. 

\subsection{Discussion}
We have investigated the phase diagram of hyper-cubic random surfaces with
an extrinsic curvature term in the action. We find two phases. The weak
coupling phase is dominated by  branched polymer structures with 
$\gamma = 1/2$. The strong coupling phase is flat. The transition between the 
two phases is first order. We show that the existence of the 
flat phase  is not in contradiction with \cite{dj86}. 

In three dimensions one can interpret the hyper-cubic surfaces also as 
membranes. The internal geometry in this model is dynamic, therefore the
surfaces are called fluid membranes. However, they are not self-avoiding. 
Self-intersection is, differently from typical membranes, allowed. 
The first order phase transition, which we reported in this work, is 
therefore an example for a phase transition  of fluid membranes at finite
coupling. For fluid membranes described by dynamical triangulations with an 
extrinsic curvature term the existence of a phase-transition 
flat phase is disputed. In \cite{extcurv_bengt} numerical evidence in favor 
for a second order phase transition was reported. The authors of 
\cite{bchhm92, abcfhhm93} can not exclude that it is  a cross-over rather 
than a phase transition. For fluid membranes with an external curvature term 
in the action a renormalization group analysis 
\cite{helf73, pol86, klein86, forst86} indicates that there should be no 
stable flat  phase at a finite coupling to the external curvature. 

We cannot take a continuum limit at the transition point because the 
transition is of first order. It is possible that the transition might become 
softer, and continuous, if one  introduces self-avoidance into the model. 
This is for example the case for a model  of Ising-spins with a 
gonihedric action \cite{sw94}. This model is dual to a surface model with  
extrinsic curvature. It does, however, not define the internal connectivity of
the surface when more than two plaquettes meet at an external link. 
Numerically a first order phase transition was found \cite{des1order}, but
with a weak self-avoidance imposed  evidence for a second order phase
transition was found \cite{des2order}. 
\clearpage

\section{The Algorithm \label{hyper_algorithm}}
The algorithm used in the simulations in this chapter is a 
Monte Carlo algorithms which was discussed in section (\ref{intro_algo}).  
In this section we want to describe the 
transformations (\ref{intro_mc_transform}) and the weight (\ref{metropolis})
specific to our new algorithm for the hyper-cubic random surface model.

\begin{figure}[h]
\hspace{3.8cm}
\psfig{file=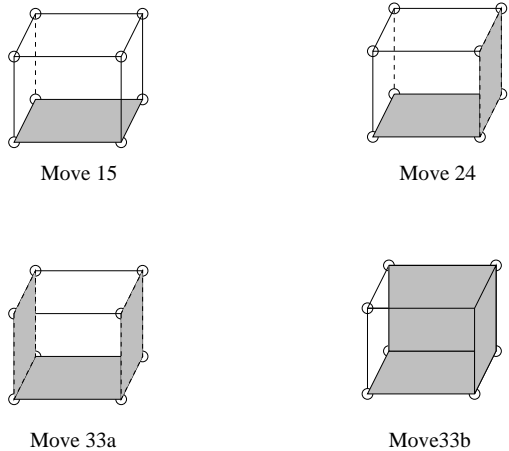,angle=270,height=6cm}
\caption{An ergodic  set of deformations}
\label{movers}
\end{figure}

The transformations are based on a set of local moves 
depicted in figure \ref{movers}. A prescription for all four pairs of moves 
is the following: choose $p$ connected plaquettes (drawn shaded in the figures)
on the surface such that they form a part of a sphere (a cube) and replace 
them by the rest of the sphere, the transparent $q = 6-p$ plaquettes in the 
figures.
We denote by $M_{pq}$ a move where $p$ is the number of
plaquettes being removed and $q$ the number of plaquettes inserted into 
the surface. Consider for example the move $M_{15}$ which removes a 
single plaquette  from the surface and fills the hole with a "house" of five 
plaquettes. In the reverse move the role of the shaded and transparent
plaquettes is interchanged. For example deformation $M_{51}$ replaces a house
with a single plaquette. Deformations $M_{33a}$ and $M_{33b}$ are self dual, 
i.e. if one deforms the "same" location twice with one of these moves the 
surface is  unchanged.  
 
\begin{figure}[t]
\hspace{4cm} 
\psfig{file=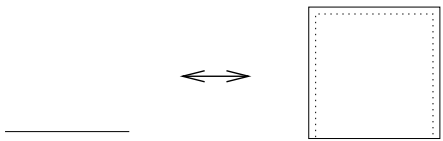,angle=270,height=1.8cm}
\caption{Spike insertion / deletion}
\label{m02}
\end{figure}
\begin{figure}[h]
\hspace{2.5cm}
\begin{minipage}[c]{4cm}
\psfig{file=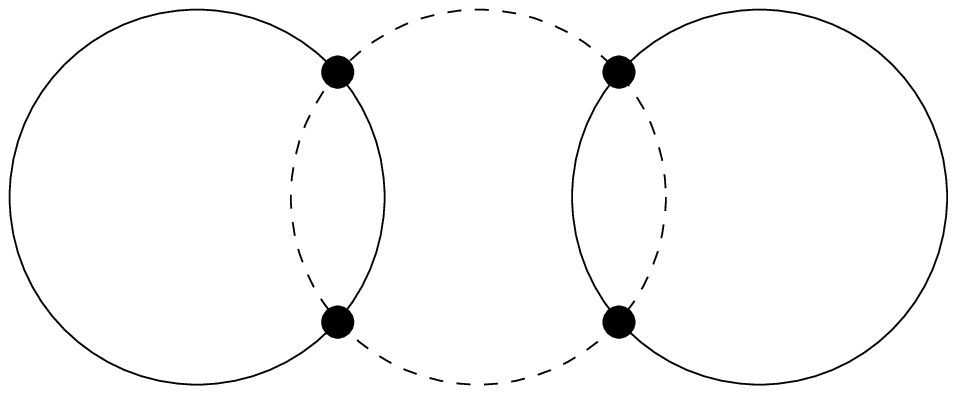,angle=270,height=1.5cm}
\end{minipage}\hspace{0.8cm}
\begin{minipage}[c]{1.2cm}
\psfig{file=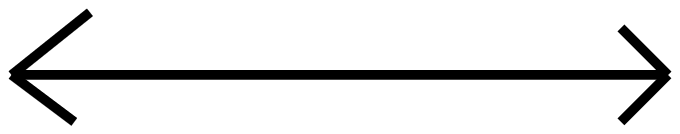,angle=270,width=1cm}
\end{minipage}\hspace{0.8cm}
\begin{minipage}[c]{4cm}
\psfig{file=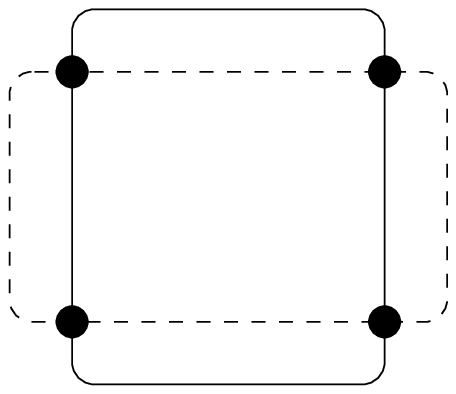,angle=270,height=1.5cm}
\end{minipage}
\caption{An example of move $M_{00}$ in the dual representation}
\label{m00}
\end{figure}

The moves $M_{pq}$ shown  in figure \ref{movers}  are known to be
ergodic \cite{bbf85}. The additional moves  which we discuss now
are therefore not necessary for ergodicity. However,
they turn out to improve the mobility of the algorithm. 
The move $M_{02}$ shown in figure \ref{m02} cuts open a link and inserts 
a spike. A spike are two plaquettes glued together along three edges.
The move $M_{00}$ is a non-local deformation in terms of the internal 
geometry but is local in the embedding space. 
It cuts open two different (internal) links, which occupy the same link in the 
embedding, and it cross-connects the plaquettes. This can probably   
better be understood in the dual representation. In the dual  representation a 
plaquette is replaced by a vertex, links are replaced by propagators and a 
vertex is dual to a closed propagator. However not all closed propagator 
loops are dual to a vertex. One can easily convince oneself that only those 
loops which turn to the right (left) at each  vertex of the loop are dual to 
a vertex on the hyper-cubic lattice. The move $M_{00}$ cuts open two such 
propagator loops and glues pairwise together the free ends of the two loops, 
provided this is allowed geometrically.  In figure \ref{m00} we show an example
of such a transformation. The diagram on the left is dual to two spikes glued
together along their open links. Under the geometric constraint that the two
spikes occupy the same unit-square in the embedding one can cut open 
one edge, say the edge to the right from the common one, and glue the now
free edges pairwise together. After doing this one ends up with a "bag in a
bag" where a "bag" are two surfaces glued along two edges with a common
corner. Note that the surfaces may not be glued together on opposite edges
because the latter configuration has a different topology, that of a torus. 

A  transformation is done in the following steps:
\begin{enumerate}
\item Choose a location $A$ for the deformation. 
\item Choose a move $M_{pq}$ from the set of  deformations
\item Choose a direction or orientation $d$. For example in move
      $M_{15}$ one can choose in which of the $2(d-2)$ possible
      directions the house should be built.  
\item Check if the move violates any  geometrical constraints. If 
      not, subject it to a Metropolis test.
\item Apply the local deformation $M_{pq}$ at $A$ in direction $d$. 
\end{enumerate}
The Monte Carlo weights $\pi _{pq}$ for deforming the surface at a specific 
place by move $M_{pq}$ can be derived from the detailed balance condition
(\ref{detailed_balance})
\eq{
\pi_{pq} \frac{1}{A}         \frac{1}{n_m} \frac{1}{n_d}  e^-{S[{\cal A} ]} 
=
\pi_{qp} \frac{1}{A + p - q} \frac{1}{n_m} \frac{1}{n_d'} e^-{S[{\cal A}']} 
\label{hyp_db_cond}
.} 
The factor $p = \frac{1}{A}$ is the probability of choosing 
 certain plaquette from the $A$ plaquettes which 
form the surface  before the move. The grand canonical moves discussed
above change the size of the surface to $A + q - p$. The probability of
choosing the correct location for the inverse transformation is 
$\frac{1}{A + p - q}$, which is  different from $p$. Therefore the 
probabilities $\pi _{pq}$ contain an area dependent pre-factor to balance  the
grand canonical fluctuations in the transformation probabilities. The next
term $1 / n_m$, the number of possible moves, however, drops out because in 
the second step of the transformation the probability to choose a certain 
transformation is the same for a move and its inverse.
The next factor takes into account the freedom one
has in choosing a direction. For example in move $M_{15}$ one
can choose from $2 (D - 2)$ directions orthogonal to the chosen plaquette. In
the inverse transformation $M_{51}$ one has no  freedom after choosing the
roof of a house to be removed. The last factor is the Boltzmann distribution
which we want to generate. Note that the action  contains contributions from
the external curvature. The weights coming from the field can be used directly
because the curvature field has no degrees of freedom of its own. 
In the following table we list the transition probabilities for the
different moves. The numbers $n_a, n_b$ give the number of allowed
deformations of the type $M_{00}$ at the specific site of the lattice. 

\begin{displaymath}
\begin{array}{|r|rcl|}\hline
M15 / M51 & \frac{\pi_{15}}{\pi_{51}} &  =  &  
               2 \frac{A}{A + 4}  (d - 2)  \exp ( \Delta S) \\
M24 / M42 & \frac{\pi_{24}}{\pi_{42}} &  =  & 
               \frac{A}{A + 2} \exp( \Delta S)  \\
M33a / M33b &  \pi_{33a} = \pi_{33b}  &  =  & \exp ( \Delta S) \\
M02 / M20 & \frac{\pi_{02}}{\pi_{20}} &  =  & 
                8 \frac{A }{A + 2}  (d - 1) \exp( \Delta S) \\
M00    & \pi_{00} &  =  &  n_{a}/ n_{b} \\
\hline
\end{array}
\end{displaymath}

\subsection{Verification \label{hyper_verify}}
\begin{table}[p]
\begin{tabular} {|l|c|c|c|c|c|c|} \hline
 & Diagram & $\sigma(E) $ & ${\cal N}(E)$  & d=2 & d=3 & $N$  \\ \hline
a & \diagram{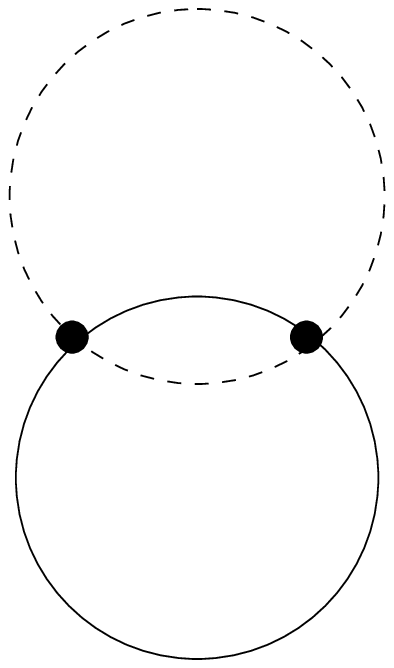}{1.0} & 8 & $4 d (d-1)$   & 1 & 3 & 3.01(2)\\
b & \diagram{n4straight.dual} {1.0} & 4 & $8 d (d-1)^2$ & 4 & 24 & 23.98(4)  \\
c & \diagram{n4crump.dual}{1.0} & 8 &$4 d (d-1)$ & 1 & 3 & 3.00(2)\\
d & \diagram{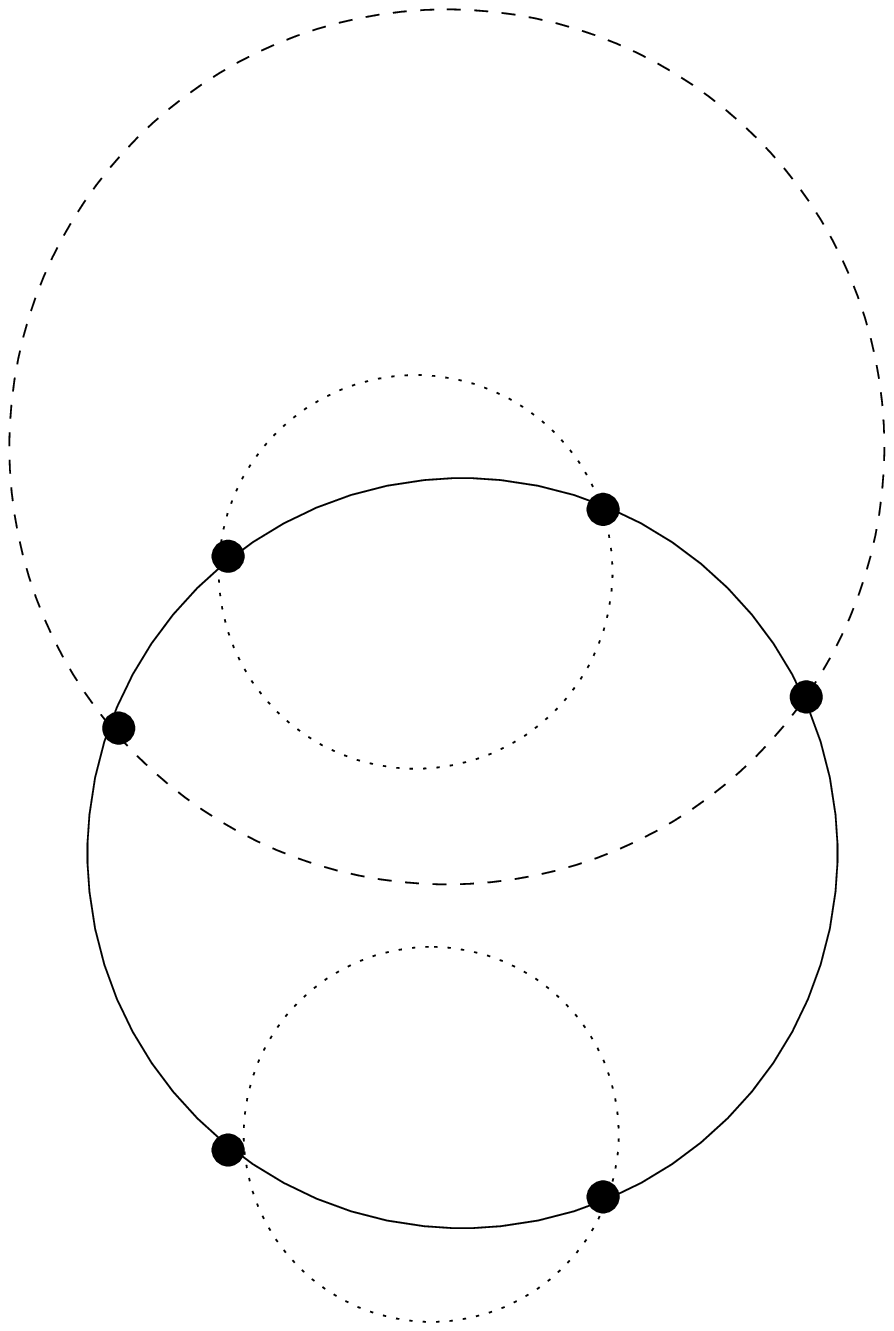}{1.7} & 4 & $16 d (d-1)^3$ &8 & 96 & 95.97(8) \\
e & \diagram{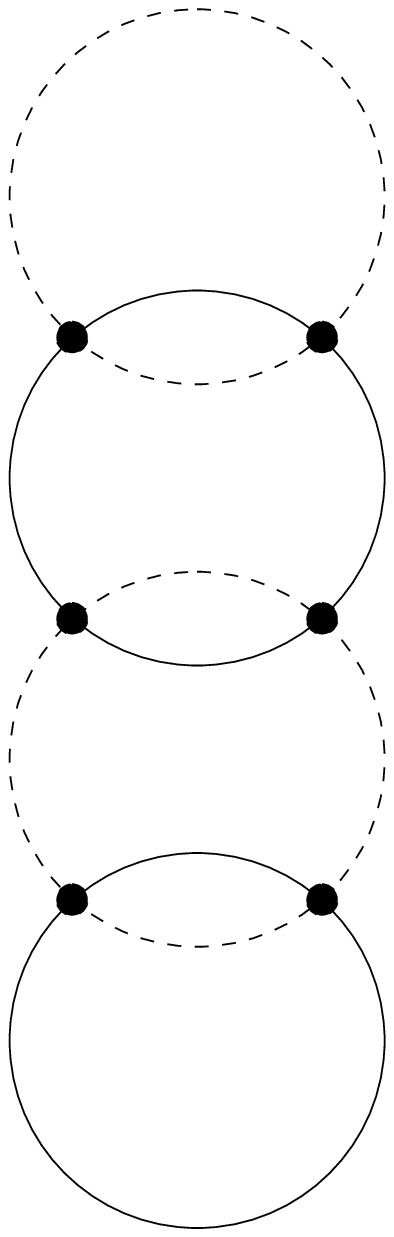}{0.9} & 2 &$16 d (d-1)^3$ & 16 & 192 & 192.000 \\
f & \diagram{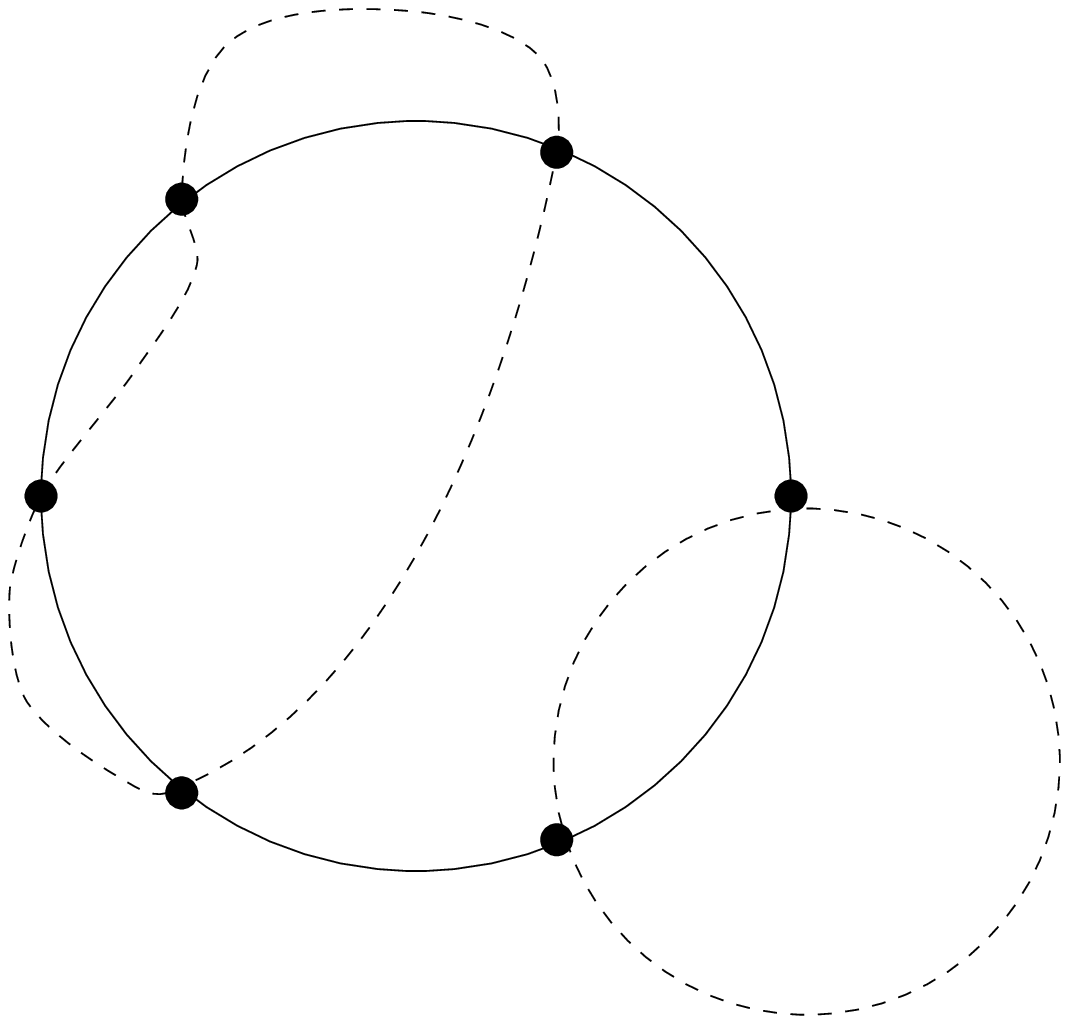}{1.7} & 1 & $8 d (d-1)^2$ & 16 & 96 & 96.00(9) \\
g & \diagram{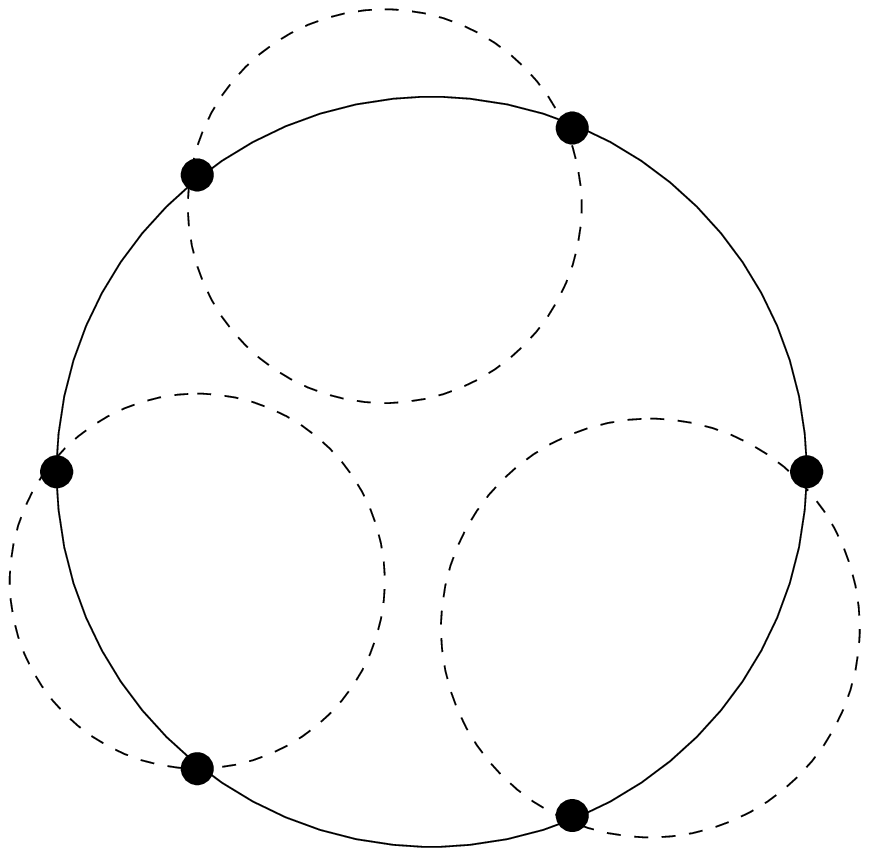}{1.4} & 6 & $16 d (d-1)^3$ & 5 1/3& 64 & 63.99(6)  \\ 
h & \diagram{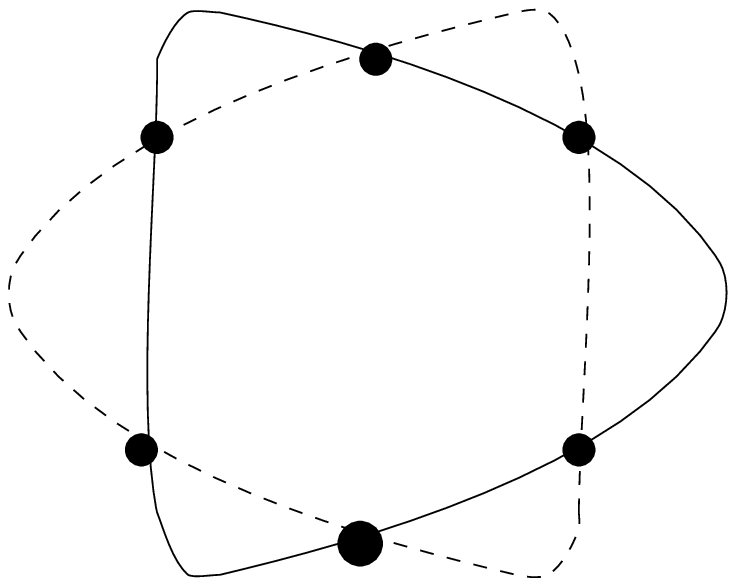}{1.3} & 12 & $4 d (d-1)$ & 2/3 & 2 & 2.09(11)\\
i & \diagram{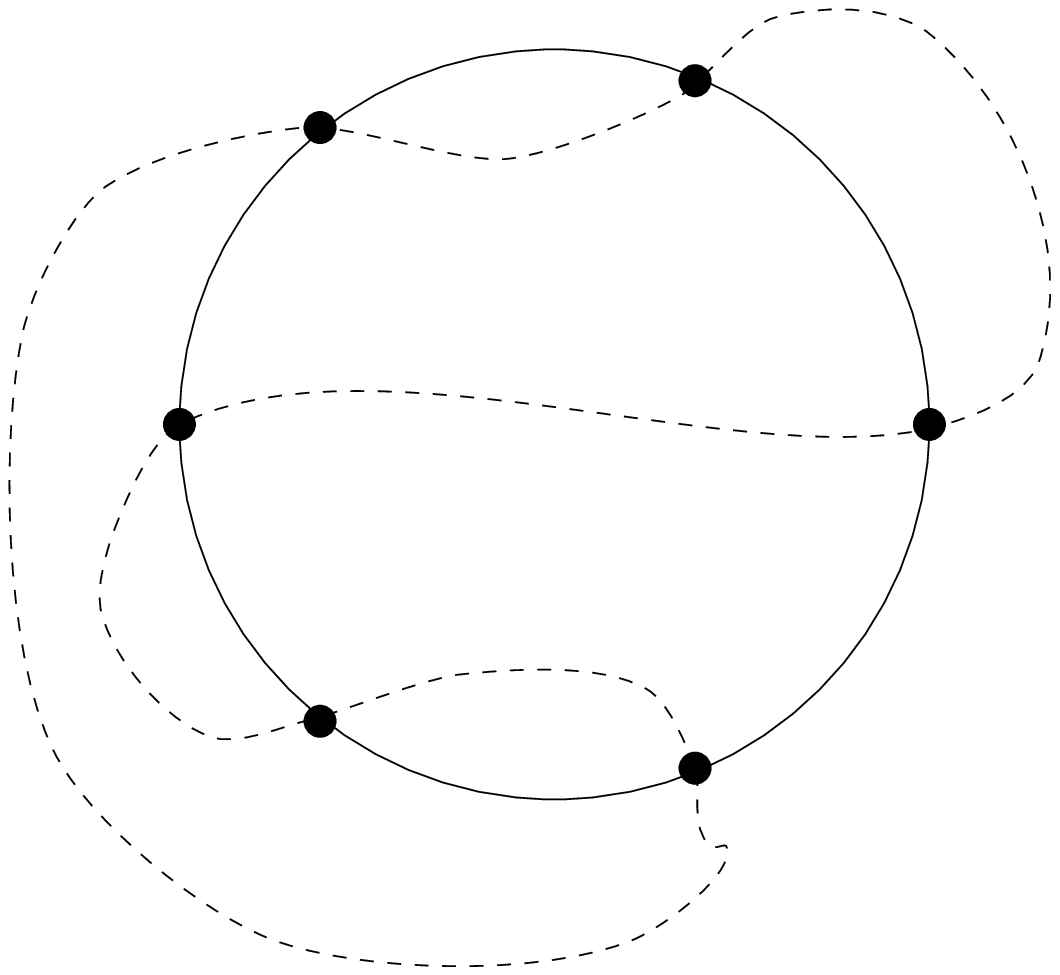}{1.3} & 4 & $4 d (d-1) $ & 2   & 6 & 5.97(8)\\
j & \diagram{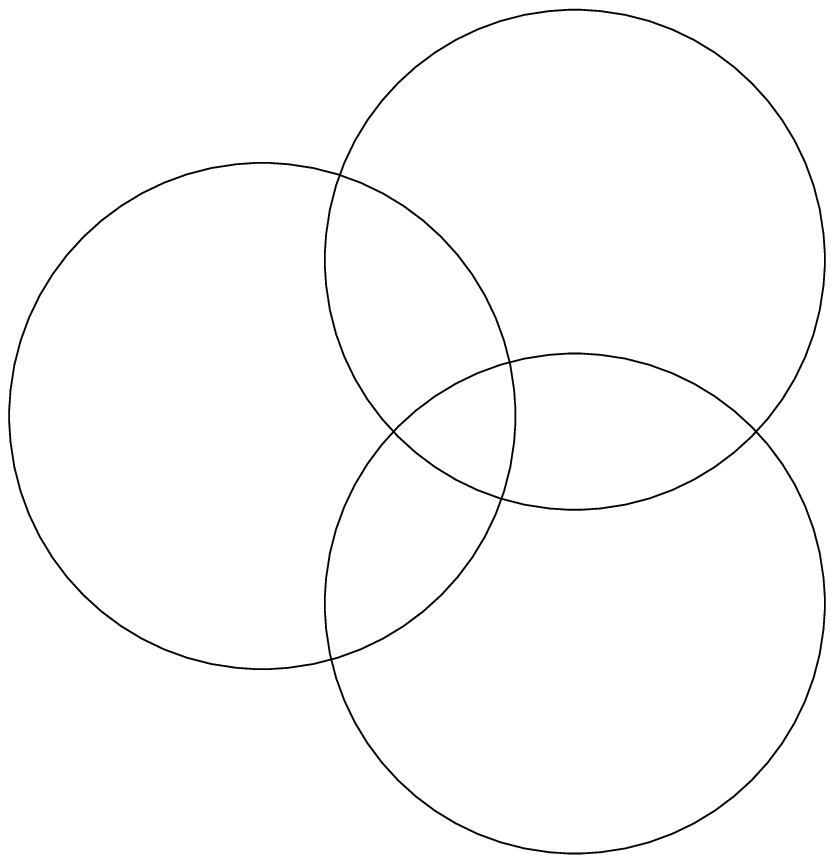}{1.3} & 24 & $8 d (d-1)(d-2)$ & 0 & 2  & 1.98(10)\\
k & \diagram{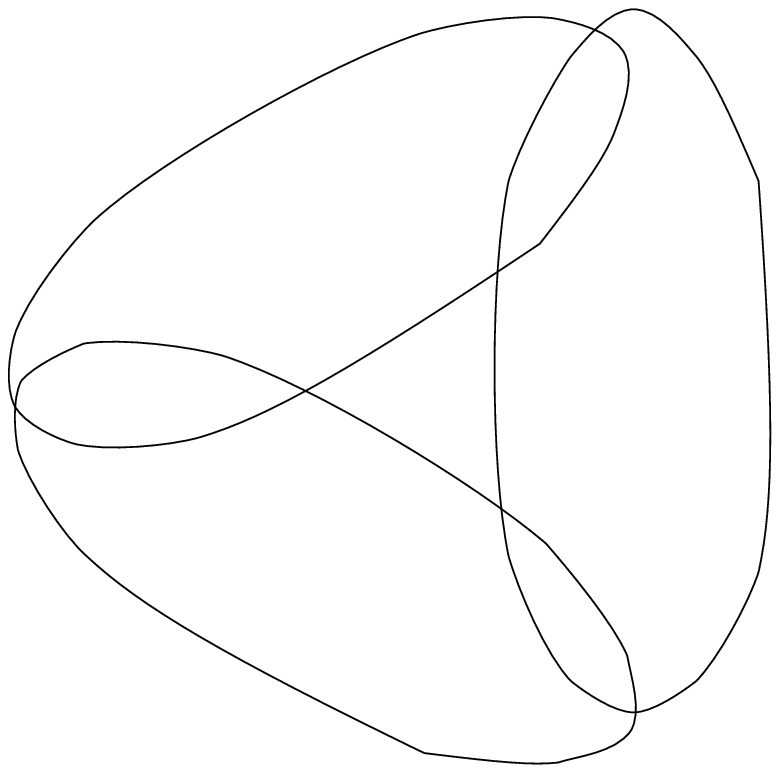}{1.1} & 6 & $8 d (d-1)(d-2)$ & 0 & 8 & 8.03(6)
\\ \hline
\end{tabular}
\caption{Diagrams of size 2,4,6, the symmetry factor $\sigma $, the number of
embeddings $\cal N$, the theoretical weight in $d=2$ and $d=3$ dimensions and
the distribution $N$ obtained numerically in $d=3$. The numerical 
distribution is normalized such that diagram $e$ has weight $192$.}
\label{hyper_diagrams}
\end{table}

To check the validity of the algorithm and its implementation we simulated
very small surfaces $E$ and compare the distribution of surfaces with a 
theoretical distribution which we have calculated explicitly for the smaller
surfaces. To avoid too many technicalities we 
present here in detail only the verification for the model without the 
extrinsic curvature term in the action. In table (\ref{hyper_diagrams}) 
we list the smallest diagrams. The number $\cal N$ is the number of possible
embeddings in the embedding space compatible with the internal geometry 
of the surface. The symmetry factor $\sigma $ are also given. Below we will 
show that the weight of a graph is ${\cal N}/\sigma $. We list the weight 
in the columns $d=2$, $d=3$ for two and three dimensions. In the last column 
the weight measured numerically with our algorithm in $d=3$ dimensions is
given. They are in perfect agreement with the theoretical values.

In writing the partition function (\ref{hyper_Z}) we were a bit sloppy
because we have silently dropped the symmetry factor $\sigma(E)$. The
justification for doing so is the fact that the typical large surface
considered so far is not symmetric and hence for almost all large
surfaces $\sigma =  1$. The small surfaces we consider now however are  
quite symmetric with $\sigma > 1$ and hence the factor becomes important. 
The precise partition function is
\eq{
{\cal Z}(\beta)= \sum _{E \in \cal S}  \frac{1}{\sigma (E)} e^{- S[E]} 
        = \sum _{\mbox{Diagrams}}  e^{-S[E]} \frac{{\cal N}(E)}{\sigma(E)} 
\label{hyper_Z_precise}
.} 
In the rightmost expression we re-express the partition function as a sum over
diagrams. The number ${\cal N}(E)$ counts the number of possible ways of
embedding a surface dual to the  diagram $E$  in the  
$D$-dimensional space.  This number is most easily counted in the geometrical 
representation. In the following we describe how to calculate $\sigma $.

\begin{table}[bt]
\centerline{
\diagram{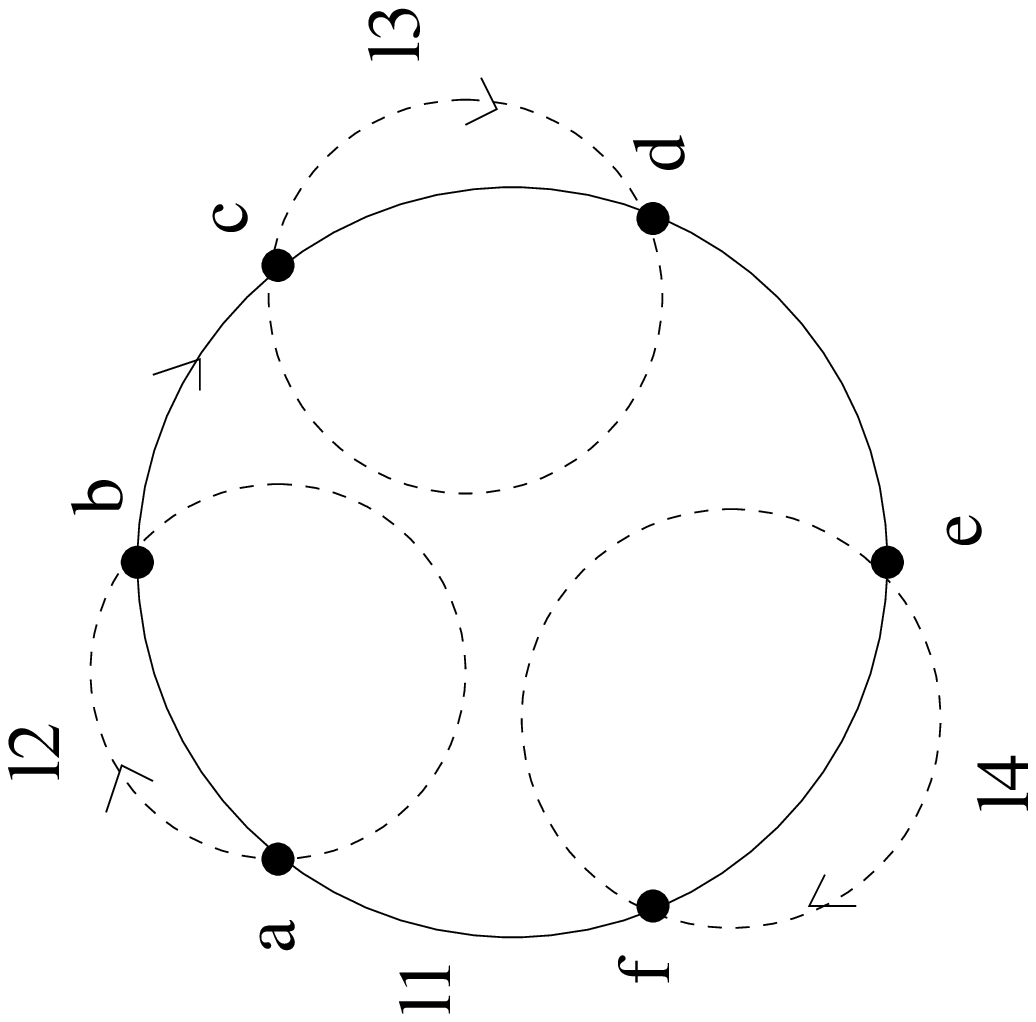}{3.0}\hspace{1cm}
\begin{tabular}{r|cccccc}\hline
a & b & b & b & f & 2 & 1 \\ 
b & c & a & a & a & 1 & 2 \\ 
c & d & d & d & b & 3 & 1 \\
d & e & c & c & c & 1 & 3 \\ 
e & f & f & f & d & 4 & 1 \\ 
f & a & e & e & e & 1 & 4 \\ 
\end{tabular}
}

\noindent
\begin{center}
\begin{tabular} {ccc}

abcdef &  1234 & ++++\\
cdefab &  1342 & ++++\\
efabcd &  1423 & ++++\\
dcbafe &  1324 & -+++\\
bafedc &  1243 & -+++\\
fedcba &  1432 & -+++\\
\end{tabular}
\end{center}
\caption{The adjacency matrix (top right) for diagram $g$ using the choice of
         labels given in the top left diagram. The lower table lists the six
         re-labelings which leave the diagram unmodified.}
\label{adjacency}
\end{table}

To depict the diagrams used to verify the algorithm we choose the dual
representation in table \ref{hyper_diagrams}. The hyper-cubic surface model is 
dual to a $M^4$-theory \cite{ek82}. The $M$ are hermitian matrices. The 
graphs are therefore fat graphs discussed in context with the matrix model 
in section \ref{matrix}. 
For  clarity, however, we show only the thin graphs, known from scalar
$\phi ^4$ theories, although they  do not contain a complete description of the
geometry. For example a $\phi ^4$ graph does not allow to define the
geometrical property of two propagators being 'opposite' to each other 
because without the indices carried on a fat
graph the propagators interacting at a vertex can not always be distinguished.
In all graphs shown in this section we use the convention to draw propagators 
which are dual to opposite links on the surface as  opposite links emerging 
from a vertex. By propagator loops we denote closed propagator loops which go 
along opposite propagators only, i.e. which do not form a right angle at any 
vertex. Geometrically this is a walk on the surface with one internal 
coordinate kept constant.  

The symmetry factor $\sigma $ for the diagrams is a discrete remnant of the 
reparametrization group. When drawing a diagram one has a
choice of how to label vertices and propagator loops. We consider oriented
surfaces, therefore one has also a freedom in the choice of the orientation 
of the loops in the dual diagram. Note that all this involves only internal 
properties of the diagram, the embedding is unimportant for this. The 
symmetry factor $\sigma $ counts the number of relabelings which lead to 
the same diagram. For not too large diagrams one can directly probe all 
possible
\eq{
N! \; l! \; 2^l 
}
relabelings for a diagram with $N$ vertices and $l$ propagator loops and count
the number of invariant relabelings. As an example consider diagram $g$. 
In table \ref{adjacency} the  
adjacency matrix for this diagram with the labels depicted in the 
left-hand figure is written down. One can easily convince oneself that the 
relabelings listed in table \ref{adjacency} leave the diagram unmodified. 
The first transformation is the identity transformation, number two has to
be interpreted as follows:
Rename vertex $a \rightarrow c, b \rightarrow d, \cdots$, loops
$1 \rightarrow 1, 2 \rightarrow 3, \cdots $ and maintain the orientation
for loops $1,2,3,4 (++++)$. We found six invariant transformations, the
symmetry factor for this diagram therefore is $\sigma = 6$.

%% file: sqg.tex
\chapter{Four dimensional simplicial quantum gravity\label{sqg_chap}}
\section{Introduction}

Quantum gravity in four dimensions, as defined by the path integral
(\ref{qg_path}) and the action (\ref{qg_cont_act}), is very difficult to deal
with. The standard tool perturbation theory does not work because the model 
is not renormalizable.  This can be seen from the fact that the coupling 
constant $G$ has a negative mass dimension. Higher order loop corrections 
will therefore always generate more counter terms.

One possible interpretation of the breakdown of perturbation theory is
that general relativity is not applicable at the Planck scale and is just 
a low energy effective theory of some unknown underlying theory. QCD, which is
a renormalizable theory, provides us with an example of this: 
the effective low energy theory for the pion is non renormalizable. 
One possible candidate for the underlying theory of quantum gravity is 
string theory. 

On the other hand,  a somewhat more conservative interpretation is that 
the divergences in the perturbative expansion do  not necessarily mean
that general relativity does not work at the Planck scale. It may be that they
appear just because of the naive way we do perturbation theory, we 
perturb around the wrong ground state. The theory may instead  
be well defined at some non-perturbative $UV$-fixed. Similarly to the nonlinear
$\sigma $ model, which is asymptotically $UV$ free \cite{p75}, the concept of
asymptotic safety of four dimensional gravity was introduced \cite{weinberg}.
It assumes the existence of a safe sub-space of the coupling space. For
couplings on the safe surface the renormalization flow leads to a save 
non-perturbative $UV$-fixed point. 
This conjecture is supported by the dimensional $2 + \epsilon$ expansion 
starting from two dimensional quantum gravity, where an $UV$-fixed point 
was found  \cite{kn90} for $\epsilon = 2$. The model considered in this 
chapter, four dimensional quantum gravity, may provide us with a candidate 
for the fixed point. 

The structure of four dimensional manifolds is  far less understood than
in two dimensions. For example, there does not exist a
topological classification of four-manifolds. In this situation the
definition of the measure $\frac{{\cal D}[g]}{Vol(Diff)}$ is very difficult. 
One major conclusion  of dynamical triangulations (see section \ref{matrix})
is the fact that the sum over discrete triangulations, which replaces 
(\ref{dt_meas}), defines the measure for the two dimensional model 
in a sensible way. This is a justification for {\em assuming} that the 
corresponding sum in the generalized four dimensional model, known as four
dimensional simplicial quantum gravity,  defines a sensible measure as well. 
The assumption implies that  the reparametrization degrees of freedom are 
removed in the discrete model. In fact, diffeomorphism invariance is 
completely lost because one can not deform smoothly the discrete structure, 
which we describe in the next section. However, the basic symmetry group of 
general relativity is reparametrization invariance. One therefore has to 
check whether this invariance is restored in the continuum limit of the 
discrete model, if it exists. 

Simplicial quantum gravity in four dimensions has been extensively studied 
by various groups. The main results are briefly summarized at the end of 
section \ref{sqg_intro}. In the beginning, the results looked quite promising.
In the recent years however it became clearer that the model as it stands has 
some problems. In numerical simulations it was found that the model has two
phases. The dominating geometries in these phases are, however, not extended
four dimensional geometries as one would expect. The weak coupling phase 
shows the behavior of branched polymers,  essentially  one dimensional 
highly branching thin tubes.  The strong coupling phase is crumpled with a
large connectivity. It has an infinite internal Hausdorff dimension. 
The transition between the two phases is first order \cite{bbkp96,b96}. 

Based on calculations which assume that mainly the conformal deformations are 
important and ignore transverse excitations of the metric, it was suggested in
\cite{ant2} that the collapse to branched polymers  can be prevented by 
coupling matter-fields to the model.  This is opposite to what happens in two 
dimensions where additional matter drives the system into a branched 
polymer phase\footnote{See also section \ref{hyper_introduction}.}. 
In section \ref{sqg_vec} we investigate the phase diagram of 
the model with multiple vector fields and show that indeed the branched
polymer phase disappears.

Another open question in four dimensional quantum gravity is what is the
importance of topological excitations.  
It is not a priori clear, whether in the path-integral formulation of
quantum gravity the sum over metrics should also run over topologies.
It seems a bit artificial to fix the topology. Let us for the moment consider
spherical topology. One may ask then why space-time can branch but is not
allowed to  join. Branches do not change the topology but if the branches 
re-join the manifold has a hole, i.e. it is not a sphere anymore.  
In two dimensions it can be shown that the naive sum over topologies 
diverges. In the double scaling limit, however, it is possible to define 
the model with the inclusion of all topological excitations. 
Using this two dimensional case as a guidance we have checked 
numerically  whether in four dimensions a similar scaling is possible.

\section{Simplicial quantum gravity\label{sqg_intro}}
The dynamical triangulations  model described in section \ref{matrix} is an 
example how to discretize the space of metrics. In this two dimensional model
the integral over equivalence classes of metrics is replaced by a sum over 
triangulations. Simplicial quantum gravity generalizes this discretization to 
three \cite{am91} and four \cite{am92, aj92} dimensions. In what follows 
we use $D=3,4$. The integral over equivalence classes of 
$D$-dimensional metrics is replaced by a sum over $D$-dimensional simplicial 
manifolds. A $D$-dimensional simplicial manifold is a simplicial 
complex $K$ with the additional constraint that  the neighborhood of each 
vertex $p \in K$ is homeomorphic to a $D$-dimensional ball. A simplicial 
complex is built from $d$-simplices, where a $d$-simplex contains $d+1$ 
vertices. A 0-simplex is a point, a 1-simplex a link, a 2-simplex 
a triangle and 3-simplex a  tetrahedron and so on.

In one and two dimensions a simplicial manifold can be built by gluing
together $d$-simplices along their $d-1$ faces. Each complex constructed in 
this way satisfies the manifold condition.  In higher dimensions the
resulting complex may however violate the manifold condition, i.e. the 
neighborhood of a vertex is not homeomorphic to the  $D$-ball. 
One can derive relations for the numbers $N_i$ of $i$-simplices on the 
manifold which are known as Dehn-Sommerville relations \cite{s27}:
\eq{N _i = \sum _{k=i}^{D} (-1)^{k+D}{k+1 \choose i+1}N_k. \label{iglo}}
These equations are not independent, they can, however, be used to eliminate
the $N_{2i+1}$ in even and $N_{2i}$ in odd dimensions. 
One can use the discrete expression for the Gauss Bonnet theorem 
(\ref{gauss_bonnet}), the Euler relation 
\eq{\sum _{i=0}^{D}(-1)^{i+d} N_i = \chi _d,  \label{euler_character}}
to eliminate one more $N_i$, the character $\chi _d$ is a topological 
invariant.  This means that in $D=3,4$ only two of the $N_i$ are independent.


So far we have defined a geometry but not a metric. A Riemannian metric is 
obtained by demanding that:
\begin{itemize}
\item The metric is flat inside a $D$-simplex
\item The metric is continuous at the $D-1$ dimensional faces of the 
$d$-simplex.
\item The $(D-1)$ dimensional faces of a $D$-simplex $S$ are flat linear
subspaces of $S$.
\end{itemize}
With this definition the metric inside a $D$-simplex $S$ is determined by the 
length of the $1$-simplices $s_i \in S$. As we will motivate below simplicial
quantum gravity considers only manifolds with fixed edge lengths $l(s_i) = a$.

To define a path integral of the form (\ref{qg_path}) for simplicial
manifolds we  sum  over all metrics, which can be represented by
such manifolds. In the Regge approach (see for example \cite{d}) one 
considers the metric space obtained by varying the edges length $a_i$ on a 
simplicial manifold with {\em fixed} connectivity. It is however quite 
difficult to respect the geometrical constraints imposed on the $a_i$'s in 
the integration.  Furthermore, a metric is not uniquely parameterized by 
the $a_i$'s. Therefore one has to deal with an unknown weight.
Consider for example the triangulation $T$ of a flat surface
with equal edges length. Moving a vertex in the plane of the surface 
obviously does not change the metric, although this degree of freedom is 
integrated over in Regge  gravity. On the other hand metrics which are very 
different from the manifold $T$, are represented by singular simplices 
which contain extremely short or long edges length. It is therefore not 
obvious whether integration over the $a_i$ really covers the whole space of 
metrics. 

Simplicial quantum gravity tries to avoid the  mentioned problems by
summing over {\em dynamical} simplicial manifolds $\cal T$. In other words 
the dynamics of the model now lies in the connectivity of the simplicial
manifold. The next step is to pick up a parameterization $a_i = a$.
This introduces an $UV$ cut-off into the model and removes 
reparametrization degrees of freedom. As in two dimensions one assumes 
that 
\eq{\label{Gdmm} \int {\cal D}[g] \rightarrow \sum _{\cal T} \frac{1}{\sigma} }
defines the measure in a sensible way.  The symmetry factor $\sigma $, a
discrete remnant of the reparametrization group, is important mainly for 
small manifolds because the typical large manifold is not symmetric. 
In the following we will drop this factor.

Before one can define gravity using the piecewise linear metric defined above
one has to discretize the Einstein Hilbert action (\ref{qg_cact}). 
The cosmological term, which is the invariant $D$-volume of
the manifold  is rewritten as the  sum of the volumes of the $D$-simplices 
the manifold is built from~: 
\eq{\int \mbox{d}^D \xi \sqrt{g} \rightarrow 
    \sum _{\mbox{$D$-simplices }} V_D = N_D V_D. \label{sqg_cosmo}}
To define the curvature on the discrete lattice one uses the
parallel transport of a vector on a closed loop on the manifold. In a curved
continuous space, the vector $V_a$ changes by 
\eq{\Delta V_a = \frac{1}{2} R^d_{abc} V_d \oint x^c \mbox{d} x^b}
On the discrete manifold one 
considers a closed loop $\alpha $ dual to a $D-2$-subsimplex $s$ 
on the manifold. The loop $\alpha $ connects the centers of the $D$-simplices
$S_i$ which have $s$ as a common sub-simplex.  By definition the space inside 
the $S_i$ is flat, $\Delta V_a = 0$ on these parts. However when $\alpha $ 
passes through a $(D-1)$ face the orientation of $V_a$ changes by the  deficit 
angle 
\eq{\Theta _D = \arccos\left ( \frac{1}{D} \right ) } 
in the plane orthogonal to $s$. The rotation of $S_a$ on the closed loop 
$\alpha $ dual to $s$ is therefore described by a single angle
\eq{R_s =  2 \pi - o(s) \Theta _D, } which 
we assign to $s$. The  order $o(s)$  is the number  of $D$-simplices 
which contain $s$ as a common sub-simplex. The integrated curvature 
is a sum over all $(D-2)$ simplices $s$ contained in the manifold~:
\eq{
\frac{1}{G_N}\int\mbox{d}^D \xi \sqrt{|g|} R  \rightarrow
\sum _s 2\pi - o(s)\Theta _D \propto c_D N_{D-2} - \frac{D(D+1)}{2} N_D
\label{sqg_curv}} 
In deriving this we used the fact that a $D$-simplex has 
$\frac{D(D+1)}{2}$ $(D-2)$ sub-simplices and introduced the average order 
\eq{c_d = \frac{2 \pi}{\Theta _D}} 
of the $s$ in flat space. For example in two dimensions one has 
$c_2 = 6$, the order of all vertices $s$ in the triangulation of a flat 
surface is $o(s) = 6$. In higher dimensions there does not exist a 
triangulation of flat space, which is  reflected by the fact that $c_d$ is 
non-integer for $D > 2$. 
Nonetheless one can ask for the average order of a vertex in a hypothetical
flat triangulation.

One gets the discrete Einstein Hilbert action by collecting
(\ref{sqg_cosmo}) and (\ref{sqg_curv}) 
\eq{
S =  \kappa _D N_D - \kappa _{D-2} N_{D-2}, 
\label{sqg_act}
}
where the coupling $\kappa _{D-2}$ is proportional to the inverse 
Newton constant $G$ and  $\kappa _D$ is a linear combination of $1/G$ and 
the cosmological constant $\Lambda $. One can consider other terms in the
action as well. For example in \cite{r2term} the model with a curvature
squared term was analyzed, this term, however, does not induce a different
critical behavior.

The grand canonical partition function is defined as
\eq{
\label{gcZ}
{\cal Z}(\kappa _{D-2}, \kappa _D) = \sum _{{\cal T } } 
    \exp (-\kappa _D N_D + \kappa _{D-2} N_{D-2}) = 
    \sum _{N_D}  \exp(-\kappa _D N_D) Z(\kappa _{D-2}, N_D)  
}
where the sum runs over $D$-dimensional simplicial manifolds $\cal T$ with 
fixed edges length $a$ and fixed topology. The canonical partition function
is defined as
\eq{
\label{cZ}
Z(\kappa _{D-2}, N_D) = \sum _{{\cal T}_{N_D}}
                              \exp ( \kappa _ {D-2} N_{D-2}),
}
where the sum runs over four dimensional simplicial manifolds ${\cal T}_{N_D}$
with $N_D$ four-simplices and fixed topology. 

\begin{figure}
\hspace{3cm}
\psfig{file=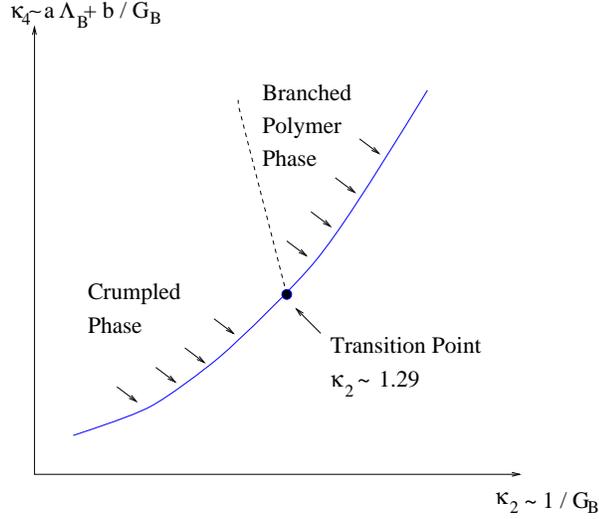,angle=270,height=7cm}
\caption{The phase diagram of four dimensional simplicial quantum gravity
  without matter.}
\label{pd_pure}
\end{figure}

We want to close this section with a summary of the results obtained
for the four dimensional model. The most important question, namely whether or
not the partition function (\ref{gcZ}) is well defined has been 
disputed.  
The exponential $\exp(-\kappa _4 N_4)$ can prevent (\ref{gcZ}) from diverging 
only if the number of simplicial manifolds ${\cal N}(N_4)$ is exponential 
bounded.  The numerical evidence is in favor of the existence of this 
bound \cite{entropy1, entropy2, entropy3}. Recently also an analytic 
argument \cite{bbcm94} was given for this. The bound defines a critical line 
$\kappa _4^c(\kappa _2)$ drawn as the solid line in the phase
diagram figure \ref{pd_pure}. For $\kappa _4 < \kappa _4^c(\kappa _2)$ 
eq. (\ref{gcZ}) diverges and the 
model is undefined. 
The continuum limit is taken by approaching the critical
line from above, $\kappa _4 \rightarrow \kappa _4^{c+}$. 

Numerically one finds two phases \cite{aj92, phase_struc2}. 
The strong coupling phase is dominated by highly connected manifolds with very 
large, possible infinite, internal Hausdorff-dimension. The dominating
geometries in the weak coupling phase have the structure of branched polymers.
The internal Hausorff dimension is $d_h = 2$, the entropy exponent 
$\gamma = \frac{1}{2}$ in this phase. 
For the model coupled to matter fields investigated so far 
\cite{matter1, matter2, matter3} only a renormalization of the 
coupling-constants was observed.

The strong coupling phase  is dominated by two singular vertices 
\cite{hi95, cat95}. By singular we mean that the local volume assigned to 
these vertices is a finite fraction of the total volume in the 
thermodynamical limit; i.e.  the local volume grows linear with $N_4$.
The appearance of these singular vertices can be understood in a constrained
mean-field theory \cite{bb97}. 
The two  singular vertices are joined together by a sub-singular link $l$; 
its local volume  grows like $N_4^{2/3}$.  The two phases are separated  by 
a first order phase transition \cite{bbkp96}, \cite{b96} at 
$\kappa _2 \approx 1.29$. 
The singular structure which dominates the crumpled phase dissolves at this
 transition. On small lattices the transition actually appears in two steps 
\cite{bbpt96, taba, ckr97}. Coming from the crumpled phase the sub-singular 
link disappears first. In the next step the singular vertices dissolve. 
For large volumes these sub-transitions seem to merge \cite{ckr97}.   
\clearpage

\section{Topology}
\subsection{Introduction}
One of the open questions in quantum gravity in dimensions larger than two  
concerns the role of topological excitations.  It seems a bit artificial to fix
the topology by hand, which is what is usually done in the simulations.  
The situation in four dimensions is  complicated
by the fact that as of date no topological classification of four manifolds is
known. Allowing an arbitrary topology will result in a very
complicated structure at the Planck scale  because curvature
fluctuations become larger at smaller scales \cite{wheeler64}. 
In \cite{hawking78} it was estimated that the dominant contribution to the 
path integral would come 
from space-times where the Euler character $\chi$ is of the order of the 
volume of the space-time in Planck units. Such a configuration with many
holes is commonly called a space-time foam. 

In \cite{bak94} a tensor model similar to the matrix model 
(section \ref{matrix}) was considered, which generates pseudo-manifolds with 
fluctuating topologies. It was shown that for large couplings $\kappa $ the 
pseudo-manifold decomposes in very small connected components while in the 
weak coupling phase there is only one connected component. However, it is not
obvious how these pseudo-manifolds, which have now unique local internal
dimension, are related to four dimensional space-time.
 
In this work check numerically, 
if topology fluctuations  are dynamically suppressed. In a theory containing 
topology fluctuations only those topologies contribute to the sum in the 
thermodynamic limit, which maximize the extensive part of the free energy. 
Other contributions are exponentially suppressed. Two dimensional quantum 
gravity provides us with an example of  to define a  theory which contains
topological excitations. In the double scaling limit which we discussed in 
section \ref{matrix}, {\em all} topological excitations are present.
The physical reason is that the leading contribution  to the free
energy is independent of the topology.  If one believes 
that {\em all} topological excitations are present also in four dimensions 
the bulk volume contribution to the free energy should also be independent of 
the topology.   We investigate
numerically three different topologies and show that  in these
cases, up to the leading order, the free energy does not depend
on the topology. We observe a topology dependence in the next to
leading order.  We study those volume corrections and analyze their 
kinematic sources.

\subsection{Model and methods}
Given the canonical partition function (\ref{cZ}), 
we define  the intensive free energy as 
\eq{ \label{freeE}
f(\kappa _2)  = \lim _{N_4\rightarrow\infty} \frac{1}{N_4} 
\log Z(\kappa _2, N_4)
.}
For large $N_4$ the free energy is assumed to have  the form:
\eq{ F(\k_2,N_4) =  N_4 f(\k_2) + \delta(\k_2,N_4), \label{F} }
where the function $\delta$ is a finite size correction, 
i.e. for any $\k_2$ 
\eq{
\lim_{N_4 \rightarrow \infty} \frac{\delta(\k_2,N_4)}{N_4} = 0 .
\label{Fn4}
}

The derivatives of the free energy are convenient to recover some basic
properties of the theory. The action density
\eq{
r(\kappa _2,N_4) = \frac{1}{N_4} \frac{\pd F}{\pd \k_2} = 
\frac{\langle N_2 \rangle}{N_4}
\label{den}
}
is normalized  such that it becomes an intensive quantity in the thermodynamic 
limit. We call the derivative with respect to $N_4$, 
\eq{
\tilde{\kappa}_4(\kappa_2,N_4) = \frac{\pd F}{\pd N_4}  .
\label{k4}
}
the pseudo-critical value of the parameter $\k_4$. 
This definition differs slightly from the one proposed in  \cite{entropy1}.
This quantity is a measure of finite size dependence of
the free energy,  i.e. of how fast it approaches the thermodynamic limit
$\tilde{\k}_4 \rightarrow \tilde{\k}_4(\k _2)$ for $N_4 \rightarrow \infty $. 
In this limit  the value $\tilde{\k}_4 (\k _2)$ defines the critical line of 
the model corresponding to the radius of convergence of the series (\ref{gcZ}).

The pseudo-critical value $\tilde{\kappa}_4(\kappa_2,N_4)$ is related to the
action density by
\eq{
\frac{\pd \tilde{\k}_4}{\pd \k_2} = r + N_4 \frac{\pd r}{\pd N_4}
\label{dk4}
}
where the second term goes to zero when $N_4$ goes to infinity. Therefore 
the critical parameter $\tilde{\k}_4$ becomes an integral of $r$ in the
thermodynamic limit. Thus, if $r$ is independent of the topology
so is $\tilde{\k}_4$ unless the integration constant depends on the topology.
To fix the integration constant it is sufficient to measure $\tilde{\k}_4$
for only one particular value of $\k_2$.

The action density (\ref{den}), the normalized average number of triangles on
the manifold, is a standard observable which is easily accessible in  
canonical simulations. For the pseudo-critical value  
$\tilde{\k}_4$ (\ref{k4}) more unconventional methods are required. 
The baby universe method,  which we used in the context with hyper-cubic 
random surfaces to extract the next to leading correction, unfortunately 
works only in the  elongated phase. Following \cite{entropy1}, we adopt here 
a  more general method based on grand-canonical simulations, which works 
equally well in all phases. We use multi-canonical simulations with a potential
$U(N_4)$. From the volume fluctuation one can extract the volume dependence of
$\tilde{\kappa} _4$. We expect  the quantity $\pd F / \pd N_4$, which we want
to measure, to smoothly approach the function $f(\kappa _2)$. 
The freedom one has in choosing  the potential $U$ can be
used to minimize the statistical error of the observable.
A Gaussian potential controlling fluctuations around a fixed 
volume $V_4$ is well suited for this problem \cite{entropy1}~:   
\eq{
U = \k_4 N_4 + \frac{\gamma}{2} (N_4 - V_4)^2
\label{u}
}
but other potentials can be used as well \cite{aj92}.
With this potential the multi-canonical partition function reads~: 
\eq{
{\cal Z} = \sum_{T}  e^{\k_2 N_2 - U(N_4)} =
\sum_{N_4} e^{F(\k_2,N_4) - \k_4 N_4 - \frac{\gamma}{2} (N_4 - V_4)^2}
}
The  distribution of the volume $N_4$ becomes Gaussian around $N_4 = V_4$ 
if one expands  $F(\k_2,N_4)$ around this point
\eq{
P(x = N_4 - V_4) \sim 
\exp( -\frac{\Gamma}{2} (x - x_0)^2 + ...)
\label{gauss}
}
The parameters $\Gamma$ and $x_0$ are
related to $F$ and $U$ by: 
\eq{
\Gamma = \gamma - \pd^2 F / \pd V_4^2, \quad 
x_0 = \big( \pd F / \pd V_4 - \k_4 \big)/\Gamma.
\label{gama_bama}
}
Both  quantities can be measured in the simulations, 
by tuning $\k_4$ close to the derivative $\pd F / \pd V_4$ and  
choosing $\gamma$  much larger than the second derivative 
$\pd^2 F / \pd V_4^2$. The parameter  $\Gamma$ can be extracted 
from the width of the distribution and $x_0$ from the
shift of the maximum from $N_4=V_4$. This gives 
an estimate for the pseudo-critical coupling at $V_4$~: 
\eq{
\tilde{\k}_4 = \pd F/\pd V_4 = \Gamma x_0 + \k_4.
}
In this formula, $\k _4$ is the coupling used in the simulation from which
$x_0$ and $\Gamma $ were extracted.  The estimator can
be improved by minimizing $x_0$. This can be done recursively by
replacing $\k_4 $ by $\tilde{\k}_4$ in subsequent simulations. 

The value for the multi-canonical coupling $\gamma $ should be chosen large
enough so that the volume fluctuations are small compared to the scale of
change of $F(N_4,\kappa _2)$. On the other hand, the algorithm used in the
simulation is grand canonical and requires volume fluctuations for
ergodicity. Therefore $\gamma $ should be small enough to allow sufficient 
$N_4$ fluctuations.   We checked that for $\gamma=0.0001$, $\Gamma$ computed 
from the width of the distribution $\Gamma = 1/\sqrt{\sigma^2(N_4)}$, 
i.e. $\Delta N_4 = 100$, was equal to $\gamma$ within the error bars. From
(\ref{gama_bama}) it follows  
that $\gamma \gg \pd^2 F / \pd^2 V_4$. Therefore the free energy 
changes very slowly across the range  of the $N_4$ distribution width as 
needed for the approximation (\ref{gauss}) to be valid.

\begin{figure}
\hspace{3cm}
\psfig{file=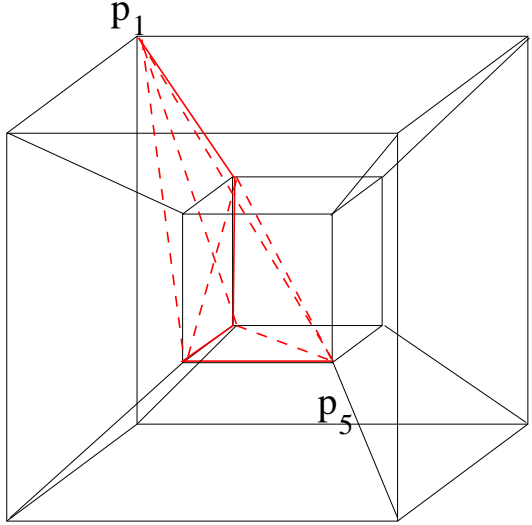,angle=270,width=7cm,height=7cm}
\caption{A four tetrahedron built into a four dimensional hyper-cube.}
\label{4dcube}
\end{figure}
Because the local moves \cite{gv92} used in the 
Monte Carlo simulation preserve the topology $t$ of the manifold, $t$ is given
by the topology of the starting configuration. In our simulation we  
simulated a spherical $S^4$ topology and two different tori, namely
$S^3\times S^1$ and  $S^1 \times S^1 \times S^1 \times S^1$.

The starting configuration for the sphere is, as usual, the $4d$-boundary
of a 5-simplex. The $S^1 \times S^1 \times S^1 \times S^1$ manifold 
can be built out of the regular $3^4$ square torus by dividing 
each elementary $4d$-cube into 4-simplices in the following way. For each 
cube we mark two points, $p_1=(0,0,0,0)$ and $p_5=(1,1,1,1)$, lying on the 
opposite ends of the main diagonal and connect them by one of the shortest
paths going along the edges of the cube. The shortest path goes through
$4$ edges, each in a different direction, and through three points,
say $p_1,p_2,p_3$. There are $24$ such paths. Each set of points 
$p_0,p_1,p_2,p_3,p_4$ forms a 4-simplex. In figure \ref{4dcube} we show as an
example one four-simplex (red) built into a four dimensional cube obtained
with this procedure. The dashed lines are additional diagonals which have to 
be inserted to complete the four-tetrahedron. The manifold condition 
requires that no vertex be connected to itself. Therefore one has to glue
together $4$ cubes in all four directions. The starting configuration
therefore contains $3^4 * 24 = 1944$ four simplices. 

The manifold $S^3 \times S^1$ can be produced from a
spherical manifold by taking away two separate 4-simplices
and gluing together the boundaries $b_{1}, b_{2}$ created by 
removing the two 4-simplices. However, two vertices in a 
4d simplicial manifold may have at most one common link. To avoid the creation
of a double connection in the gluing process, the distance between the
vertices on $b_{1}$ to
those on $b_{2}$ has to be at least three links. One can ensure this, without 
going through a tedious check, by gluing cyclically together three
copies of the original (spherical) manifold, i.e. 
$b_{1}/b_{2}', b_{1}' / b_{2}'', b_{1}''/b_{2}$. The (double-) prime is
used to distinguish the different copies.  The boundaries $b_1, b_2$ are 
chosen such that they have no common vertex.

\subsection{Numerical results \label{topology_results}}
We have performed simulations with $N_4 = 4000, 8000, 16000, 32000, 64000$
for the three topologies discussed above. The explicit form of the sub-leading 
correction $\delta(\k_2,N_4)$ depends on the phase where the measurement is 
taken. The elongated phase of simplicial gravity is dominated by  branched 
polymers \cite{aj92, bbpt96, bi96}. This means that in  the large volume 
limit one can use the ansatz 
\eq{
F(\k_2,N_4) = N_4 f_0(\k_2) + (\gamma - 3) \log N_4  + f_1(\k_2) + 
     {\cal O}\left( \frac{1}{N_4} \right ) 
\label{frel}
}
for the free energy.  The correction coefficient
$\gamma$ is assumed to depend  only on the genus $g$ of the manifold.
The logarithmic corrections disappear for the action density $r$ eq. 
(\ref{den}) 
\eq{
r = \frac{1}{N_4} \frac{\partial F(\k_2, N_4)}{\partial \k_2} = 
     f_0'(\k_2 ) + \frac{ f_1'(\k_2 )}{N_4}.
}
Therefore one should take the next order corrections $1/N_4$ 
into account. We will show that they appear for purely kinematic reasons.
Consider the limit of large positive $\k_2$ in which only manifolds 
maximizing $N_2$ contribute to the sum (\ref{gcZ}). Such triangulations can 
be obtained from  barycentric subdivisions (\ref{M15}) of 4-simplices applied
successively to a minimal starting configuration, possibly mixed with 
micro-canonical transformations (\ref{m3}), which do not change $N_2$ and 
$N_4$.
By the minimal configuration we mean the minimal volume triangulation 
which maximizes $N_2$. For the barycentric subdivisions 
one gets the relation
$N_2 = 5/2 N_4 + c^{0}$, where the constant 
\eq{
c^{0} = N_{2}^0 - 5/2 N_{4}^0 
}
characterizes the initial minimal configuration.
The number  of triangles $N_{2}$ is related to the
action density $r = N_2/N_4 = 5/2 + c^{0}/N_4$. 
This means that the constant $c^{0}$ leads to $1/N_4$ corrections of 
the action density. The contributions to the sum (\ref{gcZ}) from 
triangulations built from non-minimal ones (i.e. smaller $N_2$) 
are suppressed exponentially by $\exp - \k _2 (c^0 - c)$, 
where $c$ characterizes the non minimal starting configuration with 
$c(\k _2) < c^0$.

The minimal configuration for the sphere, the surface of a $5$-simplex, is
known and hence $c^0$ is known. For the two toroidal geometries we find the 
minimal configuration numerically. We use our Monte Carlo code in a 
cooling procedure, in which we increase $\kappa _4$, to decrease $N_4$, 
and increase $\k _2$, to maximize $N_2$. In practice we have to 
increase $\k _2$ slowly compared to $\k _4$, because increasing $\kappa _2$ 
also increases pseudo-critical coupling $\tilde{\k}_4(\k _2)$.
For the topologies we studied, we found  the following 
minimal configurations~:
\eq{
\begin{array}{rllll}
S^4 & : &
N_{4}^0 = 6, & N_{2}^0 = 20 & \Rightarrow c^{0}=5, \\ &&&& \\
S^3 \times S^1 & : & 
N_{4}^0 = 110, & N_{2}^0 = 44 & \Rightarrow c^{0}=0, \\ &&&& \\
S^1 \times S^1 \times S^1 \times S^1 & : &
N_{4}^0 = 704, & N_{2}^0 = 1472 & \Rightarrow c^{0}=288. \\
\label{minim}
\end{array}
}

For the manifold $S^3\times S^1$ the finite volume correction 
$1/N_4$ is not present at all. It is also very difficult to detect them for the
sphere $S^4$ because $c^0 = 5$ is very small compared to the volume $N_4$ 
used in the simulation. For the torus $S^1 \times S^1 \times S^1 \times S^1$
the corrections are, however,  two orders of magnitude larger and 
measurable in the volume range used in the simulations. 
This estimate of the $1/N_4$ effect is exact only for infinite
positive $\k_2$ but one expects it to work, although with a slowly
varying coefficient $c(\kappa _2)$, in the entire elongated phase. 

\begin{figure}[t]
\hspace{1.5cm}
\psfig{file=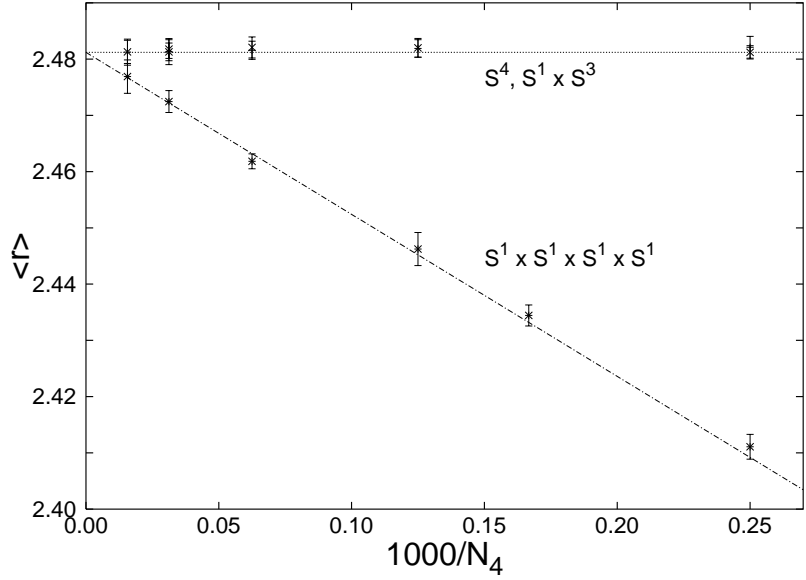,angle=270,height=11cm,rheight=8cm}
\caption{The volume dependence of the action density $r$ for three different
topologies in the elongated phase at $\k_2 = 2.0$}
\label{vcr}
\end{figure}
In  figure \ref{vcr} we show the action density $r$ measured  
in the elongated phase for $\k_2=2.0$ against $1/N_4$. 
In the same figure we display  the curve $r_\infty + c^0/N_4$ 
with $c^0=288$ and $r_\infty = 2.482$ which fits the data points very well. 
Note that that $r_\infty $ does not depend on the topology.

For the pseudo-critical coupling $\tilde{\k_4}$ we observe, with the statistics
available, no volume and/or topology dependence either. 
We find  $\tilde{\k}_4^{\infty} = 5.659(4)$ for $\kappa _2 = 2.0$ in the
infinite volume limit. 
This is compatible with the ansatz
\eq{
\tilde{\k}_4(\k_2,N_4) =   f_0(\k_2 ) + \frac{ \gamma - 3}{N_4},
\label{k4elon}
}
because $\gamma$ is known  to be of order ${\cal O}(1)$. 
With the method used, the correction could only be separated from the
statistical noise with an disproportionate amount of computer time. 

In the crumpled phase  we use a power-law ansatz 
\eq{ F(\k_2, N_4) = N_4 f_0(\k_2) + f_1(\k_2)  N_4^{\delta }  }
Taking the derivative with respect to $N_4$ one gets: 
\eq{
\tilde{\k}_4(\k_2,N_4) =  f_0(\k_2 ) + 
     \delta f_1(\k_2) N_4 ^{ \delta - 1}.
\label{k4crump}
}
For the action density (\ref{den}) one finds
\eq{
r = f_0'(\k_2 ) + f_1'(\k_2) N_4^{\delta - 1} 
\label{rcrump}
}
We checked this ansatz by fitting our numerical data for
$r(N_{4})$ to eq. (\ref{rcrump}):
\begin{center}
\begin{tabular}{|r|c|c|c|c|}\hline
Topology &  $\delta $ & $r^{\infty} = f_{0}'$ & $\log(f_{1}')$&
$\chi^{2}/\mbox{dof}$\\\hline
$S^{4}$  & 0.5 (2) & 2.039 (+0.010 / -0.013)&1.9 (1.3) & 0.78\\
$(S^{1})^{4}$ & 0.6 (2) & 2.028 (+0.008 / -0.016) & 1.0 (0.9) & 0.87 \\
$S^{1}\times S^{3}$ & 0.5 (2) & 2.038 (+0.010 / -0.021) & 1.8 (1.3) & 0.31 \\
\hline
\end{tabular}
\end{center}
and  $\k _{4}(N_{4})$ to (\ref{k4crump}):
\begin{center}
\begin{tabular}{|r|c|c|c|c|}\hline
Topology &  $\delta $ & $\k_4^{\infty}=f_{0}$&$\log(\delta f_{1})$&$\chi^{2}/\mbox{dof}$\\
\hline
$S^{4}$  & 0.6 (2) & 1.20 (2)& 1.4 (1.3) & 0.22\\
$(S^{1})^{4}$ & 0.6 (3) & 1.20 (2) & 1.4 (1.5) & 0.10 \\
$S^{1}\times S^{3}$ & 0.6 (2) & 1.20 (2) & 1.4 (1.3) & 0.18 \\
\hline
\end{tabular}
\end{center}

\begin{figure}
\hspace{1.5cm}
\psfig{file=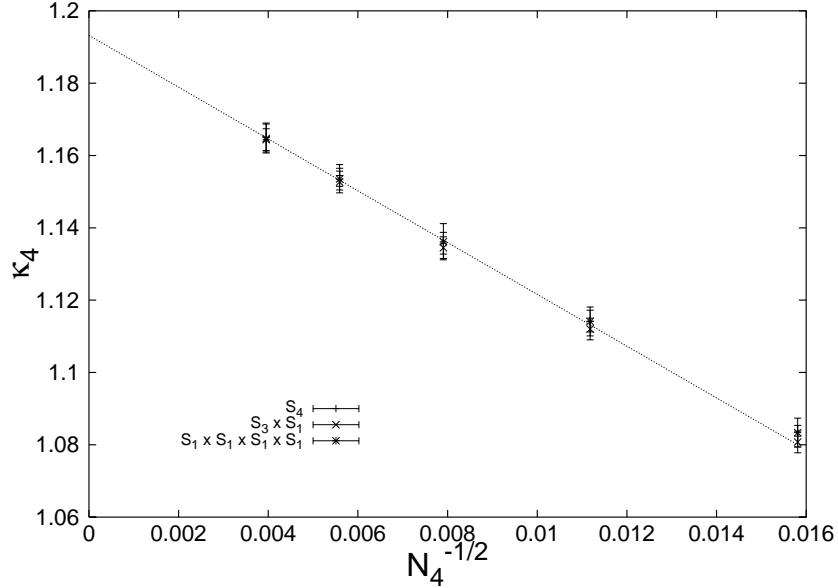,angle=270,height=11cm,rheight=8cm}
\caption{The volume dependence of the critical $\k _4$ for 
three different topologies in the crumpled phase at $\k_2 = 0.0$}
\label{k4cr}
\end{figure}

In the tables we give the logarithms of $f_1'$ and $\delta f_1$ since 
these have approximately symmetric errors.
One can see that the  values do not depend, within errors, on  
topology. In figure \ref{k4cr} we show the numerical data for $\k _{4}$ for
the three topologies and $\delta = 0.5$. 

The source of the volume correction in this phase can be understood 
by looking at the typical configurations which dominate the ensemble
for large negative $\kappa _2$. These configurations minimize the free energy 
and therefore have for fixed volume the minimal number of vertices, and
thus also the minimal number of triangles. In three dimensions  
such configurations were constructed in \cite{d}. The generalization to the
four dimensional case is obvious. One starts with a $2d$ 
triangulation of the sphere with $t$ triangles. At each triangle one builds 
a $4d$-pancake neighborhood from $q$ 4-simplices lying around the triangle in 
such a way that the links opposite to this triangle form a circle. An 
opposite link is defined as a link which does not share a common vertex
with the triangle. The next step is to put neighboring 
pancakes together by identifying these circles with
the $3d$ faces of the neighboring pancakes.
Each pancake has three such faces which lie between the
circle and an edge of the basic triangle. It also has
three neighboring pancakes. After this step one gets an $S^4$ 
sphere with $t*q$ 4-simplices and $2+t/2+q$ vertices. 
The highest connectivity $N_4 \sim (N_0-2)^2 / 2$ is reached 
when $t=2q$. The number $N_2$ of triangles can be constructed from this as 
a function of $N_4$ making use of the Dehn-Somerville relations (\ref{iglo}).
Inserting the result into (\ref{den}) one finds 
$r \propto 1/ / \sqrt (N_4)$  i.e.  $\delta = 0.5$ 
for these configurations. 

In \cite{bbkp96}, evidence was given that for spherical topology the 
phase transition is of first order. We could observe flip-flops in the
action density for the two tori as well.
This seems to show that the transition in these cases is of first order 
as well.

\subsection{Discussion}
We have  investigated the behavior of the curvature and the entropy density
for three different topologies in four dimensional simplicial quantum gravity.
We concentrated on two  $\kappa _2$ values, one  deep in the crumpled phase and
another value in the elongated phase. In both cases the value of the 
entropy density  $\kappa _4^c(\kappa _2)$ and the curvature were independent
on the topology in the large volume limit.  We have shown that $\kappa _4^c$
is an integral of the curvature. The observation that $\kappa^c _4$ is
independent from the topology in the two phases of the model therefore
suggests that the curvature and the entropy density do not depend on the
topology  for all $\kappa _2$ in the thermodynamical limit. 
This gives  support to the conjecture that it 
is possible  to define, similar to the double scaling limit in two dimensions, 
a scaling with  coherent contributions from all topological excitations. 
We can, however, not fully exclude that in the transition region the 
curvature and hence $\kappa _4$ depend on the topology but this, by chance, 
cancels out  after integrating over this region; this seems very unlikely, 
though.

Our result furthermore supports the conjecture that the entropic bound exists.
This  question was recently discussed in the literature 
\cite{ckr94, aj3, bm95}. 
The value of the entropy density $\kappa _4$ can be compared with the  
estimates given in \cite{bbcm94}, based on a summation over curvature
assignments. This estimate, does, however, not directly
give results in agreement with our numerical data for $\tilde{\k} _4$.
We conclude  that the mapping from the entropy of curvature assignments to 
the entropy of four-manifolds, which is a  basis for the calculations in 
\cite{bbcm94}, is to simple to capture fully the behavior in the 
$4D$ approach. 

We have analyzed the finite size  effects observed for the model. 
We gave kinematic arguments which explain the leading correction very 
well. 

\forget{
Finally we want to comment on the behavior of the algorithm in the
crumpled phase. The typical configuration  of simplicial gravity in this
regime has a sub-singular link and two singular vertices.
The local volume of the sub-singular link, 
i.e. the number of four-simplices which contain this link, diverges, when 
the volume of the entire configuration tends to infinity. The relaxation time, 
i.e. the number of Monte Carlo sweeps required until the
singular link appeared in the configuration, was extremely long for 
all topologies. This effect is even more pronounced if one
starts with a branched-polymer like configuration. For the 
$(S^1)^4$-torus  and small volumes (less than 64k 4-Simplices) the situation 
is even worse.  It was, at least for the run-length used in our 
numerical experiments, impossible to relax to such singular configurations. 
On the other hand, we know that they exist, because they could be
reached by shrinking down larger configurations containing a singular link.
It seems to be difficult for the algorithm to deal with two different
defects, the singular link and the hole, at the same time. 

This observation is possibly important in context  
of the question of practical ergodicity. The transformations
(see section{sqg_algorithm}) used in the simulations are known to be
ergodic, i.e. all manifolds $\sigma _a$ can be deformed into a configuration
$\sigma _b$ by a sequence of local deformations.

This does however say nothing
about the number transformations are required for this. 
In \cite{NB93} it was shown that the number of required transformations grows
faster than any recursive function if the underlying topology is  
non-recognizable. As the number of states for a fixed volume is exponentially
bounded (a recursive function) this means that in such a case the sequence of 
transformations required to deform  $\sigma _a$ into $\sigma _b$ 
necessarily has to go via manifolds with very large volume. 
}

\clearpage

\section{Simplicial quantum gravity coupled to gauge matter fields
\label{sqg_vec}}

\subsection{Introduction}
As mentioned in the introduction, the dominating geometries in the 
two phases found in four dimensional simplicial quantum gravity do not 
describe extended four dimensional structures. This, however, is required for
the model to describe the large distance behavior of space-time. 
The phase transition is of first order. Therefore one can not take the 
continuum limit at this point. In this situation it is sensible trying to 
insert terms into the action which cure these problems. 

In the crumpled phase, for example, one may try to suppress 
configurations with large order $o(t)$, i.e. singular vertices, by 
adding a term
\eq{
- \alpha \log \sum _t o(t)
}
to the action. We have  simulated the model with this term.
We found, however,  almost no back-reaction of the gravity sector on 
this term. Both  phases survive. At the transition one finds 
flip-flops in the time-series of $N_0$ which indicates that the 
transition is still first order. 

The branched polymer behavior observed in the elongated phase in fact plagues
many theories of random geometries in all dimensions. One phase of the 
hyper-cubic random surface model investigated in this thesis is a example of
such a theory. On the other hand, pure gravity in $D=2$ dimensions 
is in the non-trivial Liouville phase with string susceptibility exponent
$\gamma < 0$. In this phase it is possible to define  a sensible continuum 
limit. The  branched polymer 
behavior  returns, however, if one couples matter with central charge 
$c > 1$ to two dimensional quantum gravity. In \cite{cates, k90, dav92} 
a heuristic argument was given to explain this behavior:
When $c$ becomes larger than the $ c_b = 1$ 'barrier' metric singularities 
condense. The argument is similar to the Berezenskii-Kosterlitz-Thouless
\cite{bkt1, bkt2} argument for a transition in the 2-D $x-y$ model.
It relies on the construction of singular configurations ('spikes') and the 
observation that the action for such configurations grows logarithmically 
with the infrared cut-off.  From the continuum methods and the known dynamics 
of the conformal factor (\ref{liouville_action}) it is known that  the 
action  for a spike
\eq{
S(c) \propto  (26 - c) \log(\frac{ L} {a})
\label{2d_spike}
}
becomes smaller with larger $c$. The entropy $E \propto \log(L/a)$ of 
putting spike onto the surface on the other hand  grows logarithmically with 
the linear extent $L$ of the surface. It was shown that for $c > 1$ the 
entropy $E$ becomes larger than the action $S(c)$ and the creation of spikes 
is favored.

In \cite{jur} it was suggested that a similar phenomenon might 
occur in four dimensions. An effective action for the conformal factor
in four dimensions was derived  in \cite{ant_a, ant_b}. As in two  dimensions 
the action for a spike grows logarithmically with the linear extent $L$~:
\eq{S_E[\sigma _s] = \frac{1}{2} Q^2 q^2 \log \left ( \frac{L}{a} \right ). 
\label{4d_spike_action}}
The coupling
\eq{Q^2 = \frac{1}{180}(N_s + \frac{11}{2} N_{WF} + 62 N_V - 28) + Q_{grav}^2
\label{matter_q}}
depends on the numbers $N_s$ of scalar fields,  $N_{WF}$ of Weyl 
fermions and $N_{V}$ vector fields. The term $Q^2_{grav}$ describes the
(unknown) contribution of spin-2 gravitons. The only negative contribution
$-28$ comes from the  conformal factor itself. The most interesting
point about this result is that, differently from the two dimensional
case,  the effective action becomes larger when one adds matter to the model.
In other words, putting matter into the  model stabilizes the model and 
suppresses spikes. This suggests that in the absence of matter fields quantum 
gravity has no sensible vacuum, but this instability is lifted when the 
number of matter fields is large enough. 

It should be emphasized that in four dimensions the action 
(\ref{4d_spike_action}) for spikes can not be derived rigorously, because it is
not clear how transverse excitations of the metric should be treated. 
Nevertheless the conjecture above makes is very interesting to investigate
the influence of matter on the geometry in $4d$ simplicial quantum gravity. It
may also explain why in earlier investigations 
\cite{matter1, matter2, matter3} no non-trivial 
effects induced by the matter have  been observed.  In the framework of 
\cite{ant2} it was shown that spikes are dominant if $Q^2 < 8 /q^2$. It may
be that in the earlier investigations not enough matter fields were 
introduced so that $Q^2$ was still too small.

\subsection{The model}

From (\ref{matter_q}) one can read off that the strongest effect on the 
geometry is caused by the number of vector fields. 
Therefore we study a model 
of continuous vector gauge fields coupled to 4d simplicial gravity.
Specifically we consider non-compact  $U(1)$ gauge 
fields $A(l_{ab})$ living on the links $l_{ab}$ of the simplicial manifold and
$A(l_{ab}) = -A(l_{ba})$.  
The sum in the action
\eq{S_M = \sum_{t_{abc}} o(t_{abc}) \left[ A(l_{ab}) + 
                   A(l_{bc}) + A(l_{ca})\right]^2 
\label{sm}
}
extends over all triangles $t_{abc}$. The order $o(t_{abc})$ is equal to the
number of four simplices which contain $t_{abc}$ and is proportional to the
local volume assigned to the triangle. This factor appears to mimic the factor
$\sqrt{g}$ which  appears in a covariant action. There is no need to
introduce a coupling in front of the sum because the action has a Gaussian
form. 

In the model with $n$ gauge fields we put $n$ replicas of the
matter field on the dynamic geometry. The action
\eq{S_M^{(n)} = \sum _{i=1}^n \; ^{(i)}S_M \label{n_field_action}}  
is the sum of the one field actions (\ref{sm}) with $^{(i)}S_M = S_M$.  
Note that there is no explicit interaction between the replicas of the 
model. On a fixed lattice such a model is trivial, the behavior of a single 
replica remains unchanged. On a dynamic lattice, however, the systems are 
coupled by the back-reaction of the dynamical geometry on the matter fields.

The action for the combined system of gravity and matter contains, besides 
$S_M^{(n)}$, two more terms. The contribution of the gravity sector is given 
by the discrete Einstein-Hilbert action (\ref{sqg_act}) which in four 
dimensions reads
\eq{
S_G \,=\, -\kappa_2 N_2 + \kappa_4 N_4. 
}
where $N_k$ denotes the number of $k$-simplices. The third term is 
present for technical reasons. The algorithm  in the simulation is
inherently grand canonical and requires volume fluctuations to
satisfy  ergodicity. We use the multi-canonical potential (\ref{u})  
to simulate a
pseudo-canonical ensemble with almost fixed $N_4 \approx N_4^{(0)}$.
Measurements are taken only if $N_4 = N_4^{(0)}$. The coupling $\kappa _4$ is
tuned so that $\overline{N_4} = N_4$. 
 
The model is defined by the partition function
\eq{
Z(\kappa_2, \bar{N}_4) \,=\,  \sum_{T} W(T) \int' 
\prod _{i=1}^n \left ( \prod_{l \in T} dA^{(i)}(l) \right )  
\; \; {\rm e}^{\textstyle - S_G - S_M^{(n)} - 
{\delta \over 2} (N_4- \bar{N}_4)^2 } .
\label{part}
}
The sum extends over all distinct four dimensional simplicial manifolds $T$ 
and $W(T)$ is the symmetry factor taking care of equivalent re-labelings of 
vertices. The prime at the integral indicates that the zero modes of the gauge
field are not integrated. 

\subsection{The strong coupling expansion}
\begin{table}
{
\begin{center}
\begin{tabular}{|rr|rrll|} \hline
$N_4$ & $N_0$  & $N_g$ & $W^{(0)}$  & $W^{(1)}$ & $W^{(3)_3}$\\  \hline
6  &  6 &    1 &     1    & 1  & 1 \\
10 &  7 &    1 &     3    & $0.097638467\ldots$ 
                          & $1.03423\ldots\times10^{-4}$ \\
12 &  7 &    1 &     5    & $0.030058406\ldots$ 
                          & $1.08632\ldots\times10^{-6}$ \\
14 &  8 &    1 &    15    & $0.018550484\ldots$ 
                          & $2.83715\ldots\times10^{-8}$ \\
16 &  8 &    2 &   255/4  & $0.015777808\ldots$ 
                          & $9.68553\ldots\times10^{-10}$ \\
18 &  8 &    3 &   110    & $0.005500465\ldots$ 
                          & $1.38182\ldots\times10^{-11}$ \\
   &  9 &    3 &    95    & $0.004759295\ldots$ 
                          & $1.19996\ldots\times10^{-11}$ \\
20 &  8 &    2 &   225    & $0.002512817\ldots$ 
                          & $3.15034\ldots\times10^{-13}$ \\
   &  9 &    7 &   693    & $0.007291315\ldots$ 
                          & $8.14758\ldots\times10^{-13}$ \\
22 &  9 &   15 &  2460    & $0.005728290\ldots$ 
                          & $3.27573\ldots\times10^{-14}$ \\
   & 10 &    7 &   690    & $0.001447804\ldots$ 
                          & $6.46761\ldots\times10^{-15}$ \\
24 &  9 &   13 & 16365/2  & $0.004226212\ldots$ 
                          & $1.17586\ldots\times10^{-15}$ \\
   & 10 &   34 & 14625/2  & $0.003378959\ldots$ 
                          & $7.45244\ldots\times10^{-16}$ \\
26 &  9 &   50 & 17865    & $0.001946262\ldots$ 
                          & $2.34367\ldots\times10^{-17}$ \\
   & 10 &  124 & 39645    & $0.003936950\ldots$ 
                          & $3.97116\ldots\times10^{-17}$ \\
   & 11 &   30 &  5481    & $0.000491334\ldots$ 
                          & $4.06700\ldots\times10^{-18}$ \\
28 &  9 &   89 & 291555/7 & $0.001058334\ldots$ 
                          & $6.96159\ldots\times10^{-19}$ \\
   & 10 &  415 & 182820   & $0.004119603\ldots$ 
                          & $2.16619\ldots\times10^{-18}$ \\
   & 11 &  217 &  77057   & $0.001534637\ldots$ 
                          & $6.33995\ldots\times10^{-19}$ \\
30 &  9 &  139 &  73860   & $0.000457581\ldots$ 
                          & $1.78973\ldots\times10^{-20}$ \\
   & 10 & 1276 & 672821   & $0.003427165\ldots$ 
                          & $9.13722\ldots\times10^{-20}$ \\
   & 11 & 1208 & 564000   & $0.002507268\ldots$ 
                          & $5.17157\ldots\times10^{-20}$ \\
   & 12 &  143 &  46376   & $0.000179907\ldots$ 
                          & $2.84432\ldots\times10^{-21}$ \\  \hline
\end{tabular}
\end{center}
}
\caption{ The number of different graphs $N_g$, for a fixed 
 volume $N_4$ and fixed number of vertices $N_0$ 
 and the corresponding weights $W_{N_V}(N_4,N_0)$ for 
$0, 1$ and $3$ vector fields.    All weights are 
 normalized with the value at $N_4 = 6$.}
\label{sqg_diagrams}
\end{table}

\noindent
The first few terms of the strong coupling expansion  can be used to estimate
the exponent $\gamma $ \cite{dav2}. We used our code without matter fields 
to identify diagrams up to $N_4 = 30$ and calculated the symmetry factor
$\sigma $ for the diagrams by counting the number of invariant re-labelings
\footnote{For a detailed discussion of this method see section 
\ref{hyper_verify}.}.
Without matter the weight for a diagram is 
$W = 1 / \sigma$. To debug our computer code we identified the diagrams 
with $N_4 \le 18$ by inspection and verified that the theoretical and numerical
distributions governed by the symmetry factor agree.  

For the strong coupling expansion of the model with matter-field we have to
calculate the weight induced by the matter. The Gaussian integration which
excludes zero-modes over one species of fields leads to a determinant  $\rho $.
As the $n$ replicas of the field interact only indirect via the back-reaction
of the gravity and have no explicit interaction term in the action it is 
sufficient to calculate the weights $\rho_M$ for one field. The $n$-field 
weight is 
\eq{ W^{(n)} = \rho_M^n / \sigma.} 
The weights $\rho _M$ were calculated using Maple. 
In table (\ref{sqg_diagrams}) we summarize the results obtained with this 
procedure.

\begin{figure}[p]
\hspace{0.3cm}
\psfig{file=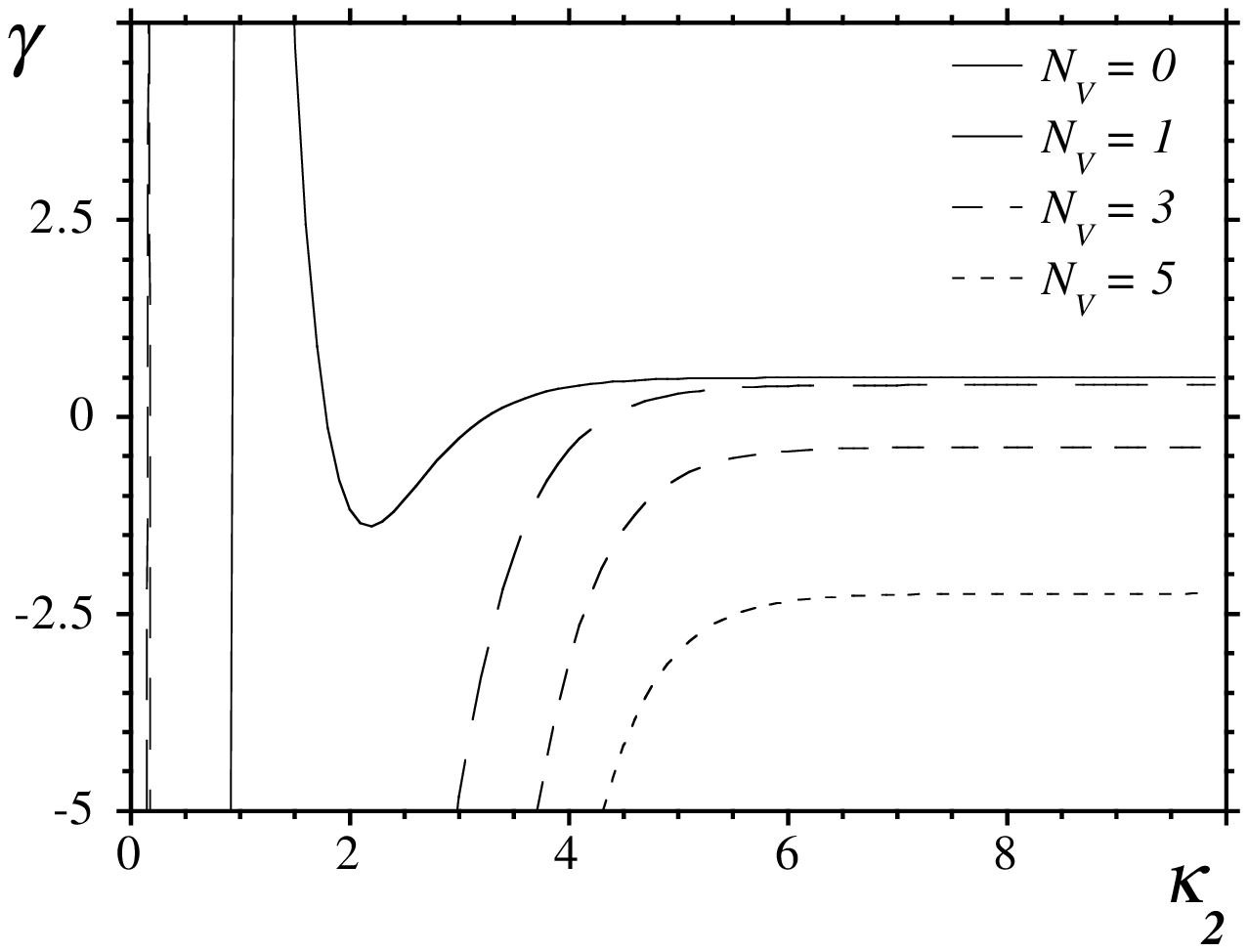,angle=0,height=7cm}
\vspace{2cm}

\hspace{1.0cm}
\psfig{file=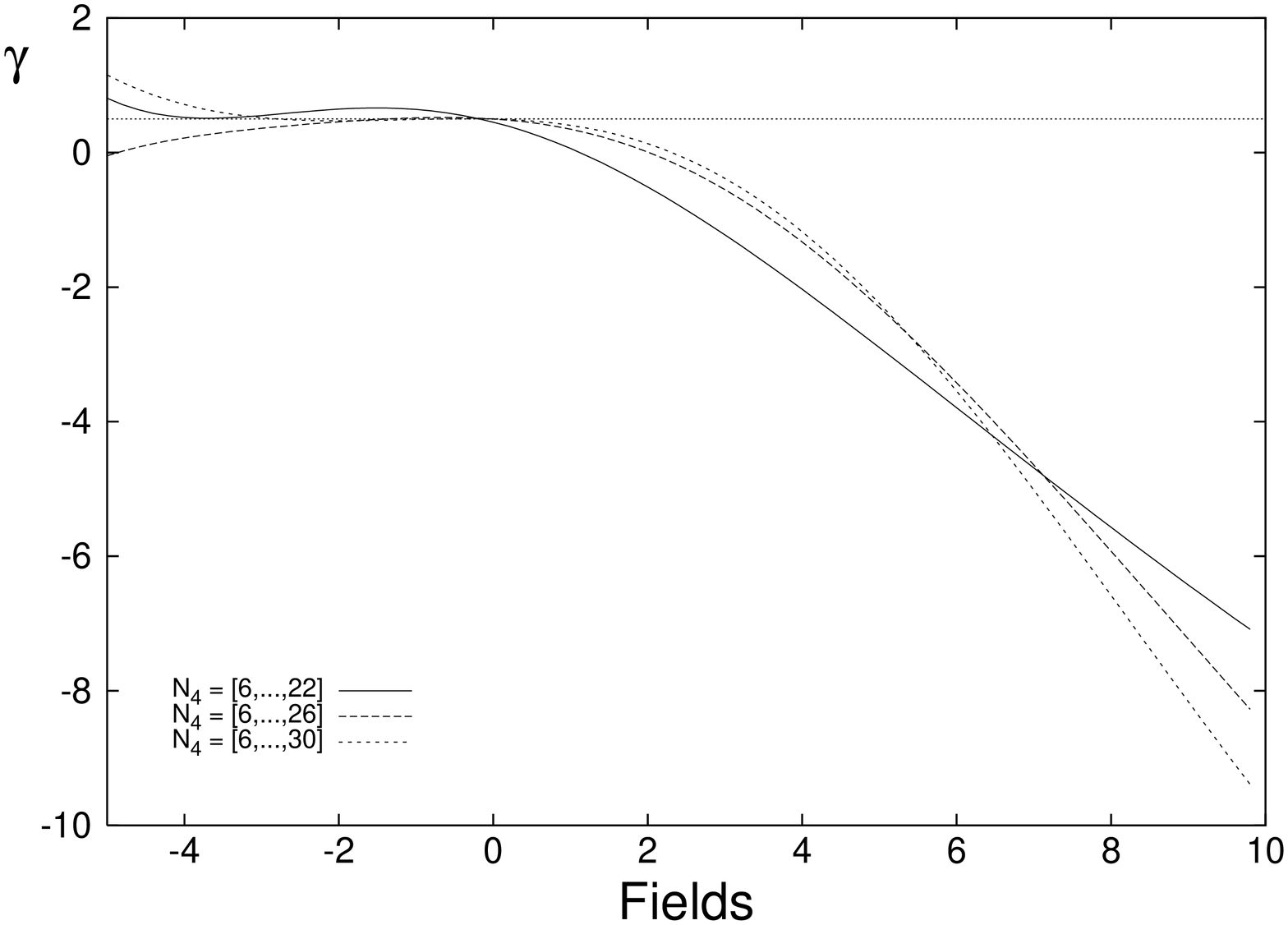,angle=270,height=9.5cm,rheight=7cm}
\caption{{\em (a)} Variations of $\gamma$ with
 $\kappa_2$, for 0, 1, 3, and 5 vector fields coupled
 to gravity.  
 {\em (b)} Variation of $\gamma $ with the number if fields for
          $\kappa _2 = 10.0$. 
          These values are obtained using the ratio
          method  to analyze the strong coupling series including
          terms  corresponding to $N_4 = 6$, 10, 14, 18, 22, 26, and 30.
          The horizontal line in the lower diagram
          is at $\gamma = 0.5$, the exponent expected for pure 
          gravity.}
\label{gamma_var}
\end{figure} 

The series was then analyzed with the ratio method \cite{dav2} to extract the 
exponent $\gamma $. In the upper plot in figure \ref{gamma_var} we show 
the behavior of $\gamma$ as a function  $\kappa _2$ obtained with this 
method.  For large $\kappa _2$ the susceptibility exponent $\gamma $ tends 
to a constant value which becomes negative when the number of fields grows.  
For pure gravity one finds $\gamma = \frac{1}{2}$ as expected. 
For small enough $\kappa _2$ the exponent $\gamma $ is large negative. 
In the intermediate region the results are unstable. This presumably
indicates a phase transition in this range of the coupling $\kappa _2$. 
In the lower plot we show the variation of $\gamma $ as a function of the 
number of vector fields at $\kappa _2 = 10.0$, i.e. in the weak coupling phase.
We used diagrams of size $[6,\ldots,22], [6,\ldots,26]$ and $[6,\ldots,30]$ 
for the three different curves. One sees that they depend slightly on 
the size of the largest diagrams used in the series expansion, however 
they seem to converge for larger diagrams. This and the numerical results
presented in the next section justify the assumption that 
the series expansion presented here describes correctly the qualitative
behavior of the exponent $\gamma $.

\subsection{Numerical results}

\begin{figure}[t]
\hspace{1.5cm}
\psfig{file=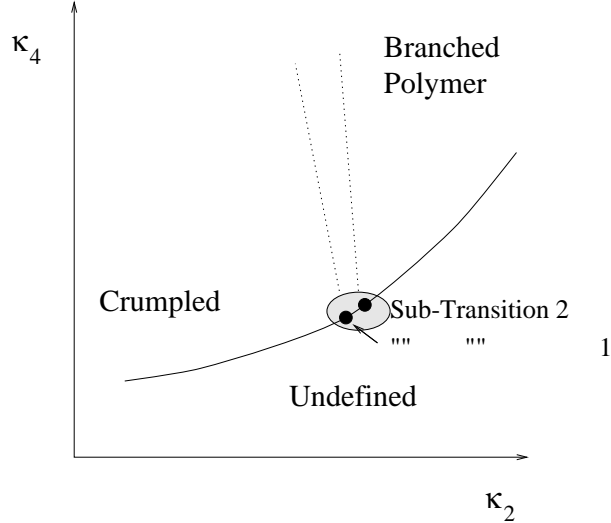,angle=270,height=7cm}

\caption{The phase diagram for the model with one vector field.}
\label{pd1field}
\end{figure}

The first question when analyzing a new model is about the  phase
diagram. The gauge fields investigated here introduce  a new parameter
into the model, namely the number $n$ of fields. We have performed Monte Carlo
simulations for the model with one and three vector fields. For the survey of
the phase structure we use rather modest lattice sizes with no more than 
16K four-simplices. It is known from the simulations of pure gravity that
finite size effects are quite strong; for example the order of the phase
transition for pure gravity reveals itself only for lattices larger than
32K. It is therefore clear that one has to be careful when drawing strong
conclusions from the numerical data presented here. Nevertheless we believe
that the qualitative picture obtained from the data is reliable. 
This is supported by the observation that the results are in good
agreement with the strong coupling expansion.

In figure \ref{pd1field} we present the phase diagram found for one vector
field. The model diverges if $\kappa _4 < \kappa _4^c(\kappa _2)$ and the
model is undefined. We observed no back-reaction in the strong coupling
phase, which is still dominated by singular structures. 
Likewise, the  weak coupling phase is still a branched polymer. In the 
transition region we observe two sub-transitions indicated by two peaks in 
the  node susceptibility
\eq{\chi_0 = (\left<N_0^2\right> - \left<N_0\right>^2)/N_4.}
The first sub-transition has a broad peak. With the statistical noise it is
difficult to locate precisely the point of the transition. Therefore our
data does not allow to judge if with growing volume the two transitions merge
to the same critical coupling. They are presumably similar to
the  two sub-transitions observed for pure gravity on small lattices\footnote{
See also the end of section \ref{sqg_intro}}, which do seem to  merge 
\cite{ckr97}.
\begin{figure}[p]
\hspace{1.5cm}
\psfig{file=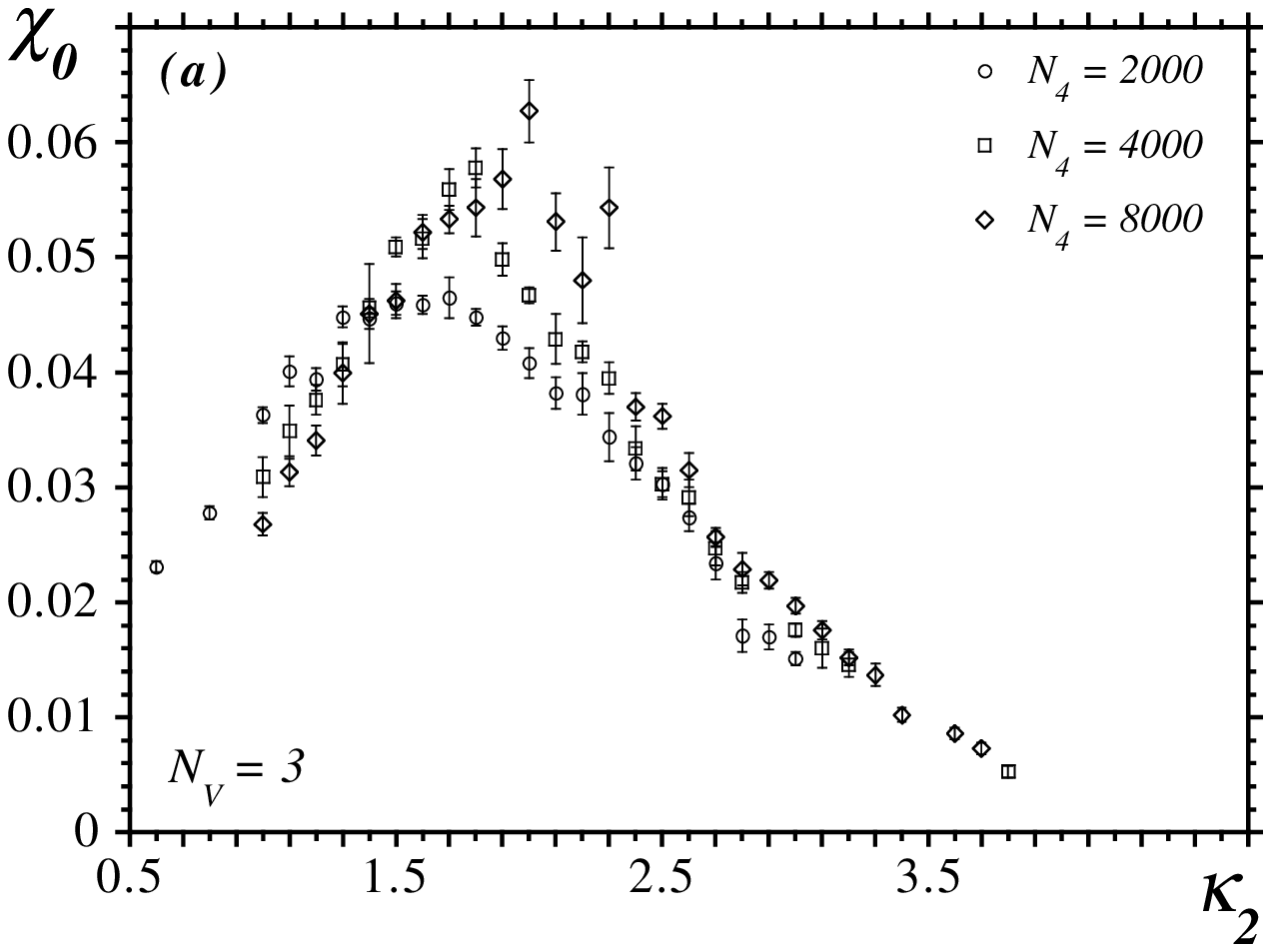,angle=0,height=7cm}
\vspace{5mm}

\hspace{1.5cm}
\psfig{file=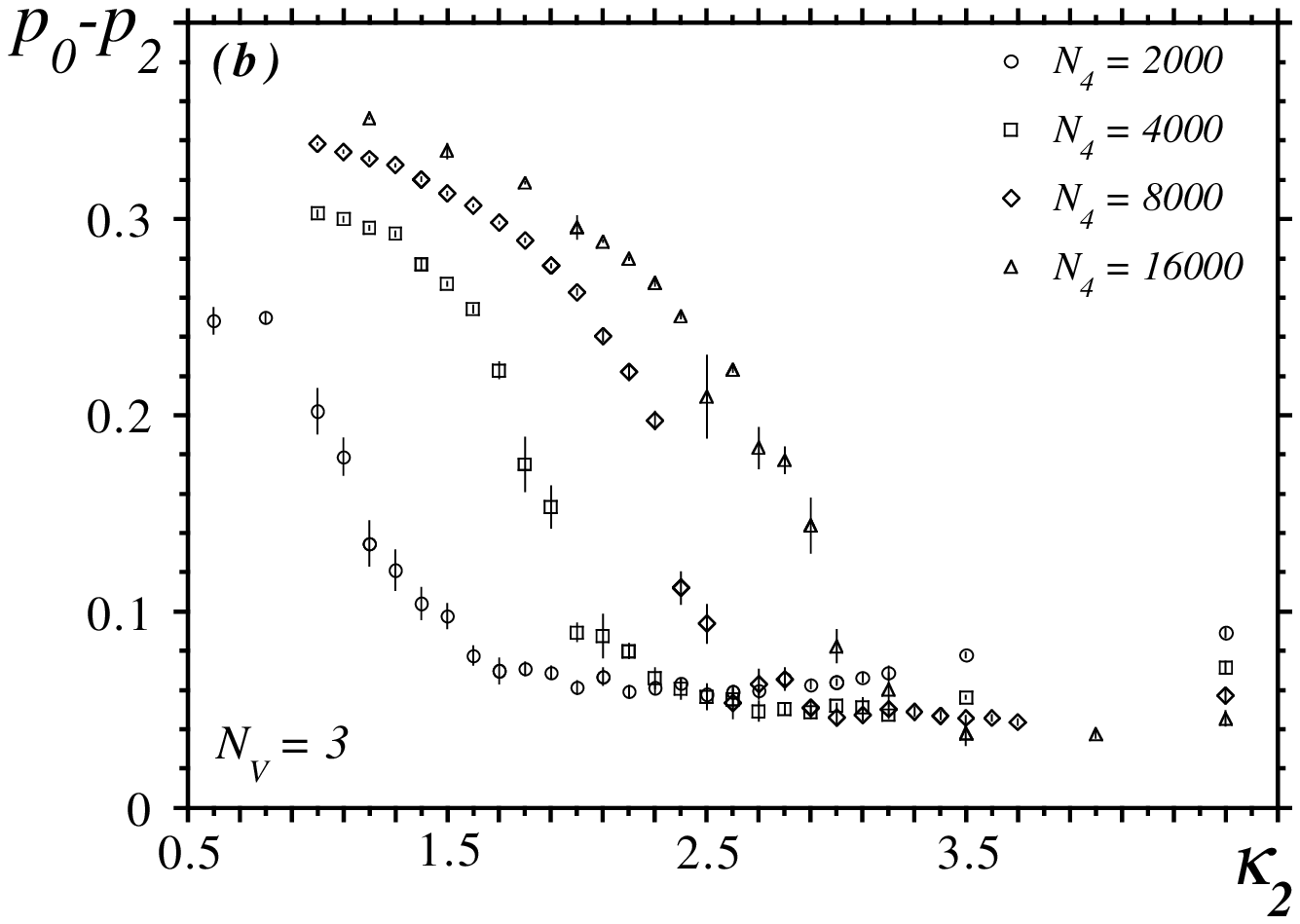,angle=0,height=7cm}
\caption{\small {\it (a)} The node susceptibility $\chi_0$,
 for three vector fields coupled to gravity, {\it vs}.\ $\kappa_2$.
 For $N_4 = 16000$ the statistics is not sufficient for a
 reliable estimate of $\chi_0$.
 {\it (b)} The corresponding change in the difference between the
 orders of the first and of the third most singular vertex, 
 $p_0$ and $p_2$, normalized with the total volume $N_4$.}
\label{vec_nchi}
\end{figure}

The picture changes when we include three vector fields. We still find singular
structures in the strong coupling phase, but the weak coupling phase has no 
longer a branched polymer behavior. The two phases are separated by a maximum 
of the node susceptibility at $\kappa _2 \approx 1.9$ which grow with the 
volume as seen in figure \ref{vec_nchi}.  To locate the critical coupling
and understand the nature of the transition  one needs better statistics and 
simulations for larger systems; this is in progress but it will, however, 
require several more months of CPU time.  

The change in the geometry  at roughly the same coupling where $\chi _0$ has
its maximum is, however,  a strong indication that we observe a real 
transition. In plot (b) of figure \ref{vec_nchi} we show the difference 
between the orders of the most singular and the third most singular vertex 
$p_1 - p_3$. In the strong coupling phase the order of the first and second
most singular vertices are roughly of the same order. They are seperated by a 
large gap from the order of the other vertices. For larger $\kappa _2$ the 
order of these two vertices merge to the order of other vertices.

\begin{figure}[h]
\hspace{1.5cm}
\psfig{file=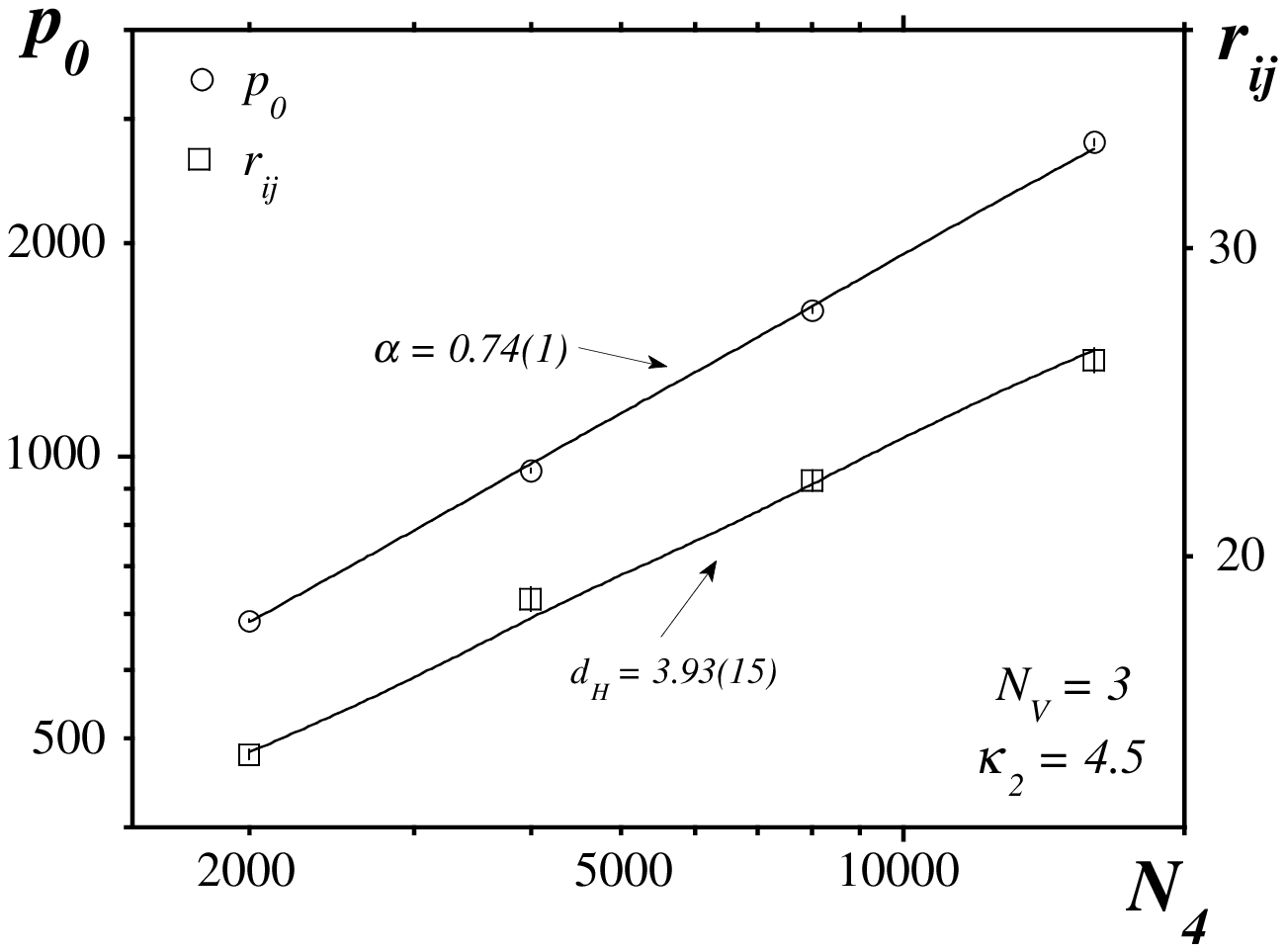,angle=0,height=7cm}

\caption{ The scaling of both the largest vertex order
 $p_0$ and  the average distance $r_{ij}$ between two simplices
 in the new weak coupling phase, for $n=3$ vector fields
 at $\kappa_2 = 4.5$ in a doubly logarithmic plot.  
 The lines are the best linear fits
 to the data.}
\label{lvo}
\end{figure}
This means that there is no singular vertex in the weak coupling phase, 
as demonstrated  by the scaling behaviour of the most singular
vertex $p_0$ in figure \ref{lvo} for $\kappa _2 = 4.5$. The local 
volume of this vertex grows sublinearly like $N_4^{3/4}$, i.e.  a vanishing 
fraction of the total volume $N_4$ in the thermodynamical limit. 

In the same plot (\ref{lvo}) we show the scaling of the average distance 
between the simplices
\cite{haus}:
\begin{equation}
\left < r_{ij} \right >_{N_4} \;=\; 
\left<\sum_{r=0}^{\infty}  r \;n(r) \right> 
\;\sim\; N_4^{1/d_H} \;,
\label{hd_fit}
\end{equation}
were $n(r)$ counts the number of simplices at a 
geodesic distance $r$ from a marked simplex. With a fit to equation 
(\ref{hd_fit}) we get for the Hausdorff dimension $d_h = 3.93(15)$.  However,
the data points do not lie on a straight line, presumably due to finite size
effects. If we exclude the data point for the smallest volume $N_4 = 2000$ we 
get $d_h = 4.68(8)$. 

\begin{table}
\begin{center}
{
\begin{tabular}{|c|c|c|} \hline
 $N_4$  &  $\kappa_2 = 4.5$  & $\kappa_2 = 6.0$ \\  \hline
 2000   &  -0.22(2) & \\
 4000   &  -0.18(3) &  -0.17(4) \\
 8000   &  -0.23(3) & \\
 16000  &  -0.30(6)   &  -0.12(6)  \\  \hline
\end{tabular}
}
\end{center}
\caption{ Measured values of the string susceptibility
 exponent $\gamma$ in the weak coupling
 (large $\kappa_2$) phase  for three vector fields}
\label{shaked}
\end{table}

We use the baby universe method \cite{baby4} to measure the entropy 
exponent $\gamma $ in the weak coupling phase. The  method was explained for 
the hyper-cubic random surface model in section \ref{shake_it}. Up to some
technicalities the method is the same for the model discussed here. 
In table \ref{shaked} we list the results obtained for three vector fields
with this method for two values of $\kappa _2$ (4.5 and 6.0) and at 
different volumes. We consistently observe negative values 
$\gamma \approx - 0.2$. This is in reasonable agreement with the 
prediction from the strong coupling expansion, $\gamma \approx -0.38$.

\subsection{Discussion}
The most interesting result is the discovery that matter fields can prevent
the collapse of 4d simplicial manifolds to branched polymers. We find 
for the first time a strong back reaction of the matter on the geometry. 
The results suggest that when 
the number of vector-fields becomes larger than one  the branched polymer 
phase disappears, and is replaced by a {\it new} phase with  negative 
$\gamma$. 

The numerical data does not allow
to determine the order of the transition unambiguously. However for one vector
field we observed indications for a first order transition but the latent heat
of the transition is smaller compared to the jump observed for pure gravity. 
This may indicate that matter softens the transition. Therefore it is not
excluded that with the correct mixture of matter-fields the transition 
is continuous.  

The new weak coupling phase observed for three vector fields is very
intriguing. The internal Hausdorff dimension if close to four, the same as
for flat space. The negative $\gamma $ values obtained numerically and with
the strong coupling expansion indicate a non-trivial behavior. 
This is a very promising result because the negative $\gamma $ may, like in
two dimensions, indicate that the whole phase is critical and a non-trivial
continuum limit can be taken. Gravitons, however, will presumably be present
only if the Newton coupling is tuned to a critical value $\kappa _2^c$. If the
transition found in our work is second order this induces infinite 
curvature-curvature correlations, which may be interpreted as gravitons. 

Our results obtained numerically and with  the series expansion support
the conjecture in \cite{ant2} mentioned in the introduction. 
As predicted we find that  additional matter stabilizes the geometry 
as opposed to the case in two dimensions where matter de-stabilizes the 
surface.

After all it should, however, not be forgotten that one still faces 
some uncertainties. It is important to check carefully the
nature of the transition. It should also be verified that the critical
coupling in the thermodynamical limit is finite. In the weak
coupling phase the curvature $N_0 / N_4 \approx 0.25$ is close to the kinematic
bound. The local volume $o(p) \propto N^{3/4}$ of the sub-singular vertices 
in this  phase take no finite fraction of the whole volume in the 
thermodynamical limit,  but the local volume still diverges. 

\clearpage

\section{Algorithm \label{sqg_algorithm}}
The algorithm used to generate the data presented in this chapter uses the
framework of dynamical Monte Carlo algorithms discussed in 
section (\ref{intro_algo}).  In this section we want to describe the 
transformations (\ref{intro_mc_transform}) and the weight (\ref{metropolis})
specific to the algorithm used to simulate four dimensional simplicial quantum
gravity. We also discuss the weights induced by the vector field used in
section \ref{sqg_vec}. 

In order to calculate the transition probabilities, remember that the 
distribution of manifolds $T$ we want to generate is given by
\eq{
p(T,\{A\}_T) \sim \exp - S_g[T] - S_A[T,\{A\}_T],
}
where the first term corresponds to the Einstein-Hilbert action. 
The action $S_A$  of the matter field living on the manifold is given
by 
\eq{
S_A[T,\{ A\}_T] = \sum _{i=1}^n 
   \sum_{t\in T} o(t) (A^{(i)}(l_{12}) + A^{(i)}(l_{23}) + A^{(i)}(l_{31}))^2.
}
The second sum runs over all triangles $t$ of the triangulation $T$,
$o(t)$ is the order of the triangle $t$, and $l_{12}$,$l_{23}$,$l_{31}$
are the three oriented edges of $t123$.  The field $A$ is an oriented 
link variable, which means $A(-l) = -A(l)$. Therefore the action is invariant
under a gauge transformation at vertex $p_0$: simultaneous translation  
$A(l) \rightarrow  A(l) + \epsilon$ of all $A(l)$ living on links emerging 
from $p_0$ does not change the action. The first sum  runs over the number 
of matter fields coupled to gravity. For the sake of simplicity, we discuss
in the following only the case $n=1$. The extension to $n > 1$ is 
straightforward.

The geometric moves $M_{(p,q)}$ which we describe in detail below  may 
generate and destroy links and hence field variables $A(i)$. In principle one 
could  update the geometry and the matter at the same time. A procedure would 
be to assign random values to new matter variables and accept or reject the 
whole transformation  with a Metropolis weight. However, if one assigns the 
$A(l_i)$ with a flat distribution the efficiency  is very low because 
then most transformations will be rejected in the Metropolis step. One might 
try to assign to the configuration $A$ a Gibbs distribution
\eq{
  \pi _{(p,q)}(A) = \frac{e^{-S(A)}} {V(A)}
\label{matter_gibbs},}
with the normalization 
\eq{V(A) = \int ' _{A} {\cal D} A e^{-S_A(A_B)}\label{vol_field}}
The integral runs over all accessible states of the new matter-fields. The
prime indicates that we do not integrate over the zero-modes of the action.
The probability  can, however, not directly be used as a heat-bath probability 
because in the grand canonical moves the normalization constant 
$V(A)$ is different for a move and its inverse. Therefore our algorithm 
updates gravity and matter in two steps. In the first step the geometric part 
of the transformation is accepted or rejected independently of the state of 
the new matter fields, which are integrated out. In the second step the fields
which were created in the geometric update are set with a heat-bath
probability. 
 
The detailed balance condition for the probability $\Pi _{(p,q)} $ to
accept the geometric move $(p,q)$ at a certain location on the manifold $A$ 
reads then 
\eq{\frac{V(A)}{n_{i}  } \Pi _{(p,q)}(A)  e^{-S_g(A) -S_s(A)} =
    \frac{V(B)}{n'_{i}} \Pi _{(p,q)}(A)  e^{-S_g(B) -S_s(B)}. 
\label{grav_metro}
}
The term $S_s$ is the action of the matter field without contributions from
terms involving the new matter field. In the geometrical update $S_s$ changes
because the  move changes the order $o(t)$ of some triangles on the 
simplicial manifold. The factor involving the $n_i$ are present to balance 
the grand canonical fluctuations of the manifold. The $n_i, i = 3,4,5$ count
the number of possible locations, where a local deformation can be applied. 
 We have discussed this factor in detail in the context with expression 
(\ref{hyp_db_cond}) for the hyper-cubic random surface model.

The matter field is not only updated in the geometrical 
transformations. The algorithm 
switches between updates in the geometrical and the matter sector. In the
matter update we choose an extensive number of links from the manifold. Each
link is sequentially  subjected to a heat-bath or an  over-relaxation step. 
In the heat--bath, link field $A$ on the link $l_{12}$ 
is chosen from the distribution~:
\eq{ P(A) \sim \exp  \left ( - \sum_{t} o(t) (A - A_{1p} + A_{2p})^2 \right ) 
           =  \exp \left (- \alpha ( A - \beta/\alpha)^2 \right )
\label{matter_heat}
}
where the sum runs over all triangles $t = t_{12p}$ which have the link
$l_{12}$ as an edge and the point $p$ as a vertex. 
The variables $\alpha, \beta$ in the rightmost expression are defined by 
\eq{ \alpha = \sum_{t} o(t) }
\eq{\beta = \sum_{t} o(t)(A_{1p} - A_{2p}).} 
The over-relaxation is a micro-canonical transformation
\eq{A \rightarrow  -A + 2\beta/\alpha,}
which does not change the action. This transformation is more efficient
compared to the heat-bath in terms of mobility, however, the move is not
ergodic. Therefore we have to mix the over-relaxation update with the 
heat-bath update.

\begin{figure}[htbp]
\centerline{{\psfig{file=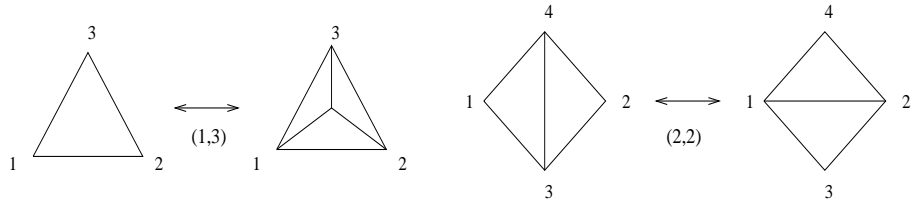,angle=270,height=2.6cm,width=12cm}}}
\caption{The $(p,q)$ transformations in two dimensions}
\label{pq2}
\end{figure}
In the following sections we describe in detail the geometric transformations 
and the weights derived for the $5$ $(p,q)$ moves in four dimensions. For the
transformations where new field variables are created we derive also the
heat-bath probability for the fields. To clarify the notation for the
geometric transformations, consider first the $3$ $(p,q)$-transformations in
two dimensions depicted in figure \ref{pq2}. We use the notation 
$tp_1p_2p_3$ to denote a triangle which contains the vertices with the labels
$p_1, p_2, p_3$.
The first transformation $M_{(1,3)}$ inserts a vertex into a triangle. 
We write this as
\eq{\mbox{t{\bf \underline{123}}}  \leftrightarrow  
\mbox{t12{\bf \underline{4}} + t13{\bf \underline{4}} + t23{\bf \underline{4}}}
.}
The inverse transformation $M_{(3,1)}$ removes a vertex. With this
procedure the number of triangles grows (decreases) by $2$, the number of
links grows (decreases) by $3$ in the triangulation. 

The second transformation $M_{(2,2)}$ flips a link:
\eq{\mbox{ t1{\bf\underline{34}} + t2{\bf\underline{34}} }
 \leftrightarrow  \mbox{
t{\bf\underline{12}}3 + t{\bf\underline{12}}4 
}  \nonumber
}
This transformation is self-dual. Using the transformation at the 'same'
location twice does not change the geometry.

In the four dimensional case presented in the following sections we denote a 
four simplex $sp_1p_2p_3p_4p_5$ by enumerating the labels of the vertices 
$p_i$.

\subsection{Move 15, 51}
The move  $M_{(1,5)}$ inserts a vertex $p6$ into a four-simplex
$s12345$. This step generates $4$ four-simplices, $10$ triangles, $5$ 
links and, together with the links, five new field variables
$A(li6) = x_i, i = 1, \ldots ,5$.  
\eq{
\mbox{
s{\bf \underline{12345}}} \leftrightarrow
s1234{\bf \underline{6}} + 
s1235{\bf \underline{6}} + s1245{\bf \underline{6}} + 
s1345{\bf \underline{6}} + s2345{\bf \underline{6}} 
\label{M15}
}

\hspace{3cm}
{\psfig{file=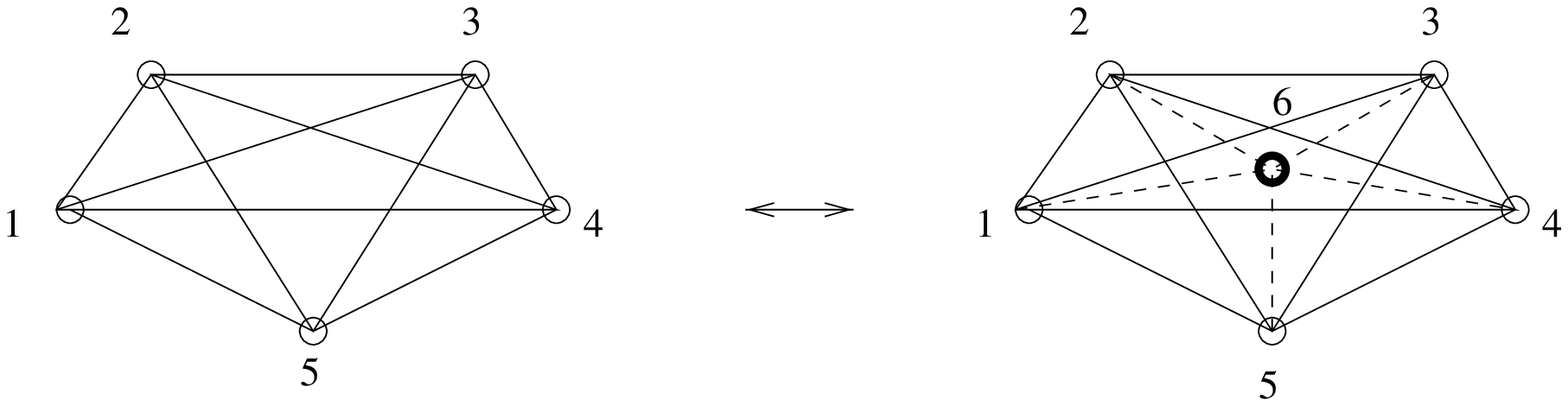,angle=270,height=2.7cm,width=6.5cm}}

The change in the geometric part of the action is
\eq{\Delta S_g = 4 \kappa _4 - 10 \kappa _2.}
The order $o(t)$ of the triangles $t \in s12345$ changes by $+1$. 
The "old" fields therefore contribute 
\eq{\Delta S_s(A) =  \sum (A_{ij} + A_{jk} + A_{ki})^2 .}
The sum runs over all triangles which change the order by $+1$ in the 
transformation $M_{(1,5)}$. The new fields contribute 
\eq{ S_A[x] =  3 \sum_{i,j,i<j} (A_{ij} + x_j - x_i)^2}
to the action. The sum runs over pairs $i<j\le 5$ of the points of $s12345$.
For the probability $\Pi_{(p,q)}$ (\ref{grav_metro}) we have to calculate the 
volume $V(x)$ (\ref{vol_field}) for the new fields.
We find
\eq{
V = \int \frac{d^5 x}{zero \ mode} e^{-S_a[x]} = 
   \frac{\pi^2}{125} \exp \left ( \frac{3}{5} \sum_{<ijk>} 
       (A_{ij} + A_{jk} + A_{ki})^2 \right )
\label{V15}
}
With (\ref{grav_metro}) we  get for the Metropolis weight 
\eq{
\Pi _{(1,5)} = \min\left\{ 1 , \frac{\pi^2}{125} \
\frac{n_4}{n'_{5}} e^{-\Delta S_g} 
\exp \left ( -\frac{2}{5} \sum_{<ijk>} (A_{ij} + A_{jk} + A_{ki})^2 \right )
\right\}
}
for the geometrical part of the move. 

The inverse transformation removes a suitable vertex from the simplicial 
manifold. No new fields are created and hence $V_b = 1$. 
The weight for this move is~:
\eq{
\Pi_{(5,1)} = \min\left \{ 1 , \frac{125}{\pi^2} 
\frac{n_{5}}{n'_{4}} e^{\Delta S_g} 
\exp \left ( \frac{2}{5} \sum_{<ijk>} (A_{ij} + A_{jk} + A_{ki})^2 \right )
\right\}
}

After the geometric part of the transformation $M_{(1,5)}$ has been accepted 
the five new link-variables $x_i = A(l_{i6})$ have to be assigned with a heat 
bath probability. One of them, say $x_5$,  may be set zero due to the gauge 
degree of freedom. The others are chosen with a Gaussian distribution~:
\eq{
\begin{array}{ll}
P(x_1,x_2,x_3,x_4,x_5=0) & \sim
\exp \left ( -3 \sum_{i,j,i<j} (A_{ij} + x_j - x_i)^2 \right )\\ & \\
 & \sim \exp \left ( -3( x^T M x - 2 w^T x) \right ) 
\end{array}
}
where  
\eq{
M =  \left[\begin{array}{rrrr} 
     4 & -1 & -1 & -1 \\
    -1 &  4 & -1 & -1 \\
    -1 & -1 &  4 & -1 \\
    -1 & -1 & -1 &  4 \end{array} \right]
}
and 
\eq{ w_i = \sum_{j,j\ne i} A_{ij}.} 

It is convenient to diagonalize the symmetric matrix $M$ to decouple the 
fields.  Denote the components of the vector in the
eigenvector bases by $y_1,y_2,y_3,y_4$. The distribution for the  $y_i$ in
this bases is given by 
\eq {
\begin{array}{rcll}
P(y_1) & \sim & \exp      ( -3(y_1 - \Omega_1)^2) & 
   \Omega_1 = \frac{1}{2} (w_1 + w_2 + w_3 + w_4) \\
P(y_2) & \sim & \exp ( -15(y_2 - \Omega_2/5)^2) & 
   \Omega_2 = \frac{1}{2} (w_1 + w_2 - w_3 - w_4) \\
P(y_3) & \sim & \exp (-15(y_3 - \Omega_3/5)^2) & 
   \Omega_3 = \frac{1}{\sqrt{2}}(w_1 - w_2) \\
P(y_4) & \sim & \exp (-15(y_4 - \Omega_4/5)^2) & 
    \Omega_4 = \frac{1}{\sqrt{2}} (w_3 - w_4) \\
\end{array}
}

This assigns a random value to the $y_i$ with the distribution given above. 
The new fields $x_i$ are obtained by a base-transformation~:
\eq{
\begin{array}{c}        
x_1 = \frac{1}{2} (y_1 + y_2 + \sqrt{2} y_3) \\
x_2 = \frac{1}{2} (y_1 + y_2 - \sqrt{2} y_3) \\
x_3 = \frac{1}{2} (y_1 - y_2 + \sqrt{2} y_4) \\
x_4 = \frac{1}{2} (y_1 - y_2 - \sqrt{2} y_4)
\end{array}     
}

\subsection{Move 24, 42}
The  move $M_{(2,4)}$ inserts one new link into the manifold. This
generates also four triangles and two four-simplices:
\eq{
\label{m2}
\mbox{
s1{\bf\underline{3456}} + s2{\bf\underline{3456}} 
$\leftrightarrow$
s{\bf\underline{12}}345 + s{\bf\underline{12}}346 
+ s{\bf\underline{12}}356 + s{\bf\underline{12}}456  
}} 

\hspace{3cm} 
{\psfig{file=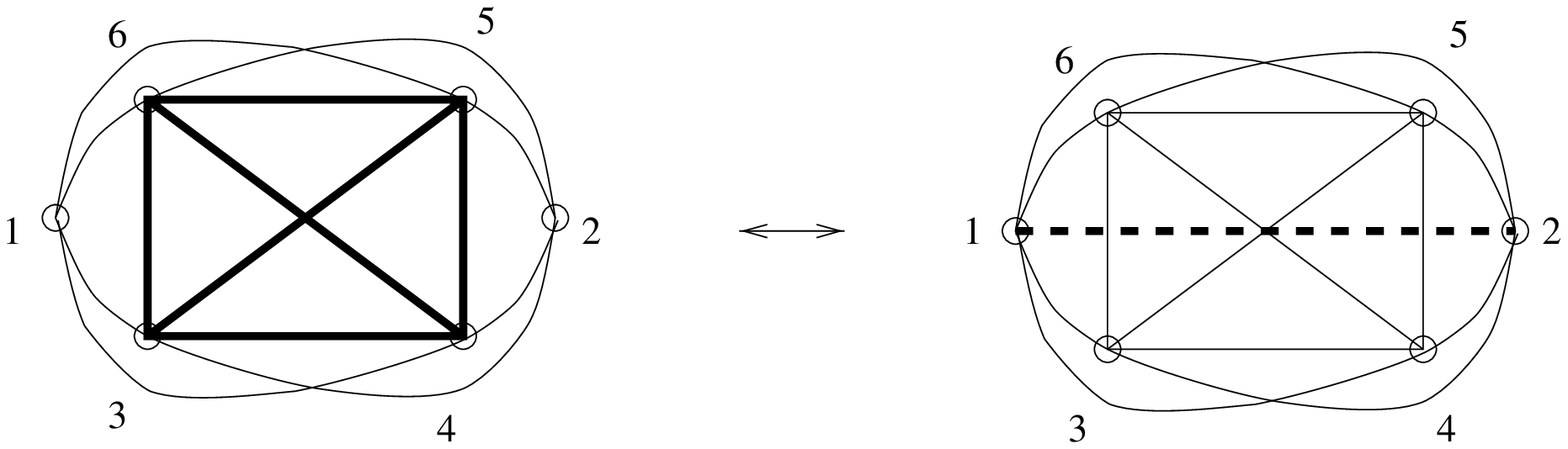,angle=270,height=2.7cm,width=7.5cm}}

\noindent
The change in the geometric part of the action is~:
\eq{\Delta S_g = 2 \kappa _4 - 4 kappa _2.}

In the calculation of the volume $V$ of the configuration space for the 
new field $x = A(l12)$ 
we use the notation (\ref{m2}): the new link $l12$ connects the vertices 
$p1$ and $p2$, the simplices $s13456$ and $s23456$ have a common tetrahedron 
$T3456$. The order of the four triangles $t \in T3456$ decrease by $1$ in the 
move $M_{(2,4)}$ while the orders of the remaining $12$ triangles in 
$s13456$ and $s23456$ increase by 1.

Denote by 
\eq{
P_{ijk} = (A_{ij} + A_{jk} + A_{ki})^2 }
the value of the plaquette $P_{ijk}$. One  gets the following
change in the action for the "old" fields~:
\eq{
\Delta S_s(A) =  - \sum_{3\le i<j<k\le 6} P_{ijk} + 
\sum_{2<i<j\le 6} (P_{1ij} + P_{2ij} )
}
The volume of the state space for the new field is
\eq{
V  = \sqrt{\frac{\pi}{12}}\exp - 3 \sum_{j=3}^{6} (A_{1j} + A_{j2})^2. 
}
Collecting these results together one finds for the transition probability
\eq{
\Pi_{(2,4)} = \min \left \{ 1, \frac{n_{14}}{n'_{3}} \times V
\times e^{-\Delta S_g - \Delta S_s}\right \}
}
and
\eq{
\Pi_{(4,2)} = \min \left \{ 1, \frac{n_{14}}{n'_{3}} \times V^{-1}
\times e^{\Delta S_g + \Delta S_s}\right \}
}
for the inverse move $M_{(4,2)}$. 

The heat-bath weight for the one new field is given by eq. 
(\ref{matter_heat}).

\subsection{Move 33}
This move flips a triangle on the manifold. 
The transformation is self-dual and does not change the number of 
(sub)-simplices. Therefore we do not have to deal with a new field:
\eq{ V = 1. }

\eq{
\label{m3}
\mbox{
s12{\bf \underline{456}} + s13{\bf \underline{456}}
+ s23{\bf \underline{456}} 
 $\leftrightarrow$ 
s{\bf \underline{123}}45 + s{\bf \underline{123}}46
+ s{\bf \underline{123}}56 }
}

\hspace{3cm}
{\psfig{file=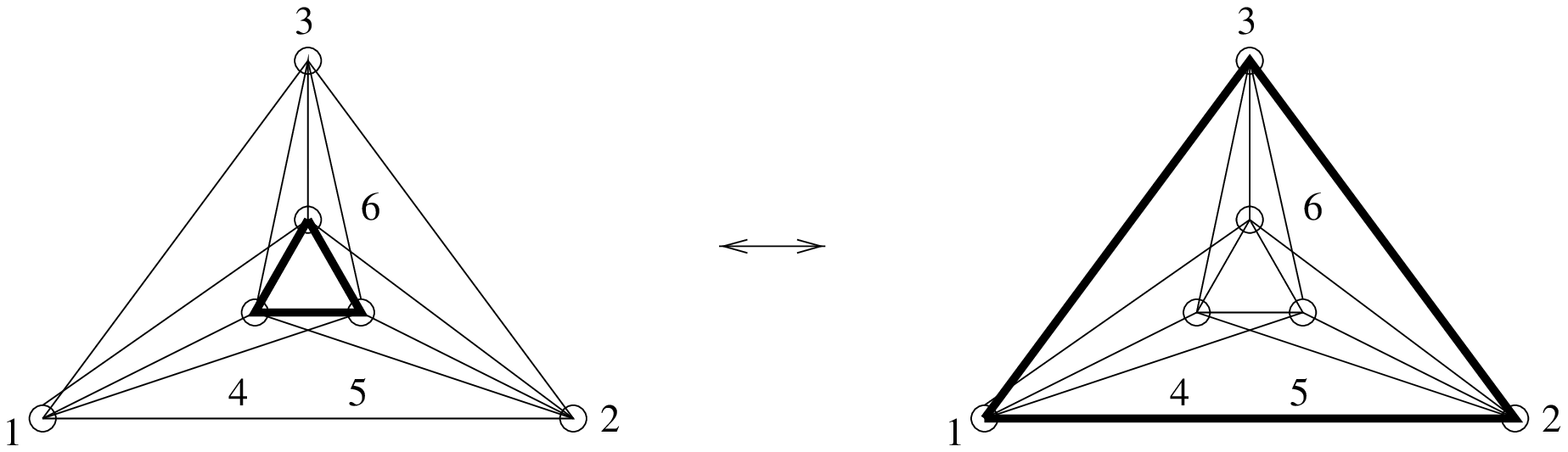,angle=270,height=2.7cm,width=7.5cm}}

In the transition probability one has to take into account the change 
$\Delta S_s$ coming from the change of $o(t)$.  From $\ref{m3}$  
one can read off 
\eq{
\Delta S_s(A) =  
   \sum_{i,j=1, i \not= j}^3 \sum _{k=4}^6 
      (A_{ij} + A_{jk} - A_{ki}) ^2 
  - \sum_{j,k=1, j \not= k}^3 \sum _{i=4}^6 
      (A_{ij} + A_{jk} - A_{ki}) ^2. 
}
The action of the gravity sector is unchanged. Therefore the transition 
probability for the geometric part of the move is
\eq{\Pi _{(3,3)} = \min \left \{1 , \frac{n_3}{n'_3} \times e^{-\Delta S_s} 
                       \right \}}

\forget{
\subsection{Verification}
\begin{table}[t]
\begin{displaymath}
\begin{array}{c|c|c|cc|cc}\\
 & W_{pg} & \rho _M & \Omega _{anal}^(1) & \Omega _{numer}^(1) 
\end{array}
\end{displaymath}
\caption{The diagrams used to verify the algorithm. In the first column of the
table the pair of moves  checked is given. The second column contains the
name of the diagrams involved. The product of the weight $W_{pg}$ for the graph
without matter and the weight $\rho _M$ gives the weight 
$\Omega _{anal}^{(n)} = W_{pg} \rho _M^n$ for $n$ fields. In the column 
$\Omega ^{(n)} _{numer}$ the result obtained numerically is given}
\label{sq_verify}
\end{table}

\begin{figure}[h]
\hspace{3cm}
{\psfig{file=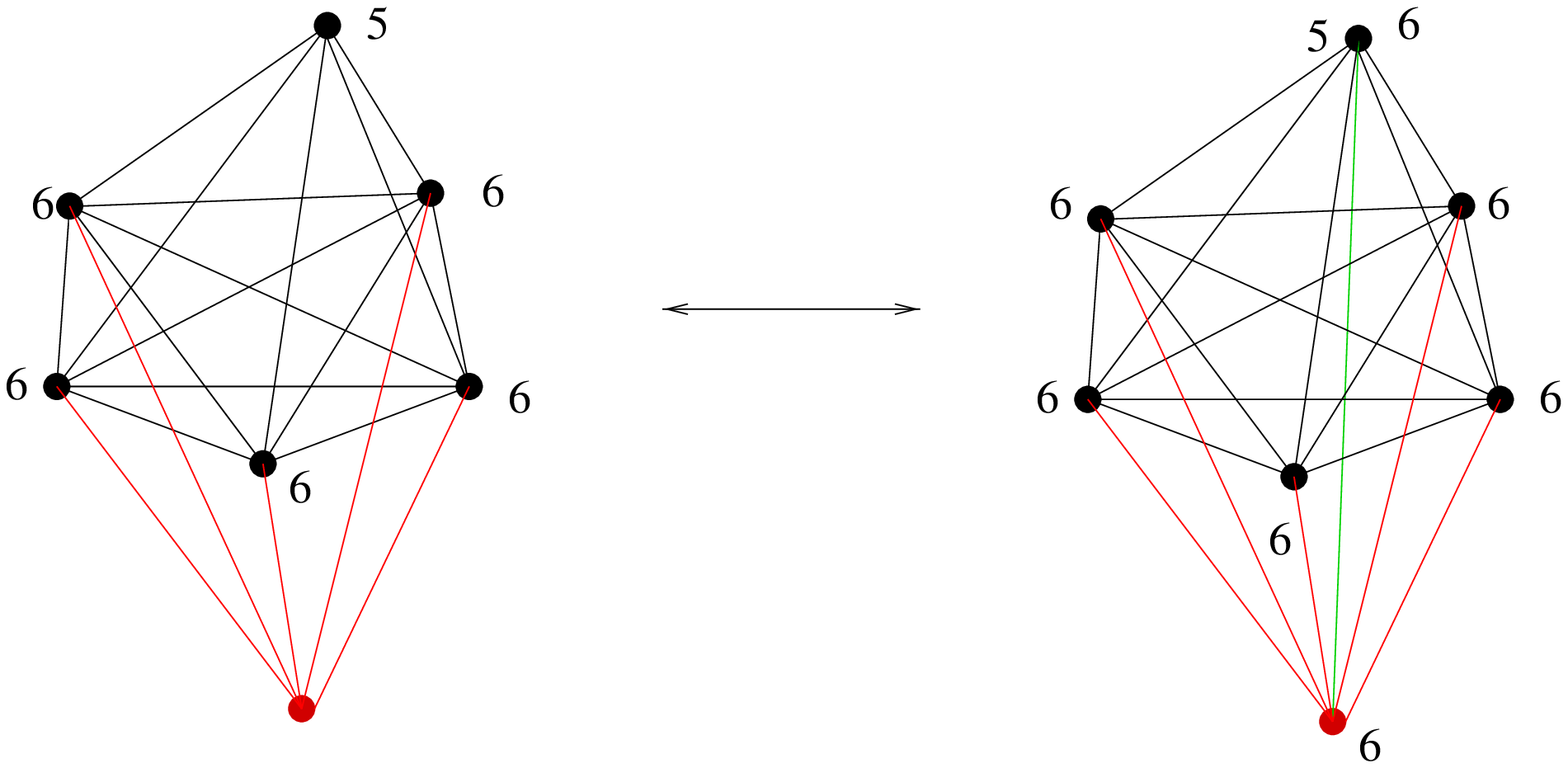,angle=270,height=4.0cm}}

\caption{Transformation $M_{(2,4)}$ connects the diagrams $c_{10}$ and $c_{14}$
by inserting the green link. The inverse transformation $M_{(4,2)}$ removes
the green link. In the left diagram one can see the effect of the pair
$M_{(1,5)}, M_{(5,1)}$ which connects $c_6$ and $c_10$: $M_{(1,5)}$
inserts the components drawn in red, the inverse $M_{(5,1)}$ removes these 
components.}
\label{pic_check}
\end{figure} 

To check the correct implementation of the transformations
we measure individually for the pairs of moves 
$(M_{(1,5)},M_{(1,5)}), (M_{(2,4)},M_{(4,5)}), (M_{(3,3)},M_{(3,3)})$
the distribution of two diagrams connected by the transformation. For example 
to check the pair  $(M_{(2,4)},M_{(4,5)})$ we considered the diagrams depicted
in figure (\ref{pic_check}). The transformation $M_{24}$ inserts the green
link which is in turn removed by transformation $M_{42}$. 
In table \ref{sq_verify} we compare the numerical distributions with the
theoretical distribution obtained in the strong coupling expansion. 
}

\forget{
In this paper we have investigated the behavior of the
entropy density and the curvature for three different
topologies in four dimensional simplicial gravity.
We employed manifolds consisting of between 4000
and 64000 simplices. We concentrated on two values of
the gravitational coupling constant $\kappa_2$,
one in the crumpled and one in the branched polymer 
phase. We found that in both the cases the value of the
entropy density and curvature in the infinite volume
limit are equal for these three topologies. This gives
further support to the conjecture that these limits exist,
a question which was discussed in the literature some time ago
\cite{ckr94}, \cite{aj3}, \cite{bm95}. 

Furthermore, it gives  support to the hypothesis that
all topologies contribute to a sum over topologies,
like in two dimensions. The value of the entropy exponent
can be compared with the  estimates given
in \cite{bbcm94}, based on a summation over all distributions
of curvature. This estimate, which surprisingly enough is exact 
for the leading term in two dimensions, does, however, not directly
give results in agreement with our numerical data for $\tilde{\k} _4$.

We further analyzed in detail the finite size effects,
and compared them to estimates from simple kinematic
arguments. This approach explain very well the finite
size corrections. It may, of course still be possible
that those effects are of a more complicated nature
near the transition.

Finally, as also seen in previous investigations,
we observe that the algorithm has very long relaxation
times in the crumpled phase. We even found that for the
$(S_1)^4$ torus, the ground state seemed not to accessible
using the approximately fixed $N_4$ algorithm, but only by
passing through states with much larger $N_4$-values.
This breakdown of practical ergodicity is still under
investigation. 
}

%% file: summary.tex
\chapter{Summary}
In this thesis we have analyzed two different models of random geometries,
namely hyper-cubic random surfaces and four dimensional simplicial quantum
gravity. Although these models are  different, e.g. the local dimension of
the models considered is two and four respectively, they have common
properties. For example both models possess a branched polymer phase. The
appearance of this phase, however, is often referred to as a 'collapse to
triviality' because the basically one dimensional structure does not describe
extended $d$-dimensional ($d=2,4$) geometries in the continuum limit. 
One problem we addressed for both models was therefore to find modifications
of the models with a different critical behavior. In both cases we find such
modifications. 

The hyper-cubic random surface model is a discretization of bosonic strings
embedded in a $D$-dimensional space.  In three dimensions the model can also 
be interpreted as a fluid membrane. We have analyzed the phase diagram of the
model with an extrinsic curvature term in the action. Numerically we find 
two phases in three and four dimensions: a weak coupling phase with a
branched polymer behavior and a flat strong coupling phase. The Hausdorff
dimension is $d_h = 4$ in the branched polymer  phase and $d_h =2$ in the 
flat phase. The transition is of first order, the critical coupling is finite.
We have shown that the existence of this transition is not in contradiction
with an earlier renormalization group analysis of the model, which under some
assumptions indicated that the model is always in the branched polymer phase
for finite coupling to the external curvature. One cannot take the
continuum limit at the transition point because the transition is of first
order. One can, however, try to smoothen this transition by introducing 
self-avoidance into the model. This is  motivated by the observation that for
the related model of Ising spins with a gonihedric action without
self-avoidance of the dual surfaces evidence for a first order and with weak 
self-avoidance evidence for a second order phase transition was found 
\cite{des1order,des2order}.

We have solved an old question concerning the hyper-cubic random surface model.
Earlier investigations indicated that the model with a certain local
constraint has a different critical behavior compared to the unrestricted
model. We have shown that  this is not the case. The true
behavior of the constrained model is masked by very strong finite size
effects. In the thermodynamical limit the constrained and the unconstrained
model have the same critical behavior. This shows the universality of the
model in the sense that the behavior of the model does not depend on details of
the discretization. 

For the model of four dimensional simplicial quantum gravity we found that 
coupling more than two gauge vector fields to gravity suppresses the branched
polymer phase. Our numerical data and a strong coupling expansion indicate 
that the branched polymer phase is replaced by a new phase of gravity.
The Hausdorff dimension in this phase is close to four, the dimension of 
flat space. We find a negative entropy exponent $\gamma $ in this phase. This 
may indicate, similar to the model in  two dimensions, that the whole phase 
is critical so that one can take a non-trivial continuum limit. 
One can hope to find gravitons in this model at the phase transition if the 
transition is of second order. However, the statistics of our numerical 
data does, at the moment, not allow to determine its order.
Our ongoing numerical
investigations of the model therefore concentrate on the nature of the phase
transition. In the future the geometrical properties of the dominating
manifolds at the transition point and in the new phase have to be analyzed. 
This will allow to answer the question of whether the dominating geometries
describe, on large scales, a flat four-dimensional Euclidean space-time. 
If this is the case, we may have succeeded in finding a stable ground state of 
four dimensional euclidean quantum gravity. The observation that matter is 
required to stabilize the ground state of quantum gravity would then suggest 
that matter may be required for the very existence of a stable universe. 

A different question which we addressed for four dimensional quantum gravity
concerns the role of topological excitations. It is not a priori clear whether 
topological excitations should be taken into account in this model. We have
checked numerically for three different topologies whether topological 
excitations are exponentially suppressed in the thermodynamical limit. 
We have shown that the leading contribution to the free energy does not 
depend on the topology and hence the topological excitations survive in the 
thermodynamical limit.  This result supports the conjecture that it is 
possible to define the model with coherent contributions from all topological
excitations at a double scaling limit similar to the two dimensional model.

\section{Acknowledgement}
I want to thank Prof. B. Petersson for proposing such an interesting research
activity. Without his guidance, technical and academic advice and patience this
thesis would never have been possible. It is a pleasure  to thank Z. Burda,
G. Thorleifsson and J. Tabaczek for the long and fruitful discussions we have
had in the recent years. I am grateful to the HLRZ J{\"u}lich for computer time
on PARAGON. The HRZ Bielefeld supported me by supplying computer hardware 
for a Pentium farm, which I used to perform large parts of the numerical
calculations.

%% file: phd.bbl
\begin{thebibliography}{99}

\bibitem{polch94} 
 J.~Polchinsky, {\em What is String Theory ?}, Proc.\ Les 
  Houches Summer School, Session LXII (1994) 287.

\bibitem{peliti} L. Peliti, {\em Amphiphilic Membranes} in {\em fluctuating
geometries in statistical mechanics and field theory}, Les Houches Summer
School (1994) 195.

\bibitem{nelson94} D. R. Nelson, {\em Defects in Superfluids, Superconductors
and Membranes} in {\em fluctuating
geometries in statistical mechanics and field theory}, Les Houches Summer
School (1994) 423.

\bibitem{KPZ} V. Knizhnik, A, Polyakov, A. Zamolodchikov, Mod. Phys. Lett A3
              (1988) 819- 

\bibitem{gp91} P. Ginsparg, {\em Matrix Models of 2d Gravity}
               Trieste HEP Cosmol. 1991, 785 (hep-th/9112013) 

\bibitem{am_hou} J. Ambj{\o}rn, {\em Quantization of geometry} in 
                {\em fluctuating geometries in statistical mechanics and 
                  field theory}, 
                 Proc.  Les Houches Summer School (1994) 77.

\bibitem{adj86} J. Ambj{\o}rn, B. Durhuus, T. Jonsson,  
                J. Phys. A21 (1977) 981.


\bibitem{adj87} J. Ambj{\o}rn, B. Durhuus, T. Jonsson,  
                Europhys. Lett. 3 (1987) 1059.


\bibitem{polyakov_book} A.M. Polyakov, {\em Gauge Fields and Strings}, 
           Harwood Academic Publishing (1987).


\bibitem{M90} T. Morris, Nucl. Phys. B341 (1990) 443.


\bibitem{L39} L.D. Landau, E.M. Lifschitz
  {\em Lehrbuch der theoretischen Physik, Band II}.
  Akademie-Verlag Berlin (1939).

\bibitem{we71} S. Weinberg,
  {\em Gravitation and Cosmology: Principles and 
       applications of the general theory 
       of relativity},  Wiley (1971).  

\bibitem{hh83}  J.B. Hartle, S.W. Hawking, Phys. Rev. D28 (1983) 2690.

\bibitem{adf85} J . Ambj{\o}rn, B. Durhuus, J. Fr{\"o}hlich,
                Nucl. Phys. B257 (1985) 433

\bibitem{d85}   F. David, Nucl. Phys. B257 (1985) 45
                F. David, Nucl. Phys. B257 (1985) 543
  
\bibitem{k85} V. A. Kazakov, Phys. Lett. B150 (1985) 182.

\bibitem{k89} V. A. Kazakov, Mod. Phys. Lett. A4 (1989) 2125.

\bibitem {m81} M.L. Mehta, Comm. Math. Phys. 79 (1981) 327.


\bibitem{do1} E. Brezin, V.A. Kazakov, Phys. Lett. B236 (1990) 144.

\bibitem{do2} M. Douglas, S. Shenker, Nucl. Phys. B335 (1990) 635.

\bibitem{do3} D. Gross, A.A. Migdal, Phys. Rev. Lett. 64 (1990) 127.

\bibitem{kk86} V.A. Kazakov, Phys. Lett. A119 (1986) 140.

\bibitem{adf86} J. Ambj{\o}rn, B. Durhuus, J. Fr{\"o}hlich, P. Orland,
                Nucl. Phys. B257 (1985) 433; B270 (1986) 457;
                            B275 (1986) 161

\bibitem {dw80} D. Weingarten, Phys. Lett. B90 (1080) 280.

\bibitem {dfj84} B. Durhuus, J. Fr{\"o}hlich, T. Johnsson, 
                 Nucl. Phys. B240 (1980) 453


\bibitem {bb86} B. Baumann, B. Berg, Phys. Lett. B164 (1985) 131\\
                B. Baumann, B. Berg, G. M\"unster, Nucl. Phys. B305 (1988) 


\bibitem{bg97}  B. Petersson, G. Thorleifsson,
                {\em Beyond the C = 1 barrier in two dimensional 
                     quantum gravity}  (hep-lat/9710077).


\bibitem{david} F. David, Nucl. Phys. B487 (1997) 633.


\bibitem {a94}   J. Ambj{\o}rn, Nucl. Phys. Proc. Suppl. 42 (1995) 3.

\bibitem {du94} B. Durhuus, Nucl. Phys. B426 (1994) 203.

\bibitem {ajt93} J. Ambj{\o}rn, S. Jain, G. Thorleifsson,
                 Phys. Lett. B307 (1993).

\bibitem {jm92} S. Jain, S. Mathur, Phys. Lett. B286 (1992) 239.

\bibitem{dj86}   B. Durhuus, T. Jonsson Phys. Lett. B140 (1986) 385.

\bibitem{FS88}  A.M. Ferrenberg, R.H. Swendsen, Phys. Rev. Lett. 61 (1988) 
                2635.

\bibitem {bbf85} B . Berg, A. Billoire, D. Foerster, 
                 Nucl. Phys. B251 (1985) 665\\
                 B. Baumann, Nucl. Phys. B285 (1987) 391


\bibitem{extcurv_bengt} J. Ambj{\o}rn, A. Irback, J. Jurkiewicz, 
                        B. Petersson, Nucl. Phys. B 393 (1993) 571.

\bibitem{bchhm92}  M. Bowick, P. Coddington, L. Han, G. Harris, E Marinari 
                   Nucl. Phys. B394 (1993) 791.

\bibitem{abcfhhm93} K. Anagnostopoulos, M. Bowick, P. Coddington, 
                    M. Falcioni, L. Han, G. Harris, E. Marinari, 
                    Phys. Lett. B317 (1993) 102.

\bibitem{helf73} W. Helfrich, J. Naturfosch. C28 (1973) 693.

\bibitem{pol86}  A. M. Polyakov, Nucl. Phys. B268 (1986) 406.

\bibitem{klein86} H. Kleinert, Phys. Lett. A114 (1986) 263.

\bibitem{forst86} D. Forster, Phys. Lett. A114 (1986) 115.

\bibitem{sw94}   G.K. Savvidy, F.J. Wegner Nucl. Phys. B413 (1994) 605.

\bibitem{des1order} D. Espriu, M. Baig, D.A. Johnston, R.P.K.C. Malmini
                    Phys. A30 (1997) 405.

\bibitem{des2order} D.A. Johnston, R.P.K.C. Malmini, 
                   Nucl. Phys. Proc. Suppl. 53 (1997) 773.

\bibitem {ek82}  T. Eguchi, H. Kawai, Phys. Lett. B114 (1982) 247.


\bibitem{p75}  A. M. Polyakov, Phys. Lett. B59 (1975) 79.

\bibitem{weinberg} S. Weinberg, in {\em General Releativity, an Einstein 
centenary survey} S.W. Hawking, W. Israel (Edts.), 
   Cambridge University Press (1979).


\bibitem{kn90} H. Kawai, N. Ninomiya, Nucl. Phys. B336 (1990) 115.


\bibitem{bbkp96}  P. Bialas, Z. Burda, A. Krzywicki, B. Petersson,
                  Nucl. Phys. B472 (1996) 293.

\bibitem{b96} B. V. de Bakker Phys. Lett. B389  (1996)  238


\bibitem{ant2}  I.~Antoniadis, P.O.~Mazur and E.~Mottola, 
                Phys.~Lett.\  B394  (1997) 49.

\bibitem{am91} M.E. Agishtein, A.A. Migdal, 
               Mod. Phys. Lett. A6 (1991) 1863.

\bibitem{am92} M.E. Agishtein, A.A. Migdal, 
               Nucl. Phys. B385 (1992) 395.

\bibitem{aj92} J. Ambj{\o}rn, J. Jurkiewicz, Phys. Lett. B278 (1992) 42

\bibitem{s27} D.M.Y. Sommerville, Proc. Roy. Soc. Lond. A115 (1927) 103.

\bibitem{d} F. David {\em Simplicial Quantum Gravity and Random
            Lattices},  Proc. Les Houches Sum. Sch. 92 (1992) 679.

\bibitem{r2term} J. Ambj{\o}rn, J. Jurkiewicz, C. F. Kristjansen,
                 Nucl. Phys. B393 (1993) 601.



\bibitem{entropy1} S. Catterall, J. Kogut and R. Renken, 
                   Phys. Rev. Lett 72 (1994) 4062.

\bibitem{entropy2} J. Ambj{\o}rn, J. Jurkiewicz,
                   Phys. Lett. B335 (1994) 355.

\bibitem{entropy3} B. Brugmann, E. Marinari, 
                   Phys. Lett. B349 (1995) 35.

\bibitem{bbcm94} C. Bartocci, U. Bruzzo, M. Carfora, A. Marzuoli,
                 J. Geom. Phys. 18 (1996) 247.



\bibitem{phase_struc2} S. Catterall, J. Kogut, R. Renken


\bibitem{matter1}
 J.~Ambj\o rn, Z.~Burda, J.~Jurkiewicz and C.F.~Kristjansen,
  Phys.~Lett.\  {\bf B297} (1992) 253; Phys.~Rev.\ D48 (1993) 3695;

\bibitem{matter2}
  J.~Ambj\o rn, S.~Bilke, Z.~Burda, J.~Jurkiewicz and B.~Petersson,
  Mod.~Phys.~Lett.\  A9 (1994) 2527;

\bibitem{matter3}
 S.M.~Catterall, J.B.~Kogut and R.L.~Renken,  Nucl.~Phys.\
  {\bf B389} (1993) 601; Nucl.~Phys.\  B422 (1994) 677.


\bibitem{hi95} T. Hotta, T. Izubuchi, J. Nashimura, 
                 Prog. Theor. Phys. 94 (1995) 263.

\bibitem{cat95} S. Catterall, G. Thorleifsson, J. Kogut, R. Renken, 
               Nucl. Phys. B468 (1996) 263.



\bibitem{bia}  P.~Bialas, Z.~Burda and D.~Johnston, Nucl.~Phys.\ 
                B493 (1997) 505.


\bibitem{bb97} P. Bialas, Z. Burda 
               {\em Collapse of 4-D random geometries} (hep-lat/9707028).


\bibitem{bbpt96} P. Bialas, Z. Burda, B. Petersson, J. Tabaczek,
                 Nucl. Phys. B495 (1997) 463.

\bibitem{taba} J. Tabaczek, M. Sc. Thesis, University of Bielefeld 1997

\bibitem{ckr97} S. Catterall,  R. Renken, J. Kogut, (hep-lat/9709007)

\bibitem{wheeler64} J. A. Wheeler, {\em Geometrodynamics and the issue of the
final state}, in {\em Relativity, Groups \& Topology}, eds. B.S. DeWitt and
C.M. DeWitt (Blackie ans Son Ltd., Glasgow, 1964)  317

\bibitem{hawking78}  S.W. Hawking, Nucl. Phys. B144 (1978) 349.

\bibitem{bak94} B. V. de Bakker, Nucl. Phys. Proc. Suppl. 42 (1995) 716.

\bibitem{gv92} M. Gross, S. Varsted,
  {\em Nucl. Phys. } B378 (1992) 367


\bibitem{bi96} P. Bialas, {\em Correlations in fluctuating geometries},
               Nucl. Phys. Proc. Suppl. 53 (1997) 739


\bibitem{ckr94} S. Catterall,  R. Renken, J. Kogut, 
                Phys. Rev. Lett. 72 (1994) 4062

\bibitem{aj3} J. Ambj\o rn and J. Jurkiewicz, Phys. Lett. 
              B335 (1994) 355.

\bibitem{bm95}  B. Br\"{u}gmann and E. Marinari, 
                Phys. Lett. B349 (1995) 35.


\bibitem{cates}  M.E.~Cates, Europhys.~Lett.\  8 (1988) 719. 

\bibitem{k90}    A.~Krzywicki, Phys.~Rev.\  D41 (1990) 3086. 

\bibitem{dav92} F.~David, Nucl.~Phys.\  B368 (1992) 671.


\bibitem{bkt1} V.I. Brezinskii, Zh. Eksp. Teor. Fiz. 59 (1970) 907
               [Soviet Physics, JETP 32 (1971) 493].

\bibitem{bkt2} M. Kosterlitz, D. Thouless, J. Phys. C6 (1973) 1181

\bibitem{jur} 
        J.~Jurkiewicz and A.~Krzywicki, Phys.~Lett.\  B392 (1997) 291.

\bibitem{ant_a} 
 I.~Antoniadis and E.~Mottola, Phys.~Rev.\  D45 (1992) 2013;

\bibitem{ant_b}
 I.~Antoniadis, P.O.~Mazur and E.~Mottola, Nucl.~Phys.\
  B388 (1992) 627.

\bibitem{dav2} 
 F.~David, J.~Jurkiewicz, A.~Krzywicki and B.~Petersson,
  Nucl.~Phys.\  B290 (1987) 218.


\bibitem{haus}
 J.~Ambj\o rn and J.~Jurkiewicz, Nucl.~Phys.\  B451 (1995) 643.

\bibitem{baby4} S. Catterall, J. Kogut, R. Renken, G. Thorleifsson, 
                Phys. Lett. B 366 (1996) 72










\forget{





\bibitem{b97} M. Bowick, {\em Random surfaces and lattice gravity},
              hep-lat/9710005 


\bibitem {NB93} A. Nabutovsky, R. Ben-Av, Commun. Math. Phys. 157 (1993) 93.



\bibitem{scaling4d}  J. Ambj\o rn and J. Jurkiewicz,
                    Nucl. Phys. B451 (1995) 643



\bibitem{comput_ergodic} 
            J. Ambj{\o}rn, J. Jurkiewicz, Phys.Lett. B 345 (1995) 435




\bibitem {ko83}  H. Kawai, Y. Okamoto, Phys. Lett. B130 (1983) 415



\bibitem {bbp97} S. Bilke, Z. Burda, B. Petersson, Phys.Lett. B395 (1997) 4

\bibitem{bibkp97} 
 S.~Bilke, Z.~Burda, A.~Krzywicki and B.~Petersson,
           Nucl.~Phys.\ {\bf B53} (Proc.~Suppl.) (1997) 743.


\bibitem{bas95}  B. De Bakker, {\em Simplicial Quantum Gravity}, PhD Thesis, 
                 ISBN 90-9008425-8



\bibitem{reduced_model} M.J. Bowick, S.M. Catterall, G. Thorleifsson 
                      Nucl. Phys. Proc. Suppl 53 (1997) 753
 




}




 
\end{thebibliography}
